\documentclass[final,3p,times]{elsarticle}
\usepackage[numbers]{natbib}

\newcommand{\osb}{\Omega_{\perp 2}}
\newcommand{\os}{\Omega_{\perp}}
\newcommand{\op}{\Omega_{\parallel}}

\def\fsz{\footnotesize}
\def\ssz{\scriptsize}
\def\tsz{\tiny}
\def\aut#1{#1}

\usepackage{latexsym}
\usepackage{amssymb}
\usepackage[mathscr]{eucal}
\usepackage{epsfig}
\usepackage{lscape}
\usepackage{amssymb}
\usepackage{amsmath}

\def\lfrac#1#2{{#1/#2}}
\def\a{\alpha}

\def\be{\begin{equation}}
\def\ee{\end{equation}}
\def\bea{\begin{eqnarray}}
\def\eea{\end{eqnarray}}
\def\bean{\begin{eqnarray*}}
\def\eean{\end{eqnarray*}}

\def\proof#1. {\par
                      \ifdim\lastskip<15pt
                      \removelastskip\penalty-200
                      \vskip5pt plus3pt minus3pt
                      \fi
                       {\def\a{#1}
                       \ifx\a\empty
                       {\noindent\bf Proof.}
                       \else
                       {\noindent\bf Proof of #1.}
                       \fi}\enspace}

\begin{document}

 \begin{frontmatter}


  \author
  {Hagen Kleinert
  }
   \ead{h.k@fu-berlin.de}
  \address
  {Institut for Theoretical Physics, Free University Berlin, Berlin, Germany}


\title{Converting Divergent Weak-Coupling  into
Exponentially Fast \\Convergent Strong-Coupling Expansions
}

\begin{abstract}
With the help of 
a simple 
variational procedure
it is possible
to convert the partial 
sums 
of order $N$
of many divergent series
expansions
$f(g)=\sum_{n=0}^\infty a_n g^n$ 
 into
partial sums
$\sum_{n=0}^N b_n g^{- \omega n}$, where $0<\omega<1$
is a parameter that parametrizes 
the approach to the large-$g$ limit. The
latter are partial sums of
a strong-coupling expansion 
of $f(g)$ 
which 
converge
against $f(g)$
   for $g$ {\em outside\/}
a certain divergence radius. The
error decreases exponentially
fast for large $N$, like $e^{-{\rm const.}\times  N^{1-\omega}}$.
We present a review of the method and various
applications.

\end{abstract}


\end{frontmatter}


\def\comment#1{}

\input gbf.sty

\section{Introduction}

Variational techniques have a long history in
theoretical physics.
On the one hand, they serve to find equations of motion
from the extrema of actions.
On the other hand
they help finding
approximate solutions of physical problems
by extremizing energies.
In quantum mechanics, the Rayleigh-Ritz variational  principle
according to which the ground state energy
of a system is bounded above by the inequality
\begin{eqnarray}
E_0 \leq \int d^3x \,\psi^*(x)\hat H \psi(x)
\label{@}\end{eqnarray}
has yielded many useful results.
In many-body physics, the
Hartree-Fock method
has helped understanding electrons in metals
and nuclear matter.
In quantum field theory the effective action approach
\cite{EFFAC}
has contributed greatly to
the theory of phase transitions.
In particular the higher effective actions
pioneered by Dominicis \cite{CDED}.

A variational method was very useful
in solving functional integrals
 of complicated quantum statistical systems, for instance the polaron problem
\cite{FEY1}.
Here another inequality plays an important role, the Jensen-Peierls inequality,
according
to which the expectation value  of an exponential of a
functional
of a functional
is at least as large as the exponential of the expectation value itself:
\begin{eqnarray}
\langle e^{-\cal O}\rangle
\geq e^{-
\langle
{\cal O}\rangle
}        .
\label{@}\end{eqnarray}

This technique was extended in 1986
to find approximate
solutions for the functional integrals of many
other quantum mechanical systems \cite{FK}.

An important progress was reached in 1993 by finding a way of
applying the technique to arbitrarily high order \cite{KLEX}.
The technique was developed furher
in the textbook
\cite{PI}.
This made it possible to perform the approximate calculation
to any desired degree of accuracy.
In contrast to the higher effective action approach,
the treatment converged
exponentially fast
also in the strong-coupling limit \cite{KBEY}.

The zero-temperature version of
this technique led to
a new solution
of
an old problem in mathematical physics,
that
the results of many
calculations
can be given only in the form of divergent weak-coupling expansions.
For instance, the energy eigenvalues
$
E $ of a
Schr\"odinger equation
of a point particle of mass $m$
\begin{eqnarray}
\left[- \hbar ^2\frac{\partial ^2}{2M}
+V({\bf x})\right] \psi({x})=E\psi ({x})
\label{@}\end{eqnarray}
moving in a
three-dimensional
potential
\begin{equation}
V(x)=\frac{ \omega ^2}{2}x^2+{g}x^4
\label{@OT}\end{equation}
can be given  as a series
in $g/ \omega ^3$
\begin{eqnarray}
E= \omega \left[
\sum_{n=0}^N a_n\left( \frac{g} \omega \right)^n\right].
\label{@}\end{eqnarray}
The  coefficients $a_n$
grow exponentially fast with $n$.
The series has a zero radius of convergence.
For the ground state
it reads
\begin{equation}
E= \omega \left[\frac{1}{2}+\frac{3}{4}\frac{g}{4 \omega ^3}-  \frac{21}{8}
\left(\frac{g}{  4 \omega ^3}\right)^2
+\frac{333}{16}
\left(\frac{g}{  4 \omega ^3}\right)^3
+\dots
  \right] .
\label{@EEQ}\end{equation}

There exist similar divergent expansions
for critical exponents
which may be calculated from
weak-coupling  expansions
of quantum field theories
and are experimentally measurable
near second-order phase transitions.
One of these is the exponents
$ \alpha $ which determines
the behavior of the specific heat
of superfluid helium
near the phase transition
to the normal fluid.
%
%
It has been measured
with extreme
accuracy in a recent satellite experiment \cite{LIPA}.
\begin{figure}
\centerline{
\epsfxsize=6cm \epsfbox{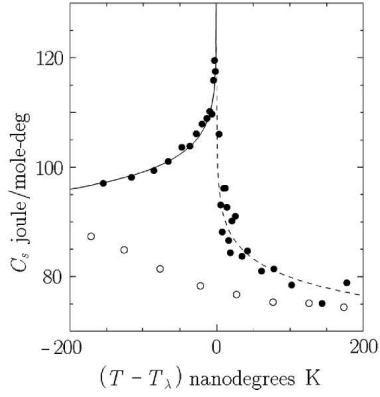}
}
\caption{\label{wofq1} Experimental data of space shuttle experiment by Lipa et al. \cite{LIPA}.
}
\label{EXP}\end{figure}
The result agrees very well with the
value of the series for $ \alpha $ as a power series in $g/m$
in the strong-coupling limit $m\rightarrow 0$ \cite{HK7L}

In many more physical examples
the
properties
are found by
evaluating divergent weak-coupling series
in the strong coupling limit.


In this lecture I shall present the main ideas
and  sketch a few applications
of  {\em Variational Perturbation Theory\/}.

\section{Quantum Mechanical Example}
\label{QME}
In order to illustrate the method
let us obtain the strong-coupling value of the
ground state energy
(\ref{@EEQ}).
We introduce a dummy variational parameter
by the substitution
\begin{equation}
 \omega \rightarrow
 \sqrt{ \Omega ^2+( \omega ^2- \Omega ^2)}\equiv
 \sqrt{ \Omega ^2+gr},
\label{@SR0}\end{equation}
where $r$  is short for
\begin{equation}
r\equiv  ( \omega ^2- \Omega ^2)/g.
\label{@}\end{equation}
This substitution
does not change the
partial sums of series
(\ref{@EEQ}):
\begin{equation}
E^{N}= \omega \sum_{n=0}^N
a_n
\left(\frac{g}{ 4\omega ^3}\right)^n
\label{@EEQp}\end{equation}
for any order $N$.
If we, however,
re-expand these partial sums in powers of $g$ at fixed $r$
up to order $N$,
and substitute
at the end $r$ by $( \omega ^2- \Omega ^2)/g$,
we obtain
new partial sums
\begin{equation}
W^{N}= \Omega \sum_{n=0}^N
a'_n
\left(\frac{g}{ \Omega ^3}\right)^n .
\label{@}\end{equation}
In contrast
to $E^N$,
these {\em do depend\/} on the variational parameter
$ \Omega $.
For higher and higher
orders, the $ \Omega $-dependence
has an increasing valley
where the dependence is very weak.
It can be found analytically
by setting the first derivative equal to zero, or, if this equation
has  no solution,
by setting the second derivative
equal to zero.
One may view this as a manifestation
of a {\em principle
of
minimal sensitivity\/} \cite{STE}.
The plots are shown in
Fig. \ref{Fig1}
for odd $N$ and even $N$.
\comment{
\begin{figure}[ptbh]
\setlength{\unitlength}{.5cm}
$\begin{array}{r}
~\\~\\~\\~\\
~\\~\\~\\~\\[-12mm]
\hspace{-5.8cm}
\input wu.tps
\\[-3cm]
\end{array} $    \\[1.5cm]
$\begin{array}{r}
~\\~\\~\\~\\
~\\~\\~\\~\\[-6mm]
\begin{centering}
\IncludeEpsImg{101.23mm}{101.23mm}{.5000}{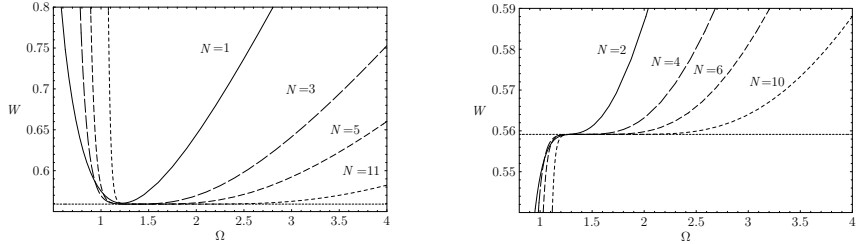}
\end{centering}
\\[-2.9cm]
\end{array} $
\caption[Typical $ \Omega $-dependence of $N$th
approximations $W_N$
 at $T=0$
]{Typical $ \Omega $-dependence of $N$th
approximations $W_N$
 at $T=0$ for increasing orders $N$.
The coupling constant has the value $g/4=0.1$. The dashed
horizontal
line indicates the exact energy.}
\label{Fig1}
\end{figure}%
}
\begin{figure}[ptbh]
\setlength{\unitlength}{.5cm}
$\begin{array}{r}
~\\[13mm]
\hspace{-5.1cm}
\centerline{\phantom{XXXXXXXXXXXXXXXXXXXXXXX}\input wug.tps }
\\[-3.5cm]
\end{array} $    \\[1.5cm]
$\begin{array}{r}
~\\~\\~\\~\\
~\\~\\~\\~\\[-6mm]
\hspace{-6cm}
\\[-2.6cm]
\end{array} $
\caption[Typical $ \Omega $-dependence of $N$th
approximations $W_N$
 at $T=0$
]{Typical $ \Omega $-dependence of $N$th
approximations $W_N$
 at $T=0$ for increasing orders $N$.
The coupling constant has the value $g/4=0.1$. The dashed
horizontal
line indicates the exact energy.}
\label{Fig1}
\end{figure}%

Even to lowest order,
the result is surprisingly accurate.
For $N=1$, the energy $ E^N$ we has the linear
dependence
\begin{equation}
E^1= \omega \left(\frac{1}{2}+\frac{3}{16}\frac{g}{ \omega ^3}\right).
\label{@}\end{equation}
After the replacement
(\ref{@SR0}) and the reexpansion up to power $g$   at fixed $r$
we find
\begin{equation}
W^1= \Omega \left(
 \frac{ 1 }{4}+\frac{ \omega ^2}{4 \Omega }+\frac{3}{16}\frac{g}{ \Omega ^4}
\right).
\label{@}\end{equation}
In the strong-coupling limit, the minimum lies at
 $\Omega \approx c (g/4)^{1/3}$ where
$c$ is some constant and the energy behaves like
\begin{equation}
W^1\approx \left(\frac{g}{4}\right)^{1/3}\displaystyle
\left(\frac{c}{4}+\frac{3}{4c^2}\right).
\label{@}\end{equation}
The minimum lies at
$c= {6}^{1/3}$
where
 $W^1
\approx (g/4)^{1/3}\displaystyle\left({3}/{4}\right)^{4/3}
{\approx (g/4)^{1/3}\times0.681420}$.
The
treatment can easily be
extended to 40 digits \cite{JK}
starting out like
$E^1{\!=(g/4)^{1/3}
\times0.667\,986\,259\,\dots}~$.

\begin{figure}[h]
\centerline{
\setlength{\unitlength}{.5cm}
\begin{picture}(10.0,6.5)
\put(-1,0){\makebox(11,6.5){\epsfxsize=5cm \epsfbox{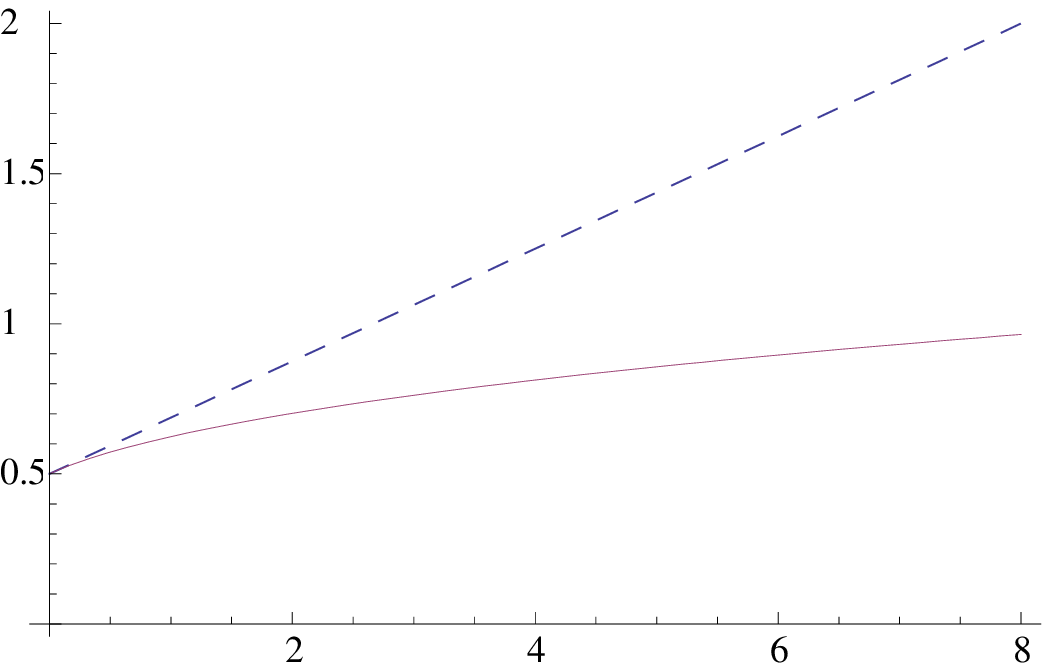}}}
\put(0,3.9){$$}
\put(9.8,.36){$\scriptstyle g$}
\put(9.7,6.1){ $\scriptstyle E^1$}
\put(9.7,3.2){$\scriptstyle{\rm min}~W^1$}
\end{picture}}
\caption[
First-order perturbative energy
$E^1$ and the variational-perturbative minimum of
$W^1$
]{First-order perturbative energy
$E^1$ and the variational-perturbative minimum of
$W^1$. The exact result follows
closely
the curve min $W^1$.}
\label{@figW1}\end{figure}

The result is shown in
for $g/4=0.1$ in
Fig. \ref{@figW1}.
If we
plot the minimum
as a function of $g$
we obtain the curve shown in Fig. \ref{@figW1}.
The curve has the asymptotic behavior $(g/4)^{1/3}\times 0.68142 $.
This grows with the {\em exact\/} power
of $g$ and has a coefficient
that differs only slightly from the
accurate value
$0.667\,986\,259\dots$
found by other approximation procedures
\cite{EJW}.

\begin{figure}[tbh]
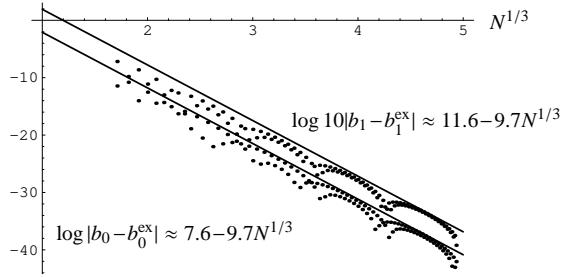

\centerline{\input plaal1n.tps}
\caption
[Asymptotic coefficients $ b _0$ and $ b _1$ as a function of the order $N$]
{Asymptotic coefficients $ b _0$ and $ b _1$ of $ W^N$ as a function of the order $N$.}
\label{CONV}\end{figure}

\begin{figure}[tbh]
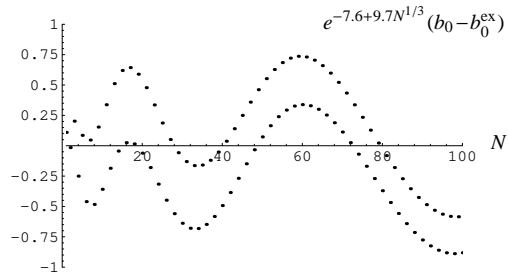

\centerline{ \input  apprsc.tps }
\caption[Oscillations of the strong-coupling coefficient $ b_0$]
{Oscillations
of the strong-coupling coefficient $ b_0$.
 }
\label{Osc}\end{figure}

The convergence of the approximations is exponential
as was shown in Refs.
\cite{KJ2,GKS,HKSC}
using the technique of order-dependent mapping
\cite{ODM}. 
If the asymptotic behavior of $E^N(g)$ and
its variational approximation  $W^N(g)$
are parametrized by
\begin{eqnarray}
W^N ( g ) & = & g^{\frac{1}{3}} \left\{ b_0 + b_1 \, g^{- \frac{2}{3}} +
b_2 \, g^{- \frac{4}{3}} + \ldots \right\},
\end{eqnarray}
the coefficients $b_0$ and $b_1$ converge with $N$ as shown in Fig. \ref{CONV}.
The approach is oscillatory (see Fig. \ref{Osc}).

\section{Quantum Field Theory and Critical Behavior}

When trying to apply the same procedure to
quantum field theory,
the above procedure needs some
important modification
caused by
the fact that
the scaling dimensions
of fields are no longer equal
to the naive dimensions
but {\em anomalous\/}.
This causes the
principle of minimal sensitivity
to fail
\cite{HK1}.
The adaption
of the variational procedure
was done 
in the textbook~\cite{KS}. Let us briefly summarize it
using an important class of field theories.

The energy is
an
O($n$)-symmetric coupling
functional of a $n$-component field
$\phi_0$ in $D$ dimensions
\begin{eqnarray}
E [ \phi_0 \, ] & = & \int d^D x \left\{ \frac{1}{2} \left[ \partial
\phi_0 ( {\bf x} ) \right]^2 + \frac{m_0^2}{2} \, \phi_0
( {\bf x} )^2
+ \frac{g_0}{4!} \left[ \phi_0 ( {\bf x} ) ^2 \right]^2 \right\}
,
\end{eqnarray}
where the parameters depend on the distance
of the
 temperature
 from the critical value
$T_c$:
\begin{eqnarray}
m_0^2 & = & {\cal O} \left( ( T - T_c )^1 \right) \, , \hspace*{1cm}
g_0 = {\cal O} \left( ( T - T_c )^0 \right)  \nonumber
\end{eqnarray}
The important critical behavior is seen
in the
correlation function which have the limiting form
\begin{eqnarray}
\left\langle \phi_i ( {\bf x} ) \, \phi_j ( {\bf x'} )
\right\rangle \sim \frac{e^{ - \lfrac{|{\bf x} -
{\bf x'}|}{\xi ( T )}}}{|{\bf x} - {\bf x'}|^{D - 2 + \eta}}.
\end{eqnarray}
where
$\eta$
is the anomalous field dimension, and $\xi$ is the
coherence length
which diverges near $T_c$ like
$\xi ( T ) \sim ( T - T_c )^{-\nu}$.

\subsection{Critical Behavior in $D- \epsilon $ Dimensions}

The field fluctuations cause divergencies which can be removed
by a renormalization of field, mass and coupling constant
to $\phi$, $m$, and $g$. This is most elegantly done
by assuming the dimension
of spacetime to be $D=4- \epsilon $, in which case
the renormalization factor are
\begin{eqnarray}
g_0 & = & Z_g ( g , \epsilon ) \, Z_{\phi} ( g , \epsilon )^{-2} \,
\mu^{\epsilon} \, g ,\label{@g0g} \\
m_0^2 & = & Z_m ( g , \epsilon ) \, Z_{\phi} ( g , \epsilon )^{-1}\, m^2,
 \label{@m0m}\\
\phi^2_0 & = & Z_{\phi} ( g , \epsilon ) \, \phi^2.\label{@phi0}
\end{eqnarray}
The factors have weak-coupling expansions:
\begin{eqnarray}
Z_g ( g , \epsilon ) & = & 1 + \frac{n + 8}{3 \epsilon} \, g +
\left\{ \frac{(n+8)^2}{9 \epsilon^2} - \frac{5 n + 22}{9 \epsilon} \right\}
g^2 + \ldots ~, \\
Z_{\phi} ( g , \epsilon ) & = & 1 - \frac{n + 2}{36 \epsilon}
\, g^2 + \ldots~, \\
Z_{m} ( g , \epsilon ) & = & 1 + \frac{n + 2}{3 \epsilon} \, g + \left\{
\frac{(n + 2)(n + 5)}{9 \epsilon^2} - \frac{n + 2}{6 \epsilon} \right\}
g^2 + \ldots~. \nonumber
\end{eqnarray}

The dependence of these on the scale
parameter $\mu$ defines the renormalization group functions
\begin{eqnarray}
\beta ( g , \epsilon ) & = & \mu \, \left. \frac{d g}{d \mu}\right|_0 ~~~ =
- \epsilon \left\{ \frac{\partial}{\partial g} \ln \left[ g Z_g ( g ,
\epsilon ) Z_{\phi} ( g , \epsilon )^{-2} \right] \right\}^{-1} , \\
\gamma_m ( g ) & = & \frac{\mu}{m} \, \left. \frac{d m}{d \mu} \right|_0
~= - \frac{\beta ( g , \epsilon )}{2} \, \frac{\partial}{\partial g} \, \ln
\left[ Z_m ( g , \epsilon ) Z_{\phi} ( g , \epsilon )^{-1}
\right], \\
\gamma ( g ) & = & - \frac{\mu}{\phi} \, \left. \frac{d \phi}{d \mu} \right|_0
\,= \frac{\beta ( g , \epsilon )}{2} \, \frac{\partial}{\partial g} \, \ln
Z_{\phi} ( g , \epsilon )\label{@gamma0} .
\end{eqnarray}
At the phase transition
$g_0$ goes to the strong-coupling
limit $g_0\rightarrow \infty$.
In this limit the renormalized coupling $g$ tends to a constant $g^*$,
called the fixed point of the theory.

From the
renormalization group functions
in the strong-coupling limit
one finds the physical observables
at the critical point
\begin{eqnarray}
\eta & = & 2 \gamma ( g^* )  = \frac{n + 2}{2 ( n + 8 )^2} \, \epsilon^2
+ \, \ldots~,\label{@etae} \\
\nu & = & \frac{1}{2 \left[ 1 - \gamma_m ( g^* ) \right]} = \frac{1}{2}
+ \frac{n + 2}{4 ( n + 8 )} \epsilon  + \frac{(n + 2) ( n + 3) ( n + 20 )}{8
( n + 8 )^3} \, \epsilon^2 + \ldots~, \\
\omega & = & \beta' ( g^* , \epsilon ) = \epsilon -
\frac{3 ( 3 n + 14 )}{( n + 8 )^2} \, \epsilon^2 + \, \ldots ~.
\end{eqnarray}
The quantity $ \epsilon $ is the so-called {\em anomalous dimension\/}
of the field $\phi(x)$.

The $ \epsilon$-expansions
are divergent and are typically
evaluated at the  physical value $ \epsilon =1$
where $D=3$ by various resummation
procedures
\cite{Resumm}.

In variational perturbation theory the procedure is different.
One rewrites the power series
of Eq.~(\ref{@g0g})
as
of
the renormalized coupling  $g_0$:
\begin{eqnarray}
g ( g_0 ) & = & g_0 - \frac{n + 8}{3 \epsilon} \, g_0^2 +
\left\{ \frac{( n + 8 )^2}{9 \epsilon^2} + \frac{9 n + 42}{18 \epsilon}
\right\}\,g^3_0 + \ldots~.
\label{@serg}
\end{eqnarray}
For the dependence
of the renormalized mass on the bare coupling one finds from
Eq.~(\ref{@m0m})
\begin{eqnarray}
\frac{m^2(g_0)}{m^2_0} & = & 1 - \frac{n + 2}{3 \epsilon} \, g_0 +
\left\{ \frac{(n + 2)(n + 5)}{9 \epsilon^2} + \frac{5 (n +2)}{36 \epsilon}
\right\} \, g_0^2 + \ldots ~.
 \label{@serm}
\end{eqnarray}
and
for the anomalous dimension
from
Eq.~(\ref{@phi0}),
(\ref{@gamma0}), and
(\ref{@etae}):
\begin{equation}
 \eta (g_0)=\frac{n+2}{18}g_0^2-\frac{(n+2)(n+8)}{216}\left(1-\frac{8}{ \epsilon }\right)
g_0^2+\dots~.
\label{@}\end{equation}

Due to the anomalous dimension $ \eta \neq 0$,
the dependence of the approximations
on the variational parameter
develops no longer a
horizontal flat valley (see Appendix A).
Instead,
the valley turns out to have
a slope
which can only be
removed by
introducing
another parameter $q$
in to substitution rule
(\ref{@SR0}).
We rewrite the series
in $g$ as a series in $g/ \kappa ^q$,
 and replace
$ \kappa $ by
\begin{equation}
  \kappa \rightarrow
\sqrt{ K ^2+ (\kappa ^2-K^2)}\equiv \sqrt{ K ^2+gr},
\label{@SR1}\end{equation}
by
\begin{equation}
r=( \kappa ^2-K^2)/g.
\label{@}\end{equation}

As before we re-expand the partial sums of the series
in powers of $g$
 at
fixed $r$ up to power $g^N$
to obtain $W^N$.
After this we set $ \kappa \rightarrow 1$ and
plot $W^N$ as a function of $K$.
By varying $q$ we can make the valley of minimal $ K $-dependence
horizontal
\cite{HK1}.

The asymptotic behavior
of the variational parameter $K(g_0)$
and the critical exponent
as a function  of $g_0$, called  generically $f(g_0)$,
is now  in general
\begin{eqnarray}
 K  ( g _0) & = & g^{\lfrac{1}{q}} \left\{ c_0 + c_1 \, g_0^{- \lfrac{2}{q}} +
c_2 \, g_0^{- \lfrac{4}{q}} + \ldots \right\} \nonumber \\
f ( g_0 ) & = & g^{\lfrac{p}{q}} \left\{ b_0 + b_1 \, g_0^{- \lfrac{2}{q}} +
b_2 \, g_0^{- \lfrac{4}{q}} + \ldots \right\},
\end{eqnarray}
\begin{figure}[bt]
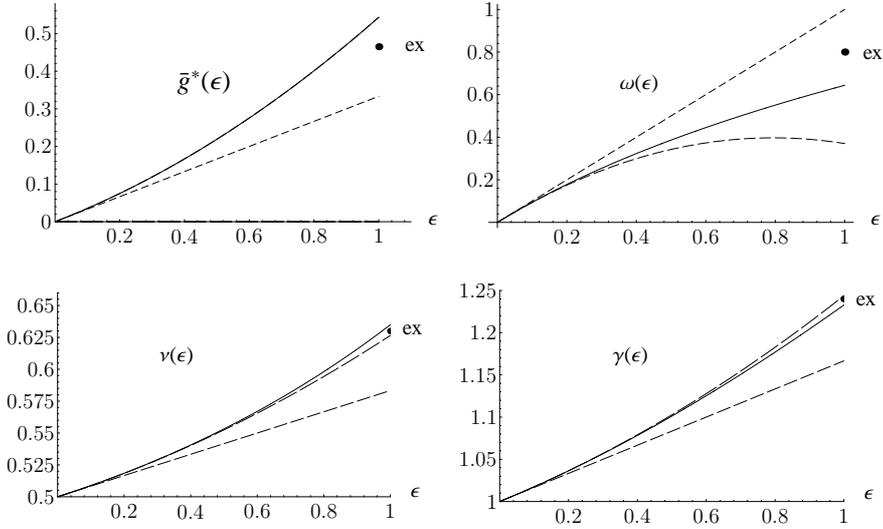

\vspace{-.5cm}
\hspace{-9.1cm}
\centerline{
\phantom{xxxxxxxxxxxxxxxxxxx}
\phantom{xxxxxxxxxxxxxxxxxxx}
\raisebox{0mm}{
\input gr.tps ~~~~~ $\!\!\!$
\input omf.tps  }} ~\\$\!\!\!$
\phantom{xxx.}\hspace{-9.7cm}
\centerline{
\phantom{xxxxxxxxxxxxxxxxxxx}
\phantom{xxxxxxxxxxxxxxxxxxx}
\raisebox{0mm}{
\input nucr.tps ~~~\,
~~\input gamcr.tps } } ~\\[0cm]
\caption
{
Strong-coupling values of
the renormalization group functions for $n=1$
(the so-called
Ising universality class).
}
\label{omf}\end{figure}

In the proof of the exponentially fast convergence
in Refs.
\cite{KJ2,GKS,HKSC}.
it was shown that the  approach of the correct result proceeds as a function of the
highest order $L$  of the partial sum as
$e^{-c L^{1- 2/q }}$.

In this way we find
from (\ref{@serg})
the strong-coupling behavior  \cite{rem1}
\begin{eqnarray}
g ( g_0 ) & = &
 g^* + b_1\, g_0^{- \frac{\omega}{\epsilon}} + \ldots,
\end{eqnarray}
The exponent
$ \omega $
 is the
 famous Wegner exponent
\cite{Wegner}.
Further we find
from (\ref{@serm})
\begin{eqnarray}
\frac{m^2(g_0)}{m^2_0}
& = & b_0 \, g_0^{- \frac{2}{\epsilon} \gamma^*_m} + \ldots~,
\end{eqnarray}
where the parameter
$ \omega $ and $\gamma^*_m$
are found from the strong-coupling limits
\begin{eqnarray}
\frac{ \omega }{ \epsilon }
=-1-g_0\left[ \frac{g''(g_0)}{g'(g_0)}\right] _{g_0\rightarrow \infty},~~~~~\gamma^*_m  =  -
\frac{\epsilon}{2} \left[ \, \frac{d \ln m^2 ( g_0 ) / m^2_0}{d \ln g_0} \right]_
{g_0 \rightarrow \infty}
  .~
\end{eqnarray}
This parameter
determines also the divergence of the
coherence length in the critical behavior
$\xi ( T ) \sim ( T - T_c )^{-\nu}$:
\begin{equation}
\nu  =1/(2-\gamma^*_m ).
\label{@}\end{equation}

The results are
\begin{equation}
\omega  = \frac{\epsilon}{2 \sqrt{1 + \frac{3
( 3 n + 14 )\epsilon}{( n + 8 )^2}} - 1},~~~~~~~~~
\nu  =  \frac{1 + \frac{5}{2(n + 8)} \epsilon}{2 \left[
1 - \frac{n -3}{2 ( n + 8 )} \, \epsilon - \frac{3 ( n + 2)(3 n + 14)}{2
(n+8)^3} \epsilon^2 \right] }.
\label{@}\end{equation}
They are plotted in Fig. \ref{omf} as a function of $ \epsilon $.

Instead of an expansion in $D=4- \epsilon $
dimensions on may also treat expansions obtained by Nickel \cite{NICK}
directly in $D=3$ dimensions.

\subsection{Three-Dimensional Treatment}

If one plots the strong-coupling limits
of the series obtained from the partial sums of order $L$
as a function of
$x(L)=e^{-cL^{1- \omega }}$ to account for the
theoretical approach to the asymptotic limit,
one finds for  various $n$ \cite{rem2}:
\begin{figure}[tbhp]
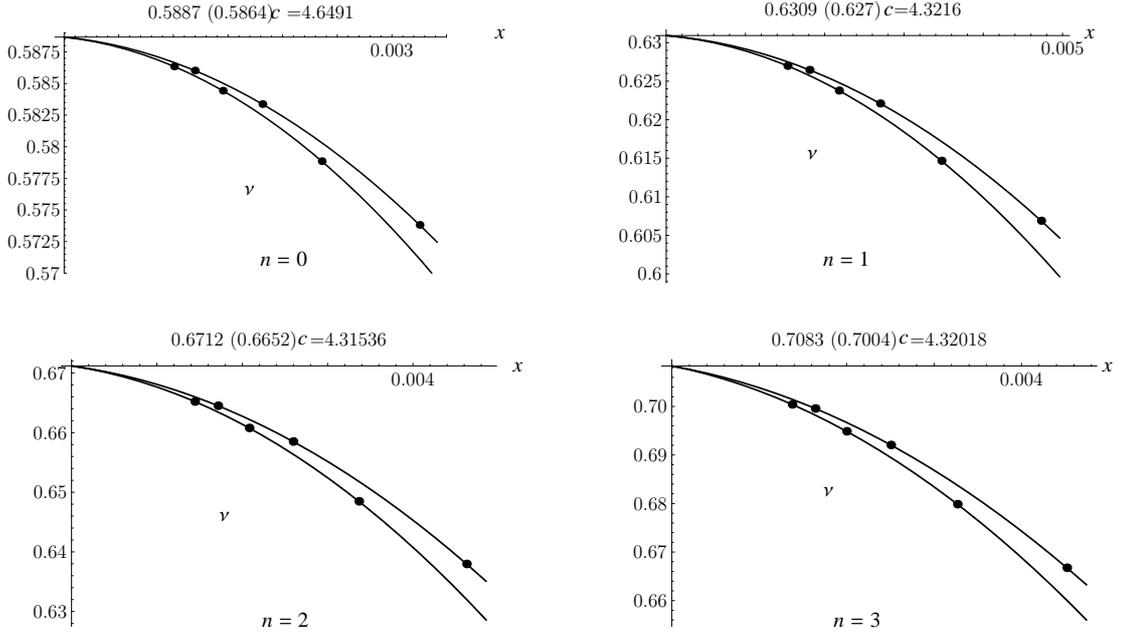

\unitlength.1mm
\unitlength0.354mm
\unitlength0.47mm
~\\[0cm]
\phantom{xxx}
\centerline{\hspace{-2cm}
\raisebox{0mm}{ \input crnu0g.tps }
}
    \vspace{4.3cm}
 \unitlength1mm
\caption
{
Strong-coupling values
for the critical exponent $ \nu ^{-1}(x)$
as a function of
$x(L)=e^{-cL^{1- \omega }}$
}
\label{nu0}\end{figure}%

\begin{figure}[h]
\centerline{
\setlength{\unitlength}{.5cm}
\begin{picture}(10.0,6.5)
\put(-1,0){\makebox(11,6.5){\epsfxsize=12cm \epsfbox{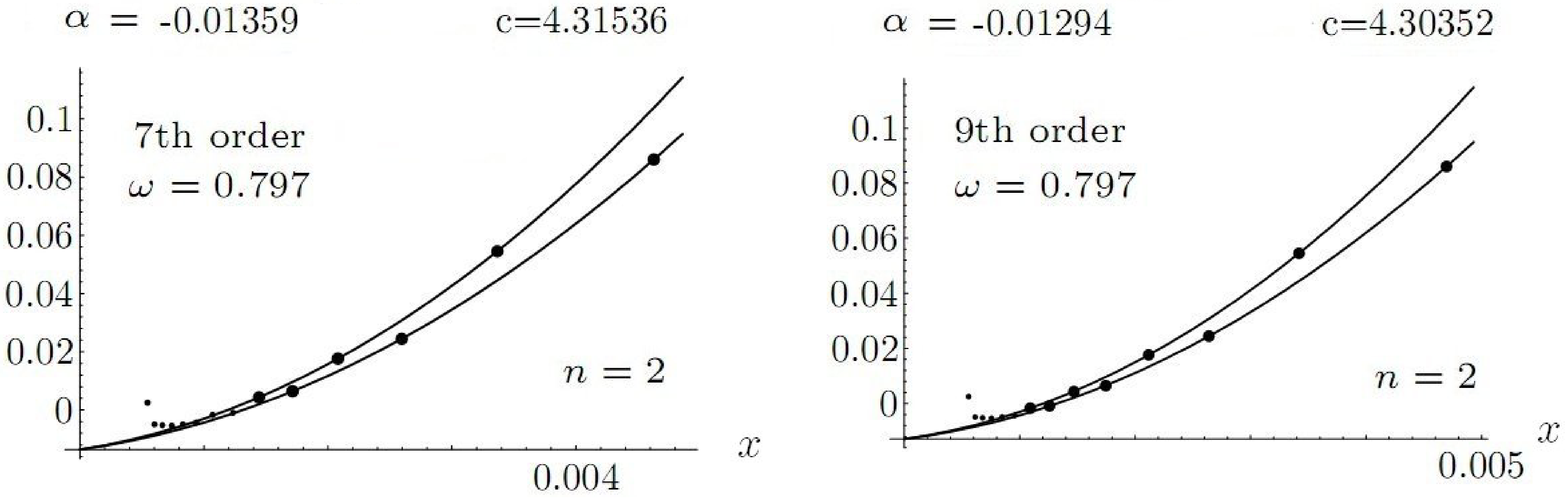}}}
\put(0,3.9){$$}
\end{picture}}
\caption[Strong-coupling limits of $ \alpha $
as a function of $x=e^{-cL^{1- \omega }}$
for 7th and 9th order in perturbation theory
]{Strong-coupling limits of $ \alpha $
as a function of $x=e^{-cL^{1- \omega }}$
for 7th and 9th order in perturbation theory.
The latter limit $ \alpha \approx-0.0129$ agrees well with
the satellite experiment \cite{LIPA}.
}
\label{9th}\end{figure}

\begin{figure}[tbhp]
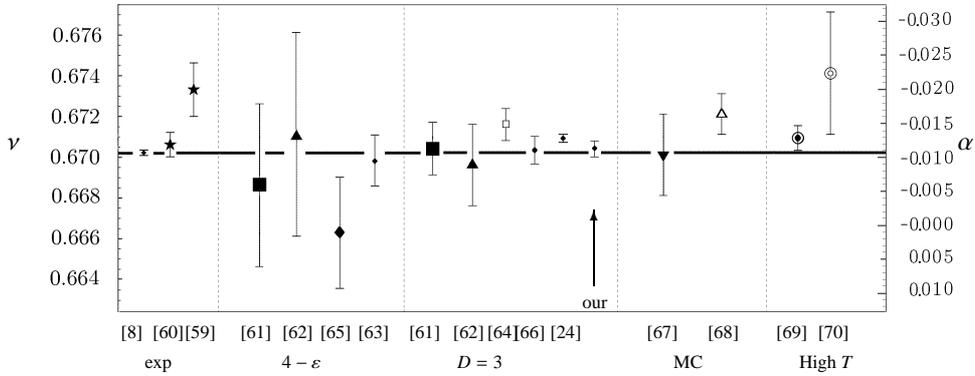

\hspace{-4.6cm}
\centerline{
\input nu2my.tps }
\caption[]{Survey of experimental and theoretical values for $ \alpha $.
The latter come from resummed perturbation expansions
of $\phi^4$-theory in $4- \varepsilon $ dimensions, in three
dimensions, and
 from high-temperature expansions of XY-models on a lattice.
The sources are indicated below.}
\label{EXPC}\end{figure}

For the critical exponent $ \alpha $ characterizing the
behavior of the specific heat $C\approx|T-T_c|^{- \alpha }$ of superfluid
helium
near the critical temperature $T_c$,
the strong-coupling limit is
\cite{HKSC}.
\begin{equation}
 \alpha \approx2-3\times 0.6712\approx-0.0136.
\label{@}\end{equation}
If we extrapolate the
 asymptotic behavior
 expansion coefficients
of $ \nu $ up to the 9th order according using the theoretically known
large-order behavior
this result can be improved to
$ \alpha \approx-0.0129$
\cite{HK7L}
 (see Fig. \ref{9th}).
This value agrees perfectly with the space shuttle value \cite{LIPA}
$ \alpha =-0.01285\pm0.00038$.
The experimental result extracted from Fig. \ref{EXP}
and the various theoretical
numbers obtained from
 the
divergent
perturbation series for $ \alpha $ are summarized in Fig.
 \ref{EXPC}.

 \section{Shift of the Critical Temperature
in Bose-Einstein Condensate by Repulsive Interaction}
\label{BECa}
A free Bose gas
condenses
at a critical temperature
\begin{equation}
T_c^{(0)} =  \frac{2 \pi}{ M} \left[\frac{n}{\zeta(3/2)}
\right]^\frac{2}{3}\;,
\label{T0}
\end{equation}
where $n$
is the particle density.
A small relative shift
of  $T_c$ with respect
to $T_c^{(0)}$ can be calculated from the general formula
\begin{equation}
\frac{ \Delta T_c}{T_c^{(0)}}
=-\frac{2}{3}\frac{ \Delta n}{n^{(0)}},
\label{@shift0}\end{equation}
where $n^{(0)}$ is the particle density in the free condensate
and $ \Delta n$ its change
at  $T_c$
caused by a small
repulsive point interaction
parametrized by an $s$-wave  scattering length $a$.
For small $a$, this behaves like
  \cite{gordon,second}
\begin{equation}
\frac{ \Delta T_c}{T_c^{(0)}}= c_1 a
n^{1/3} + [ c_2^{\prime} \ln(a n^{1/3}) +c_2 ] a^2
n^{2/3} + {\cal O} (a^3 n).
\label{@c1}\end{equation}
where
$c'_2=-64 \pi \zeta(1/2)/3
\zeta(3/2)^{5/3}\simeq 19.7518$
can be calculated perturbatively,
whereas
$c_1$ and $c_2$
require nonperturbative
techniques
since infrared divergences
at $T_c$ make them
 basically strong-coupling results.
The standard technique
to reach this regime
 is based  on  a
resummation of perturbation expansions
using  the renormalization group
 \cite{zj,KS}, first applied in
this context by Ref.~\cite{st}.

Using quantum field theory,
the temperature shift
can be found from  the formula
\begin{equation}  \!\!\!\!\!\!\!\!\!\!\!
\frac{ \Delta T_c}{T_c^{(0)}}
\approx-\frac{2}{3}\frac{MT_c^{(0)}}{n}
\left\langle { \Delta \phi^2}\right\rangle
=-\frac{4\pi}{3}\frac{(MT_c^{(0)})^2}{n}
4!\left\langle \frac{ \Delta \phi^2}{u}\right\rangle\,a
=-\frac{4\pi}{3}(2\pi)^2\frac{1}{[\zeta(3/2)]^{4/3}}
 4!
\left\langle\frac{ \Delta \phi^2}{u}\right\rangle \,an^{1/3},
\label{@}\end{equation}
corresponding in Eq.~(\ref{@c1}) to
\begin{equation} \!\!\!\!\!\!\!\!
 c_1 \approx-1103.09
\left\langle\frac{ \Delta \phi^2}{u}\right\rangle
 .
\label{@NumF}\end{equation}
\begin{figure}[b]
\centerline{
\epsfxsize=6cm \epsfbox{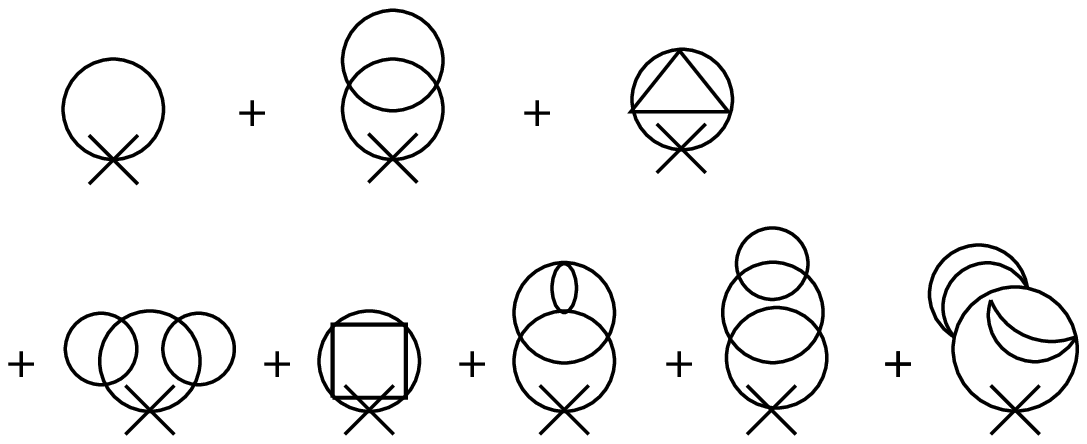}
}
~~\\[-.7cm]
\caption{Diagrams contributing to the
expectation value
 $\langle \phi^2 \rangle$.
}
\label{@FGS}\end{figure}

A calculation
of the Feynman diagrams in Fig.~\ref{@FGS}
yields  the following
 five-loop perturbation expansion  for the expectation value
$\langle \phi^2/u\rangle $ \cite{ramospr,braat}
\begin{eqnarray}
\!\!\!\!\!\!\!\!\!\!\!\!\!\!
\left\langle\frac{ \phi^2}{u}\right\rangle
&=& F\left({u}\right)\equiv -\frac{N  }{4\,\pi } \frac{m}{u}-
  a_2\frac{N\,
     \left( 2 + N \right) }{18\,
     {(4\pi)}^3}
\frac{u}{m}
 +
  a_3\frac{N\,
     \left( 16 + 10\,N + N^2 \right) }{108\,{(4\pi)}^5}
\left(\frac{u}{m}\right)^2
\nonumber \\&&
   -\left[ ~~ a_{41}
\frac{N( 2 + N )^2}
{324 \,(4\pi)^7}
 +
  a_{42}
\frac{N\,
     \left( 40 + 32\,N + 8\,N^2 +
       N^3 \right) }{648\,
     {(4\pi)}^7}
 +
  a_{43}
\frac{N\,     \left( 44 + 32\,N + 5\,N^2 \right)       }
{324\,           (4\pi)^7}
   \right.\nonumber \\&&\left.
 ~~~+
  a_{44}
\frac{N\,
     {\left( 2 + N \right) }^2}
     {324\,{(4\pi)}^7}
 +
  a_{45}\frac{N\,
     \left( 44 + 32\,N + 5\,N^2 \right)
       \,u^4}{324\,m^3\,
     {(4\pi)}^7}
\right]
\left(\frac{u}{m}\right)^ 3
+\dots.
\label{@eq9}\end{eqnarray}
 where
$a_2\equiv
\log (4/3)/2\approx0.143841 $
and the other constants are only known numerically
\cite{braatexp}:
\begin{eqnarray} \!\!\!
\begin{tabular}{ll}
$a_3~ = 0.642144,~
a_{41} = -0.115069,~
a_{42} = 3.128107,~
a_{43} = 1.63,~
a_{44} = -0.624638,~
a_{45} = 2.39.$
 \end{tabular}
\label{@10}\end{eqnarray}
Writing the above expansion  up to the $L$th term
 as
$F_L(u)=\Sigma_{l=-1}^Lf_l(u/4\pi m)^l$, the
expansion coefficients for the relevant number of components
 $N=2$
are
\cite{braatexp}:
\begin{eqnarray}  \!\!\!\!\!\!\!\!
f_{-1}=
-126.651\times 10^{-4}
,~~f_0=0,~~
f_1=-4.04837 \times  10^{-4},~~
f_2=  2.39701\times {10}^{-4},~
f_3=-1.80\times {10}^{-4}.
\label{@fexpp}\end{eqnarray}
\comment{
For comparison with large-$n$
calculations we also give the expansion in this limit:
\begin{eqnarray}  \!\!\!\!\!\!\!\!
\left\langle\frac{ \Delta \phi^2}{{u}}\right\rangle=F^{n=\infty}_3
&\equiv&
-4.02719\,{10}^{-6}\,u +
  1.90442\,{10}^{-8}\,u^2 -
  9.85172\,{10}^{-11}\,u^3   \nonumber \\
&+&
  \left( -8.05437\,{10}^{-6}\,
      u + 1.90442\,{10}^{-7}\,u^2 -
     2.27976\,{10}^{-9}\,u^3 \right)\frac{1}{n}
\nonumber \\&
+&   \left( 3.04707\,{10}^{-7}\,u^2 -
     1.2443\,{10}^{-8}\,u^3 \right)  \frac{1}{n^2}
.
\label{@}\end{eqnarray}
}
We need the
value of the series
$F_L(u)$ in the critical limit $m\rightarrow 0$,
which is obviously equivalent
to the
strong-coupling limit of $F_L(u)$.
As mentioned above, this limit should be
most accurately found
with the help of
variational perturbation theory \cite{SC3,SCE,KS}.

If the series were of quantum mechanical
origin,
we could have found this limit
by applying the
square-root trick (\ref{@SR0})
of Ref.
\cite{PI}.
In the present situation
where we are only interested in the extreme strong-coupling
limit, we would form the sequence of truncated expansions
$F_L(u)$ for $1,2,3$
 and replace each
term
\begin{equation}
(u/m)^l\rightarrow  K^l[1-1]^{-l/2}_ {L-l}
\label{@rep}\end{equation}
where the symbol
$[1-1]^{r}_ {k}$
is defined as the
 binomial expansion of $(1-1)^r$ truncated
after  the
$k$th term
\begin{equation}
[1-1]^{r}_ {k}\equiv \sum_{i=0}^{k}\left( r \atop i\right)(-1)^i=(-1)^k
\left( r -1\atop k\right).
\label{@}\end{equation}
The resulting  expressions
must be optimized in the variational
parameter $K$.
They are listed
 in Table \ref{vexp}.
\begin{table}[tbhp]
\caption[]{Trial functions
for the naive quantum-mechanical variational perturbation expansion
 }
\begin{tabular}{lll}
&\fsz $W^{\rm QM}_1={-0.0596831}{K^{-1}} - 0.0000322159\,K,$\\
&\fsz $W^{\rm QM}_2={-0.0497359}{K^{-1}} - 0.0000483239\,K +
  1.51792\,{10}^{-6}\,K^2,$\\
&\fsz $W^{\rm QM}_3={-0.0435189}{K^{-1}} - 0.0000604049\,K +
  3.03584\,{10}^{-6}\,K^2 -
  .908\,{10}^{-7}\,K^3.$
\end{tabular}
\label{vexp}\end{table}
The approximants
$W^{\rm QM}_{1,2,3}$
have  extrema
$W^{\rm QM ext}_{1,2,3}
\approx -0.00277,\,+0.00405,\, -0.0029,$
corresponding, via (\ref{@NumF}), to $c_1\approx3.059,\,-4.46,\,3.01$.
These values have previously been obtained in Ref.~\cite{ramospr}
in a much more complicated way via a so-called $ \delta $-expansion.
Note the negative sign of the second approximation
arising from the fact that an extremum exists only at negative $K$.
According to our rules of variational perturbation theory
one should, in this case,
 use the saddle point at positive $K$
which would yield
 $W_2^{\rm QM}=
-0.00153$
 corresponding to
$c_1\approx1.69$ rather than -4.46, leading to the
more reasonable
 approximation sequence
$c_1\approx3.059,\,1.69,\,3.01$, which shows no sign of convergence.
In
$W^{\rm QM}_{3}$,
there is also a pair of complex extrema
from which the authors of Ref.~\cite{ramospr}
extract the real part Re $\tilde W
_{3\,\rm complex}
^{\rm QM}
\approx-0.00134$
 corresponding to
$c_1\approx1.48$,
which they state as
 their
final result.
There is, however, no acceptable
theoretical justification for such a choice
\cite{HK1}.

This lack of convergence is not astonishing
since we are dealing
with field theory,
where the dimensions are anomalous and the naive
principle of minimal sensitivity
breaks down
(contrary to ubiquitous statements in the literature \cite{ubi}).
The valley in the dependence
on the variational parameter is no longer horizontal
\cite{HK1}.

The correct procedure goes as follows:
We form the logarithmic derivative
of the expansion (\ref{@eq9}):
\begin{eqnarray}
 \beta\left({u}\right) \equiv \frac{\partial \log F(u)}{\partial \log u}
=-1 +
2\frac{f_1}{f_{-1}}\left(\frac{u}{m}\right)^2
+3\frac{f_2}{f_{-1}}
\left(\frac{u}{m}\right)^3
+\left(
4\frac{f_3}{f_{-1}}
-2\frac{f_1^2}{f^2_{-1}} \right)
\left(\frac{u}{m}\right)^4+\dots~.
 \label{@betaf}\end{eqnarray}
In order for $F(u)$ to go to a constant
in the critical limit
$m\rightarrow 0$,
this function must go to zero
in the strong-coupling limit
$u\rightarrow \infty$.
Writing the expansion
as
$\beta _L
\left({u}\right)= -1+
\Sigma
_{l=2}^L\, b_l(u/4\pi m)^l
$, the
coefficients are
\begin{equation}
 b _2=0.0639293,~~~
 b_3=-0.056778,~~~
 b _4=0.0548799.
\label{@}\end{equation}
The sums
$ \beta_L(u)$ have to be evaluated
for
 $u \rightarrow \infty $
allowing for the universal anomalous dimension $ \omega $
by which the
physical observables
of $\phi^4$-theories approach the scaling limit \cite{zj,KS}.
The approach to the critical point  $A+B(m/u)^{ \omega '}$
where $ \omega '=
 \omega /(1-\eta /2)$ \cite{rem}.
The exponent
 $ \eta $ is the small anomalous dimension of the field
 while $ \omega $ again
 the
 Wegner exponent
\cite{Wegner}
of
renormalization group theory $ \Delta \equiv  \omega  \nu $.
Here it
appears
in the variational
expression for the
 strong-coupling limit
which is found  \cite{SC3,SCE} by replacing
$(u/m)^l$ by $ K^l[1-1]^{-ql/2}_ {L-l} $,
where $q\equiv 2/ \omega' $.
Thus we obtain the
variational expressions
\begin{eqnarray}
W^ \beta _3&=&
-1 +
\left( \frac{2\,{f_1}}
      {{f_{-1}}}
+
     \frac{2\,{f_1}\,q}
      {{f_{-1}}} \right)
  K^2
  +\frac{3\,{f_2}
}
   {{f_{-1}}}
K^3
\\
W^ \beta _4&=&-1+
\left( \frac{2\,{f_1}}
       {{f_{-1}}}
 +
      \frac{3\,{f_1}\,q}
       {{f_{-1}}} +
      \frac{{f_1}\,q^2}
       {{f_{-1}}} \right)
   K^2
  +
\left( \frac{3\,{f_2}}
       {{f_{-1}}} +
      \frac{9\,{f_2}\,q}
       {2\,{f_{-1}}} \right)
  K^3
 +
   \left( \frac{-2\,{{f_1}}^2}
       {{{f_{-1}}}^2} +
      \frac{4\,{f_3}}
       {{f_{-1}}} \right) \,K^4
  \label{@}\end{eqnarray}
The first has a vanishing extremum at
$ \omega' _3=0.592$, the second has
neither an extremum nor a saddle point.
However, a complex pair of extrema
lies reasonably close to the real axis at $ \omega' _4=0.635\pm0.116$,
 whose real part is
not far from
 the true
 exponent of approach  $ \omega'_\infty \approx0.81$ \cite{zj,KS},
to which $ \omega' _L$  will converge for order  $L\rightarrow \infty$
 \cite{SC3}.
Given these $ \omega' $-values,
we now form
the variational expressions
 $W_L$
from  $F_L$ by the replacement
$(u/m)^l\rightarrow   K^l[1-1]^{-ql/2}_ {L-l} $,
which are
\begin{eqnarray}
\hspace{-5mm}
W_2&=&   f_{-1}\left(1-\frac{3}{4}q+\frac{1}{8}q^2\right)K^{-1}+f_1 K,\\
\hspace{-5mm}
W_3&=&   f_{-1}\left(1-\frac{11}{13}q+\frac{1}{4}q^2-\frac{1}{48}q^3\right)K^{-1}
+f_1\left(1+\frac{q}{2}\right) K+f_2K^2,~~~\label{@18}\\
\hspace{-5mm}
W_4&=&   f_{-1}\left(1-\frac{25}{24}q+\frac{35}{96}q^2
-\frac{5}{96}q^3
+\frac{1}{384}q^4
\right)K^{-1}
+f_1\left(1+\frac{3}{4}q+\frac{1}{8}q^2\right) K+
f_2(1+q)K^2
+f_3K^3.~~~\label{@19}
\label{@}\end{eqnarray}

\begin{figure}[bht]
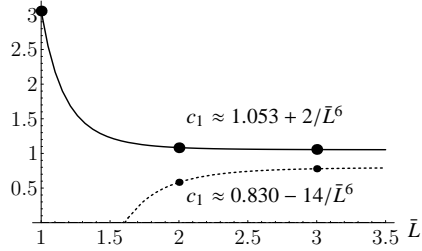

\vspace{2cm}
\input plapprno.tps
\caption[]{The three approximants for $c_1$
plotted against the order of variational
approximation  $\bar L\equiv L-1=1,2,3$,
and  extrapolation to the infinite-order limit.}
\label{figure}\end{figure}

The lowest function $W_2$ is optimized with the  naive
growth parameter $q=1$
since to this order
no anomalous value can be determined
from the zero of the beta function (\ref{@betaf}).
 The optimal result is $W_2^ {\rm opt}=-\sqrt{\log[4/3]/6}/8\pi^2\approx-0.00277$
corresponding to
$c_1\equiv 3.06$.
The next function $W_3$ is optimized with the
above determined
$q_3=2/ \omega'_3$ and yields
$W_3^ {\rm opt}\approx-0.000976$
corresponding to
$c_1\equiv 1.078$.
Although
  $ \omega' _4$ is not real
we shall insert its real part
into $W_4 $ and find
$W_4^{\rm opt}\equiv -0.000957$
corresponding to
$c_1\equiv 1.057$.
The three values of $c_1$
for $\bar L\equiv L-1=1,2,3$ can well be fitted
by a function
$c_{1}\approx1.053+2/\bar L^6$
(see Fig.~\ref{figure}).
Such a fit is suggested
by the general large-$L$ behavior
$a+be^{-c\,\bar L^{1- \omega' }}$ which was
derived in
Refs.~\cite{PI}.
Due to the smallness
of $1- \omega' \approx0.2$, this can be replaced by
$\approx a'+b'/\bar L^{s}$.

Alternatively, we may optimize the functions $W_{1,2,3}$ using the
known precise
value  of
$q_\infty=2/ \omega' _\infty\approx2/0.81$.
Then $W_2$ turns out to have no optimum, whereas
the others yield
$W^{\rm opt}_{3,4}
\approx
-0.000554,\,
-0.000735,$
  corresponding via Eq.~(\ref{@NumF})
to $c_1=0.580,\,0.773$.
If these two values
 are fitted
by the same inverse power of $\bar L$, we find
$c_{1}\approx0.83- 14/\bar L^6$.
From the extrapolations to
infinite order we estimate
$c_{1,\infty}\approx
0.92\pm0.13$.

\comment{It is possible to derive also an estimate from below.
For this we try to optimize all three functions $W_{123}$
with the precise $q_\infty\approx2/0.2$.
The first has no optimum, while $W_{2,3}\approx-0.0005289,\,0.000706$
corresponding to $c_1\approx0.583,\,0.779$.
These values are also shown in Fig.~(\ref{figure}).
They can be fitted to reach the same large-$L$ limit by a curve
$0.82-4.66/x^{4.5}$.
}

This
 result
is to be compared with latest Monte Carlo data
which estimate
$c_1\approx1.32\pm0.02$
\cite{arnold1,russos}.
Previous theoretical
estimates are
 $c_1\approx2.90$ \cite {baymprl},
$2.33$ from a $1/N$-expansion \cite {baymN}),
$1.71$ from a next-to-leading order
in a
$1/N$-expansion
 \cite {arnold},
$3.059$ from
an inapplicable $ \delta $-expansion \cite {prb}
to three loops,
and $1.48$ from
the same $ \delta $-expansion
to five loops, with a questionable
evaluation at a complex extremum \cite {ramospr} and some
wrong expansion coefficients (see \cite{braatexp}).
Remarkably, our result lies close
to the
average between the latest
and the
first
Monte Carlo result
 $c_1\approx
0.34\pm0.03$  in Ref.~\cite{Grue}.

As a cross check of the reliability of our theory
consider the result in the limit $N\rightarrow \infty$.
Here we must  drop the first term in the expansion
(\ref{@eq9})
which vanishes at the critical point
(but would diverge
 for  $N\rightarrow \infty$ at finite $m$).
The remaining  expansion
coefficients
of $
\left\langle{ \phi^2}/{u}\right\rangle/N$
in powers of $Nu/4\pi m$
are
\begin{eqnarray}  \!\!\!\!\!\!\!\!
f_1=-6.35917 \, 10^{-4},~~~
f_2=4.7315 \,{10}^{-4},~~
f_3=-3.84146\,{10}^{-4}.
\label{@fexpp}\end{eqnarray}
Using the $N\rightarrow \infty$ limit of $ \omega' $
which is equal to $1$ implying $q=2$ in Eqs.~(\ref{@18})
and (\ref{@19}),
we obtain
 the two variational approximations
\begin{eqnarray}
W_2^\infty= -0.00127183 K+0.00047315K^2,~~
W_3^\infty=  -0.00190775K+0.00141945K^2-0.000384146K^3,
\label{@}\end{eqnarray}
whose optima yield
the approximations $c_1\approx1.886$ and $2.017$,
converging rapidly towards the
exact large-$N$ result $2.33$ of Ref.~\cite {baymN}, with a 10\% error.

Numerically,  the first two $1/N$-corrections
found from a fit to
large-$N$ results
obtained by using the known large-$N$ expression for
$ \omega' =1-8( 8/3\pi^2N)+2(104/3-9\pi^2/2)(8/3\pi^2N)^2$
\cite{omN} produce a
finite-$N$  correction
factor $(1-3.1/N+30.3/N^2+\dots)$,
to be compared with
$(1-0.527/N+\dots)$ obtained in Ref.~\cite{arnold}.

Since the large-$N$ results can only be obtained
so well
without the use of the first term
we repeat the evaluations
of the series at the physical value
$N=2$ without the first term, where the variational expressions for $f$ are
\begin{eqnarray}
W_2& =&f_1\left(1+\frac{q}2\right)K+f_2K^2,\nonumber \\
W_3&=&f_1\left(1+\frac{3}{4}q+\frac{1}{8}q^2\right)K+f_2\left(1+q\right)K^2+f_3K^3.
\label{@}\end{eqnarray}
The lowest order optimum
lies now at
 $W_2^{\rm opt}=-f_1^2(2+q)^2/16f_2^2$, yielding
 $c_1\equiv 0.942$
for the exact
$q=2/0.81$.
To next order,
an optimal turning point of $W_3$
yields
 $c_1\approx 1.038$.

At this order,
we can
 derive
a
variational
expression for the determination of $ \omega '$
using the analog of
Eq.~(\ref{@betaf}) which reads
\begin{eqnarray}
 \beta\left({u}\right) \equiv \frac{\partial \log F(u)}{\partial \log u}
=1 +
\frac{f_2}{f_{1}}\,\frac{u}{m}
+\left(2
\frac{f_3}{f_{1}}
-\frac{f_2^2}{f_{1}^2}
\right)
\left(\frac{u}{m}\right)^2
+\dots~.
 \label{@betafn}\end{eqnarray}
  After the replacement (\ref{@rep})
we find
\begin{eqnarray}
W^ \beta _3&=&1+
\frac{{f_2}(1+q/2)}
       {{f_{1}}}K
 +
\left(
      2\frac{{f_3}}
       {{f_{1}}}
      -\frac{\,{f_2^2}}
       {{f_{1}^2}}
\right)
   K^2
 +\dots
  \label{@}\end{eqnarray}
whose vanishing extremum determines $ \omega '=2/q $ as being
\begin{equation}
 \omega '_3=\left(2 \sqrt{2f_1f_3/f_2^2-1}-1\right)^{-1}\approx0.675,
\label{@}\end{equation}
leading to
$c_1\approx1.238$
from an optimal turning point of $W_3$.
There are now too few points to perform  an extrapolation
to infinite order.
From the average of the two highest-order results
we obtain our
final estimate: $c_1\approx 1.14\pm0.11$,
 such that
 the critical temperature shift is
\begin{equation}
\frac{ \Delta T_c}{T_c^{(0)}}
\approx
(1.14\pm0.11)\,
an^{1/3}.
\label{@res}\end{equation}
This lies reasonably close to
 the Monte Carlo number $c_1\approx1.32\pm0.02$.

\section{Membrane Between Walls}

\begin{figure}[h]
\centerline{
\setlength{\unitlength}{.5cm}
\begin{picture}(10.0,6.5)
\put(0,0){\makebox(11,6.5){\epsfxsize=5cm \epsfbox{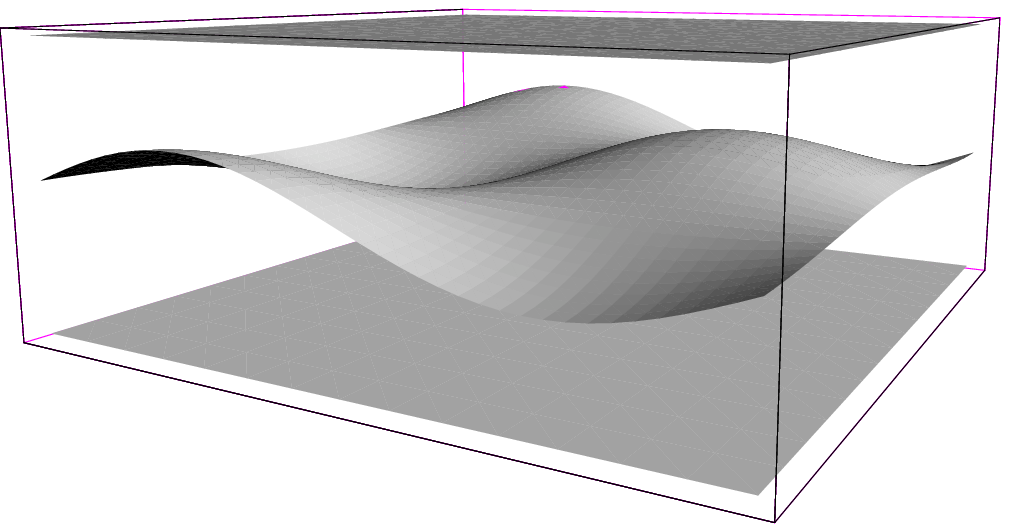}}}
\put(0,3.9){$\scriptstyle z$}
\put(3.5,1.2){$\scriptstyle x$}
\put(9.8,1.7){$\scriptstyle y$}
\put(11.2,5.5){$\!\!\!\!\scriptstyle d/2$}
\put(10.7,4.25){~$\scriptstyle 0$}
\put(10.7,3.0){\!\!\!$\scriptstyle -d/2$}
\end{picture}} \vspace{-1.5em}
\caption[Membrane fluctuating  between walls
with distance $d$]{
Membrane fluctuating  between walls
with distance $d$.}
\label{@fig1}\end{figure}

As another example consider
a tension-free membrane of bending stiffness
$ \kappa $ between hard walls \cite{HKM}
(see Fig.~\ref{@fig1}).

\begin{figure}[h]
\centerline{
\setlength{\unitlength}{.5cm}
\begin{picture}(8,7.5)
\put(0,0){\makebox(8,7.5){\epsfysize=3.5cm \epsfbox{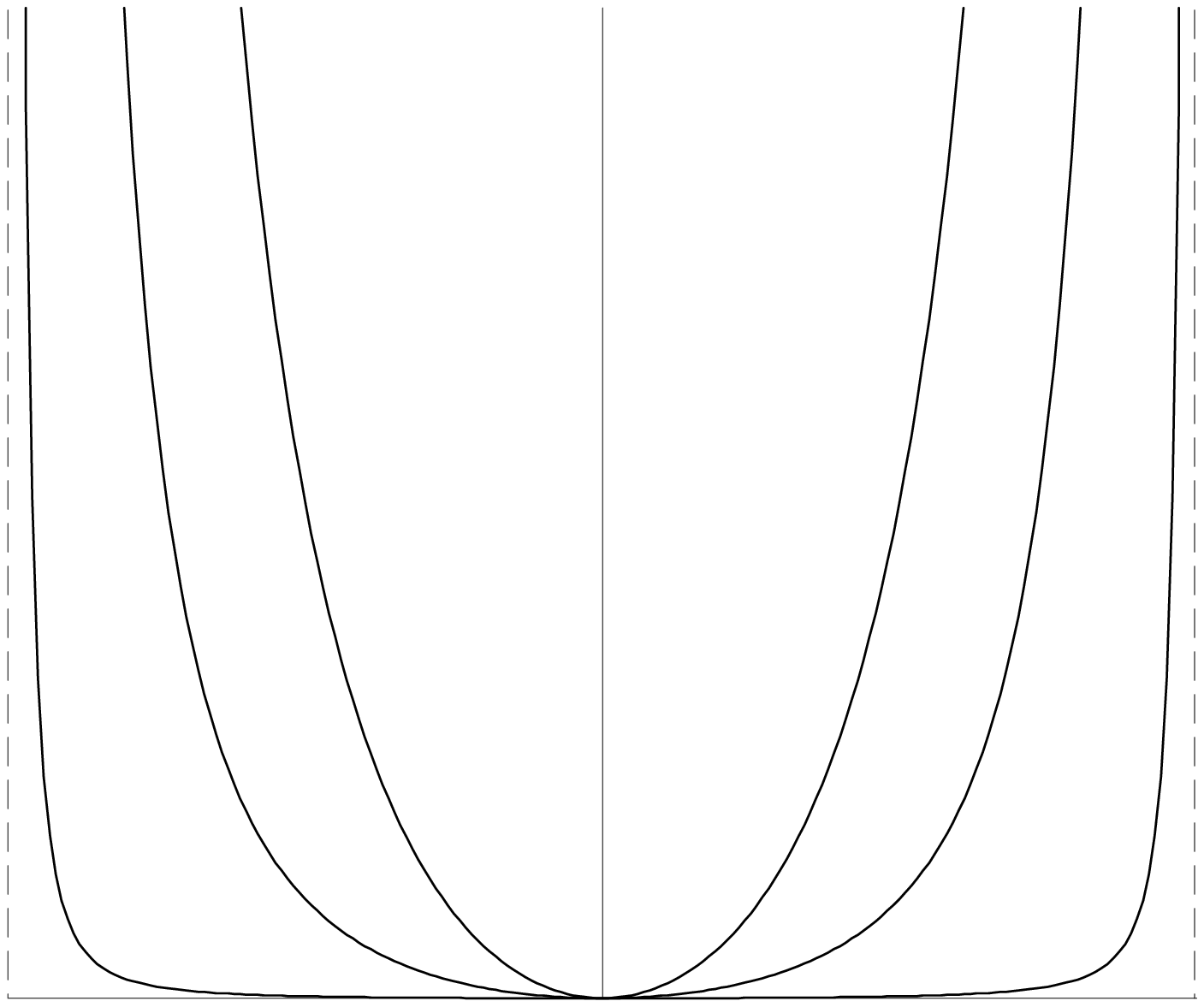}}}
\put(-3.1,2.1){$m\to 0$}
\put(-0.99,2.25){\vector(1,0){0.8}}
\put(2.0,6.7){$\scriptstyle V(h)$}
\put(7.6,-0.5){$\scriptstyle d/2$}
\put(5.8,-0.5){$\scriptstyle h$}
\put(3.83,-0.5){$\scriptstyle 0$}
\put(-1.0,-0.5){$\scriptstyle -d/2$}
\end{picture}} \vspace*{.3cm}
\caption[Softened hard-wall potential which becomes
infinitely hard in the limit $m\rightarrow 0$
]{  Softened hard-wall potential which becomes
infinitely hard in the limit $m\rightarrow 0$}
\label{@figXXX}
\end{figure}

Its thermal fluctuations
are described by a functional integral
over a Boltzmann factor
\begin{eqnarray}
Z=\prod_x\int_{-d/2}^ {d/2} {\cal D}h\,e^{-E/k_0T},
\label{@FUZ}\end{eqnarray}
where $h(x)$ is the height function
of the membrane
and $E$ is the bending energy
\begin{equation}
E=\frac{ \kappa }{2}\int d^2x \left[ \partial ^2 h(x)\right] ^2.
\label{@}\end{equation}
This functional integral
has not been solved exactly,
in spite of its simplicity.
It can, however, be approximated by
the functional integral
\begin{eqnarray}
Z=\prod_x\int_{-\infty}^ {\infty} {\cal D}h\,e^{-[E+V(x)]/k_0T},
\label{@}\end{eqnarray}
in which the height fluctuates
between $-\infty$ and $\infty$
in a potential (see Fig. \ref{@figXXX})
\begin{equation}
V(x)=m^4 \frac{d^2}{\pi^2 }\tan^2 \left(\frac{\pi h}{d}\right).
\label{@VV}\end{equation}
This problem can be solved perturbatively
yielding
$Z=e^{-Af}$, where $A$ is the area of the membrane and
$f$ has, to order $N$, the series
\begin{equation}
{f^N}=\frac{m^2}{2}\left[1+\frac{1}{8}+\frac{\pi^2}{m^2d^2}
\frac{1}{64}
 +\dots+
\left(\frac{\pi^2}{m^2d^2}\right)^N  a_N
\dots\right] .
\label{@Ser}\end{equation}
The hard-wall limit $m\rightarrow 0$ amounts to the strong-coupling limit
 of this series.

\begin{figure}[tbh]
\vspace{-1em}
\centerline{
\setlength{\unitlength}{.65cm}
\begin{picture}(10.0,6.5)
\put(-2,0){\makebox(11,6.5){\epsfxsize=9.1cm \epsfbox{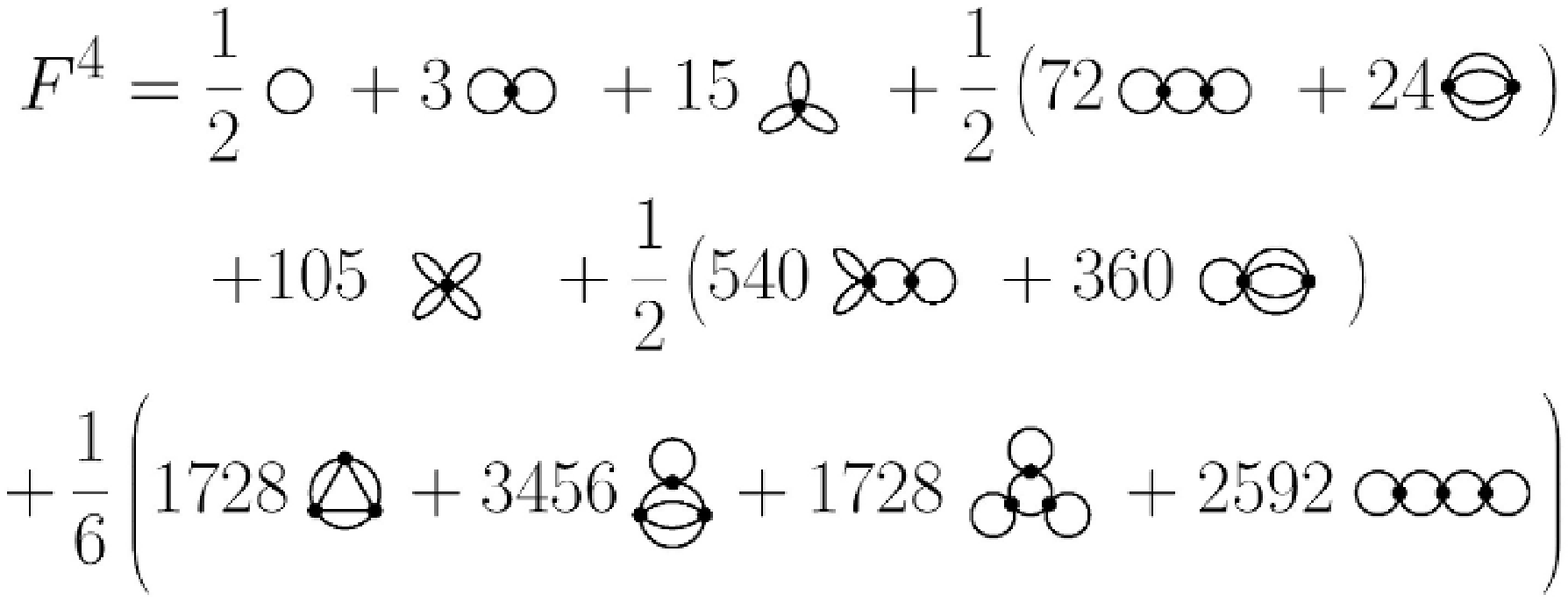}}}
\put(12.4,1.4){.}
\end{picture}}
\vspace{-1em}
\caption[Feynman diagrams
in the perturbative expansion
of the free energy of the Membrane between walls
]{
Feynman diagrams
in the perturbative expansion
of the free energy of the Membrane between walls up to the order $N=4$.
}
\label{@figFD}\end{figure}

We expand the potential
(\ref{@OT})
into a power series
\begin{eqnarray}
V ( h ) = m^4 \frac{h^2}2 +
  m^4 \,\frac{\pi^2}{d^2}  \left\{
\frac{1}{3} \,h^4 + \frac{17}{90} \,\frac{\pi^2}{d^2}\, h^6
+
\frac{31}{315} \,\frac{\pi^4}{d^4} \, h^8
+\frac{691}{14175} \,\frac{\pi^6}{d^6} \, h^{12}
+\frac{10922}{467775} \,\frac{\pi^8}{d^8} \, h^{16}
+ \ldots \right\}.
\label{@VVV}\end{eqnarray}
If we denote
the interaction terms
by
\begin{equation}
V^{\rm int }=\frac{ \kappa m^4}2\sum_{k=1}^\infty  \varepsilon _k \left(\frac{\pi}{d}h\right)^{2k},
\label{@}\end{equation}
and calculate the Feynman diagrams
shown in Fig. \ref{@figFD},
The functional
integral (\ref{@FUZ})
can be expressed as an exponential
$Z=e^{-Af}$, where $A$ is the area of the membrane and
\begin{equation}
f^N=\frac{m^2}{2}\left[1+\frac{1}{8}+\frac{\pi^2}{m^2d^2}
\frac{1}{64}
 +\dots+
\left(\frac{\pi^2}{m^2d^2}\right)^N  a_N
\dots\right] .
\label{@Ser}\end{equation}
Using the Bender-Wu recursion
relations \cite{BEWU}, we can express
the coefficients
in terms of $ \varepsilon _K$
as
{
\begin{eqnarray}
&&f^N=\frac{m^2}{2}+\frac{3\pi^2}{4d^2} \varepsilon _4
-\frac{\pi^4}{8d^4}\left(21 \varepsilon^2 _4 -
15 \varepsilon _6\right)
+\frac{\pi^6}{16d^6}\left(
333
 \varepsilon _4^3 \!-\!
360 \varepsilon _4 \varepsilon _6
\!+\!105 \varepsilon _8\right)\nonumber \\
&&\!-\frac{\pi^8}{128d^8}\left(
30885
 \varepsilon _4^4
\!-\!44880 \varepsilon^2_4 \varepsilon _6
\!+\!6990 \varepsilon _6^2
\!+\!1512 \varepsilon _4 \varepsilon _8
\!+\!3780 \varepsilon _{10}
\right)\!+\!\dots\,.
\label{@}\nonumber \end{eqnarray}}%
The hard-wall result is obtained in the limit
$m \rightarrow 0$, which is
the
 strong-coupling limit
of the series (\ref{@Ser}).

\section{Variational Perturbation Theory of Tunneling}

\comment{
So far all applications have been restricted
to
divergent expansions
that have alternating signs and are
 Borel-summable.
The great advantage of variational perturbation theory
is that it also yields correct results
for
non-Borel-summable
 series.
This is possible by
 an analytic continuation
 of
the variation procedure.
 We demonstrate the power of the method
by an application to the exactly known partition function
of the anharmonic oscillator
in zero space-time dimensions.
In one space-time dimension
we
derive
the imaginary part of the
 ground state energy of
 the anharmonic oscillator for {\em all\/} negative
 values of
 the coupling constant $g$, including the
non-analytic tunneling regime at
small $-g$.
As a
highlight
 of the theory
one can
retrieve
 the divergent
perturbation expansion from
the  action
of the critical bubble
and the contribution
of the higher  loop
fluctuations around the bubble.
\section{Introduction}
}

None of
 the presently
known resummation schemes \cite{Resumm,KS}
is able to deal with non-Borel-summable series.
Such series arise
in
 the
 theoretical description
of many
important physical phenomena,
in particular tunneling
processes.
In the path integral, these are dominated by
non-perturbative contributions coming from
nontrivial classical solutions
called {\em critical bubbles\/} \cite{Langer,PI}
or {\em bounces\/} \cite{Coleman},
and fluctuations around these.

A non-Borel-summable
series can become
Borel-summable
if
 the expansion parameter, usually some coupling constant $g$,
is continued to negative values.
In this way,
non-Borel-summable
series  can be evaluated with any desired accuracy
 by an
analytic continuation
of
 variational perturbation
 theory
 \cite{PI,KS} in
 the complex $g$-plane.
This implies that
 variational perturbation
 theory
can give us information on
non-perturbative properties of the theory.

\comment{Variational perturbation
 theory
has a long history
\cite{refs,finiteg,KleinertJanke,Guida}. It is based on the introduction of a
dummy variational
 parameter $ \Omega $
on which
the full perturbation expansion does not depend,
while the truncated expansion
does. An optimal $ \Omega $ is selected
by the principle of minimal sensitivity \cite{STE},
 requiring the
quantity of interest to be stationary as function of the
variational parameter. The optimal
$ \Omega $
 is usually
taken from a zero of
 the derivative with
respect to $ \Omega $.
If the first derivative has no zero,
a zero of the second derivative is chosen.
For Borel-summable series, these zeros are always real, in contrast
to statements
in
the literature
 \cite{Neveu,RULES}
which have proposed
the use of complex
zeros. Complex zeros produce
in general  wrong results
for Borel-summable series, as was recently shown
in Ref.~\cite{HK1}.
There it was
also shown that there
does exist
 a wide range of applications
of complex zeros if one
wants to resum
 non-Borel-summable series, which have so far remained intractable.
These arise typically
in tunneling problems,
and we
shall
see
 that variational perturbation
theory provides us with an
efficient method for evaluating
these series,
 rendering
their real
and imaginary parts
with any desirable accuracy,
if only enough perturbation coefficients are available.
An important problem which had to be solved
is the specification of the proper
choice
the optimal zero from  the many possible candidates
existing in higher orders.
A non-Borel-summable series
is associated with a function which has
 an essential singularity at the origin
in the complex $g$-plane,
which is the starting point of a
 left-hand cut. Near the
 tip of the cut,
the imaginary part of the function  approaches zero
rapidly like $\exp({-\alpha/|g|})$
for $g \to 0-$.
If the variational approximation
is plotted against $g$ with
an enlargement factor $\exp({\alpha/|g|})$,
oscillations
become visible near $g=0$.
The choice of the optimal
 complex zeros
of the derivative with respect to the variational parameter
is fixed
 by the
requirement
of obtaining, in each order,
the least oscillating
imaginary part when approaching
 the tip of
 the cut. We may call this selection rule the
{\em principle of minimal
sensitivity and oscillations\/}.
In Section \ref{@sec2},
we shall explain and test
 the new principle
 on the exactly known partition function
$Z(g)$ of the anharmonic oscillator
in zero space time dimensions.
%
In Section \ref{@sec3}, we apply the method to the
critical-bubble regime of small $-g$ of the anharmonic oscillator
and find the action of the critical bubble and the
corrections caused by the  fluctuations around it.
In Section \ref{@sec4} we present
yet another
method of calculating the
properties of the
critical-bubble regime.
This method
is
restricted to quantum mechanical systems.
Its results for the anharmonic oscillator
 give
 more evidence
for the correctness of the general method
of Sections \ref{@sec2} and \ref{@sec3}.
}
\subsection{Test of
 Variational Perturbation Theory
for  Simple Model of
Non-Borel-summable Expansions}
\label{@sec2}
The partition function
$Z(g)$ of the anharmonic oscillator
in zero space-time dimensions is
\begin{align}
\label{FOKKER}
Z(g) = \frac{1}{ \sqrt{\pi}}\; \int_{-\infty}^\infty
 \exp{(-x^2/2-g\ x^4/4)}\ dx=\frac{\exp{(1/8g)} }
{\sqrt{4\pi g}}K_{1/4}(1/8g)\ ,
\end{align}
where $K_{ \nu }(z)$ is the modified Bessel function.
For small $g$, the function $Z(g)$ has a divergent
Taylor
series expansion, to be called
{\em weak-coupling expansion\/}:
\begin{align}
\label{FP-WEAK}
Z_{\rm weak}^{(L)}(g) = \sum_{l=0}^L\; a_l\;g^l, \qquad \!\!\!\!\!\mbox{with }\ a_l=(-1)^l\ \frac{\Gamma(2l+1/2)}{l!\sqrt \pi} .
\end{align}
For $g<0$, this is
non-Borel-summable.
For large  $|g|$ there exists a convergent {\em strong-coupling expansion\/}:
\begin{align}
\label{FP-STRONG}
Z_{\rm strong}^{(L)}(g) =g^{-l/4}\ \sum_{l=0}^L\; b_l\;g^{-l/2}, \qquad\!\!\!\!\!
 \mbox{with }\ b_l=(-1)^l\ \frac{\Gamma(l/2+1/4)}{2 l!\sqrt \pi}.
\end{align}
%
As is obvious from the integral representation (\ref{FOKKER}),
$Z(g)$
obeys
the second-order differential equation
\begin{align}
\label{DGL}
16 g^2 Z''(g)+4(1+8g)Z'(g)+3 Z(g)=0,
\end{align}
which has two independent solutions.
One of them is
 $Z(g)$, which  is finite for $g>0$ with $Z(0)=a_0$.
 The
weak-coupling
coefficients $a_l$ in
(\ref{FP-WEAK})
can be obtained
by inserting
into
 (\ref{DGL})
 the Taylor series
and comparing coefficients.
The result is the
recursion relation
\begin{align}
\label{REC}
a_{l+1}=-\frac{16l(l+1)+3}{4(l+1)}\, a_l.
\end{align}

A similar recursion relation
can be derived for the strong-coupling coefficients $b_l$ in Eq.~(\ref{FP-STRONG}).
We observe that
 the two independent solutions
$Z(g)$
of (\ref{DGL})
behave like $Z(g) \propto g^\alpha$ for $g \to \infty$ with the powers
$\alpha=-1/4$ and $-3/4$. The
function
(\ref{FOKKER})
has
 $\alpha=-1/4$.
It is convenient to remove the leading power
from $Z(g)$ and define a function $\zeta(x)$
such that  $Z(g) = g^{-1/4}\ \zeta(g^{-1/2})$.
The Taylor coefficients of $\zeta(x)$
 are the strong-coupling coefficients $b_l$ in Eq.~(\ref{FP-STRONG}).
The function $ \zeta (x)$  satisfies
 the differential equation
and initial conditions:
\begin{align}
\label{DGL2}
4\zeta''(x)-2x\zeta'(x)-\zeta(x) &=0,~~~~\mbox{with}~~
\zeta(0)=b_0 ~~\mbox{and}~~ \zeta'(0)=b_1 .
\end{align}
The
Taylor coefficients $b_l$  of $\zeta(x)$
satisfy the recursion relation
\begin{align}
\label{REC2}
b_{l+2}=\frac{2l+1}{4(l+1)(l+2)}b_{l} \,.
\end{align}
Analytic continuation of $Z(g)$ around $g=\infty$ to
 the left-hand cut gives:
\begin{align}
\label{CUT}
Z(-g)&=(-g)^{-1/4} \zeta((-g)^{-1/2})\\
&=(-g)^{-1/4} \sum_{l=0}^\infty b_l(-g)^{-l/2} \exp{\left[-\frac{i\pi}{4}(2l+1)\right] } \qquad \mbox{for $g>0$},
\end{align}
so that we find an imaginary part
\begin{align}
\label{CUTi}
{\rm Im} \, Z(-g)&= -(4g)^{-1/4}  \sum_{l=0}^\infty b_l(-g)^{-l/2} \sin{\left[
-\frac{i\pi}{4}(2l+1)\right] }\\
&=-(4g)^{-1/4}\   \sum_{l=0}^\infty \beta_l(-g)^{-l/2}  \ ,
\end{align}
where
\begin{align}
\label{CUT2}
\beta_0&=b_0,~~~~ \beta_1=b_1,~~~~ \beta_{l+2}=-\frac{2l+1}{4(l+1)(l+2)}\beta_{l}\ .
\end{align}
It is easy to show that
\begin{equation}
\sum_{l=0}^\infty \beta_l x^l=\zeta(x)\exp{(-x^2/4)},
 \label{@}\end{equation}
so that
\begin{align}
\label{CUT-STRONG}
{\rm Im}\, Z(-g)=-\frac{1}{\sqrt 2}\, g^{-1/4}\  \exp{(-1/4g)}\
 \sum_{l=0}^\infty \; b_l\;g^{-l/2}\ .
\end{align}
From this we may re-obtain
the weak-coupling coefficients $a_l$ by means of the dispersion relation
\begin{align}
\label{DISP}
 Z(g)=& -\frac{1}{\pi}\int_0^\infty \frac{{\rm Im}\, Z(-z)}{z+g}\,dz\\
=& \frac{1}{\pi \sqrt 2}\ \sum_{j=0}^\infty \ b_j \int_0^\infty
\frac{  \exp{(-1/4z)}\  z^{-j/2-1/4 }}{z+g}\,dz.
\end{align}
Indeed, replacing $1/(z+g)$ by $\int_0^\infty \exp{(-x(z+g))}\,dx$,
and expanding  $\exp{(-x\,g)}$ into a power series, all integrals can be evaluated to yield:
\begin{align}
\label{DISP-2}
 Z(g)=& \frac{1}{\pi}\ \sum_{j=0}^\infty \ 2^j b_j \ \sum_{l=0}^\infty \ (-g)^l \Gamma(l+j/2+1/4)\ .
\end{align}
Thus we find
for  the weak-coupling coefficients $a_l$ an expansion in terms of
the strong-coupling coefficients
\begin{align}
\label{DISP-3}
 a_l=& \frac{(-1)^l}{\pi}\ \sum_{j=0}^\infty \ 2^j b_j \ \Gamma(l+j/2+1/4).
\end{align}
Inserting $b_j$ from
Eq.~(\ref{FP-STRONG}), this becomes
\begin{align}
\label{DISP-3}
 a_l=
 \frac{(-1)^l}{2\pi^{3/2}}\ \sum_{j=0}^\infty \  \frac{2^j (-1)^j}{j!}\ \Gamma(j/2+1/4)\Gamma(l+j/2+1/4)
=(-1)^l\ \frac{\Gamma(2l+1/2)}{l!\sqrt \pi}\ ,
\end{align}
coinciding with
(\ref{FP-WEAK}).

Let us now apply
variational perturbation theory
to the weak-coupling expansion
(\ref{FP-WEAK}).
We have seen in Eq.~(\ref{CUT}), that
the
 strong-coupling expansion
can easily be continued analytically
to negative
$g$.  This continuation can, however, be used for an evaluation only for sufficiently
large $|g|$
where
the
 strong-coupling expansion
converges.
In the
 tunneling regime
near the tip of the left-hand cut,
the expansion diverges.
Let us
show
that
an evaluation of
the weak-coupling expansion
according to the rules of variational perturbation theory
continued
into the complex plane
 gives extremely good results on the entire
left-hand cut with a fast convergence even near the tip
at $g=0$.

 The $L$th variational approximation to $Z(g)$ is given by
(see \cite{HKSC,SC3,SCE,KS})
\begin{equation}
\label{FP-VAR}
Z_{\rm var}^{(L)}(g,\Omega)
 = \Omega^{p} \ \sum_{j=0}^L \left(\frac{g}{\Omega^q}\right)^j \epsilon_j(\sigma),~~~
\end{equation}
with
\begin{equation}
\label{FP-s}
\sigma\equiv \Omega^{q-2}(\Omega^2-1)/g\,,
\end{equation}

where $q=2/\omega=4$, $p =-1$ and
\begin{align}
\label{FP-EPS}
\epsilon_j(\sigma) = \sum_{l=0}^j a_l \binom{(p-lq)/2}{j-l} (-\sigma)^{j-l}\,.
\end{align}
In order to find a
valley of minimal sensitivity, the
zeros of the derivative of $Z_{\rm var}^{(L)}(g,\Omega)$ with respect to $\Omega$ are needed. They are given by the
zeros of the polynomials in $ \sigma $:
\begin{align}
\label{FP-DERIV}
P^{(L)}(\sigma) = \sum_{l=0}^L a_l (p-lq+2l-2L) \binom{(p-lq)/2}{L-l} (-\sigma)^{L-l}
=0,
\end{align}
since it can be shown
\cite{KJ2,HKSC}
that the derivative depends only on $ \sigma $:
\begin{align}
\label{FP-DERIV-II}
\frac{d Z_{\rm var}^{(L)}(g,\Omega)}{d\Omega} =\Omega^{p-1} \left(\frac{g}{\Omega^q}\right)^L P^{(L)}(\sigma)\,.
\end{align}
\begin{figure}[htp!]
\begin{center}
\setlength{\unitlength}{.5cm}
\begin{picture}(19,6.5)
\put(0.5,0.5){\scalebox{.4}[.4]{\includegraphics*{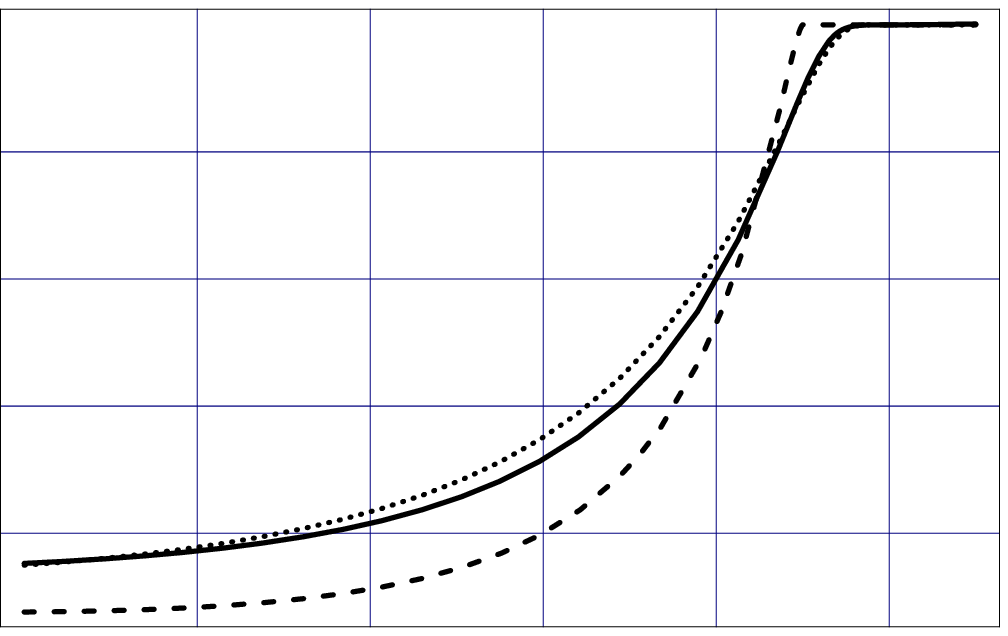}}}
\put(10,0.5){\scalebox{.4}[.4]{\includegraphics*{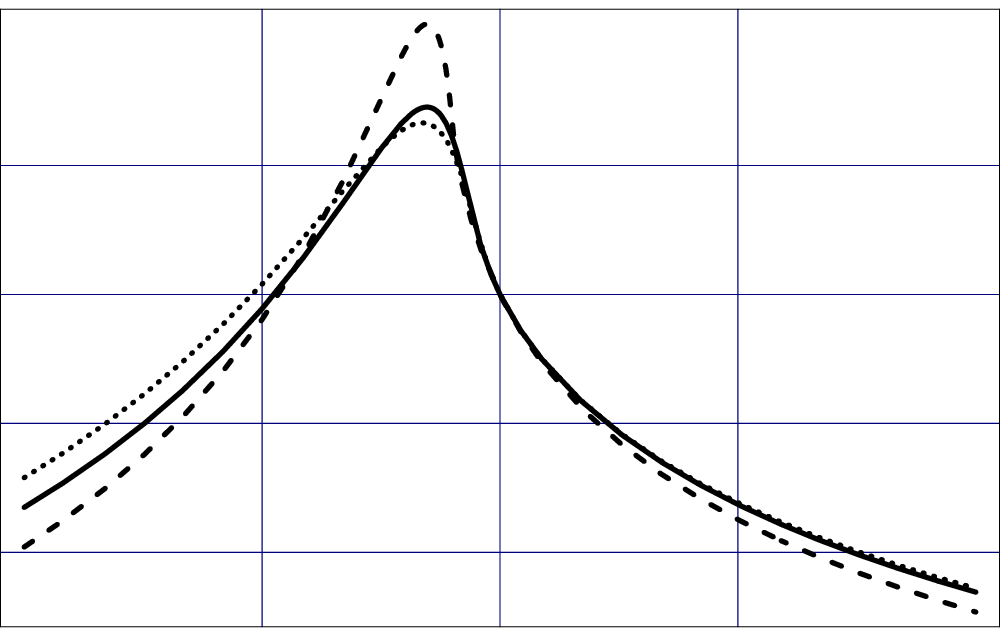}}}
\put(8.5,0.2){\tsz$g$}
\put(1.73,0.2){\tsz$-.8$}
\put(3.73,3.8){\tsz$Z(g)$}
\put(14.73,3.8){\tsz$Z(g)$}
\put(4.6,0.2){\tsz$-.4$}
\put(7.64,0.2){\tsz$0$}
\put(18,0.2){\tsz$g$}
\put(11.8,0.2){\tsz$-.5$}
\put(14,0.2){\tsz$0$}
\put(15.85,0.2){\tsz$.5$}
\put(-.1,3.25){\tsz$-.2$}
\put(-.1,1.2){\tsz$-.4$}
\put(9.8,3.15){\tsz$1$}
\put(9.7,1){\tsz$.8$}
\end{picture}
\caption[O2]{
Plot of the 1st-
and 2nd-order
calculation for the
non-Borel-summable region of $g<0$, where the function has a
cut with non-vanishing imaginary part:
imaginary (left) and
 real parts (right) of $Z_{\rm var}^{(1)}(g)$ (dashed curve) and
$Z_{\rm var}^{(2)}(g)$ (solid curve) are plotted against $g$
and compared
with the
 exact values of the partition function (dotted curve).
The root of (\ref{FP-s}) giving the optimal variational parameter $ \Omega $ has been chosen to reproduce
the weak-coupling result near $g=0$.
}
\label{O2}
\end{center}
\end{figure}
Consider in more detail  the lowest non-trivial order with $L=1$.
 From Eq.~(\ref{FP-DERIV}) we obtain
\begin{align}
\label{Zero}
\sigma= &\frac{5}2,~~~~~~ {\rm corresponding~to}~~~~ \Omega= \frac{1}{2}\Big(1 \pm \sqrt{1 + 10g}\Big)\ .
\end{align}
In order to ensure that  our method  reproduces
 the weak-coupling result for small $g$, we have to take the positive sign in front of the square root. In
Fig.~\ref{O2} we have plotted $Z_{\rm var}^{(1)}(g)$ (dashed curve)
and $Z_{\rm var}^{(2)}(g)$ (solid curve) and compared these
with the exact result (doted curve) in the tunneling regime.
The agreement is quite good
even at these low orders \cite{Tunn}.
Next we study the behavior of $Z_{\rm var}^{(L)}(g)$ to higher orders $L$. For
selected coupling values
in the
non-Borel-summable region, $g=-.01,\ -.1,\ -1,\ -10$,
 we want to see the error as a function of the order.
We want to find  from this model system
the rule for selecting systematically the
best zero of $P^{(L)}(\sigma)$ solving Eq.~(\ref{FP-DERIV}),
which leads to
the optimal value
of the variational parameter $\Omega$.
For this purpose  we plot the variational results of all zeros. This
is shown in Fig.~\ref{NB-I}, where the logarithm of the deviations
from the exact value is plotted against the order $L$.
The outcome of different zeros cluster  strongly near the best value.
Therefore, choosing any zero out of the middle of the cluster is reasonable, in particular, because it does not depend on the knowledge of the exact solution,
so that this rule may be taken over to realistic cases.
\begin{figure}[htb]
\begin{center}
\setlength{\unitlength}{.5cm}
\begin{picture}(19,12.5)
\put(0.5,0.5){\scalebox{.4}[.4]{\includegraphics*{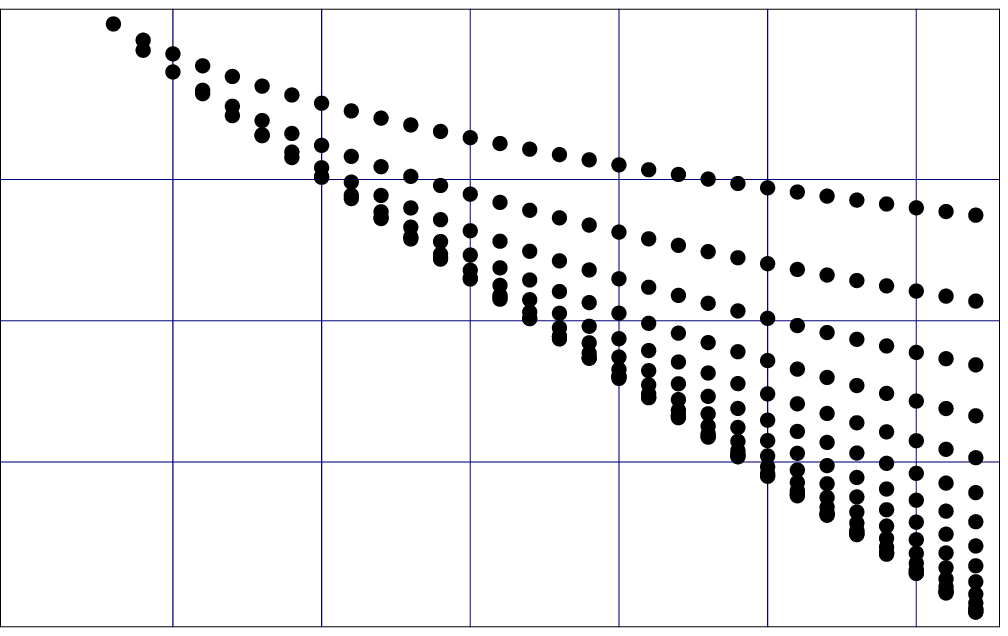}}}
\put(0.5,7){\scalebox{.4}[.4]{\includegraphics*{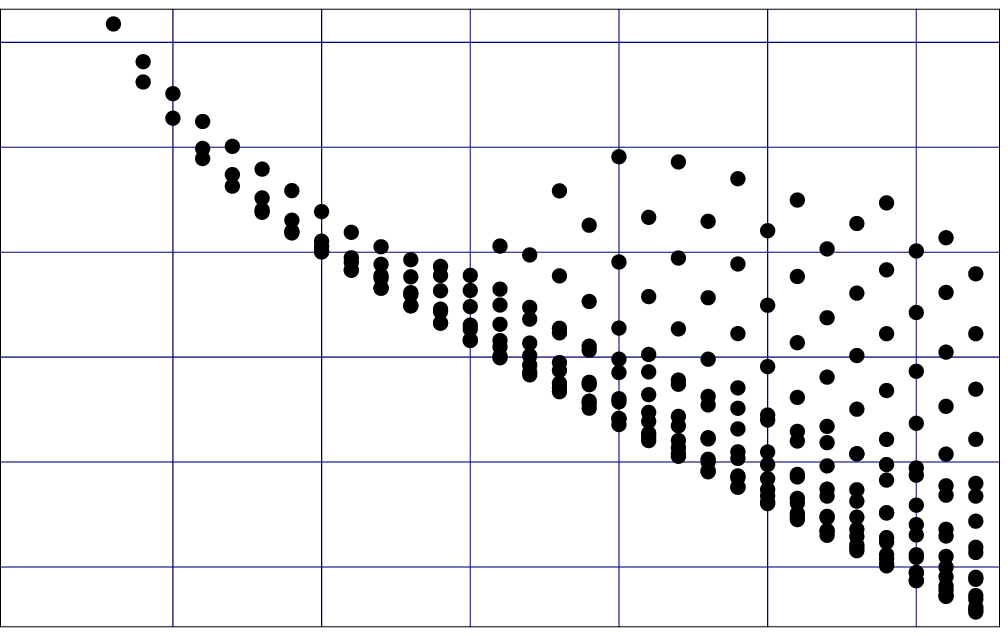}}}
\put(10,0.5){\scalebox{.4}[.4]{\includegraphics*{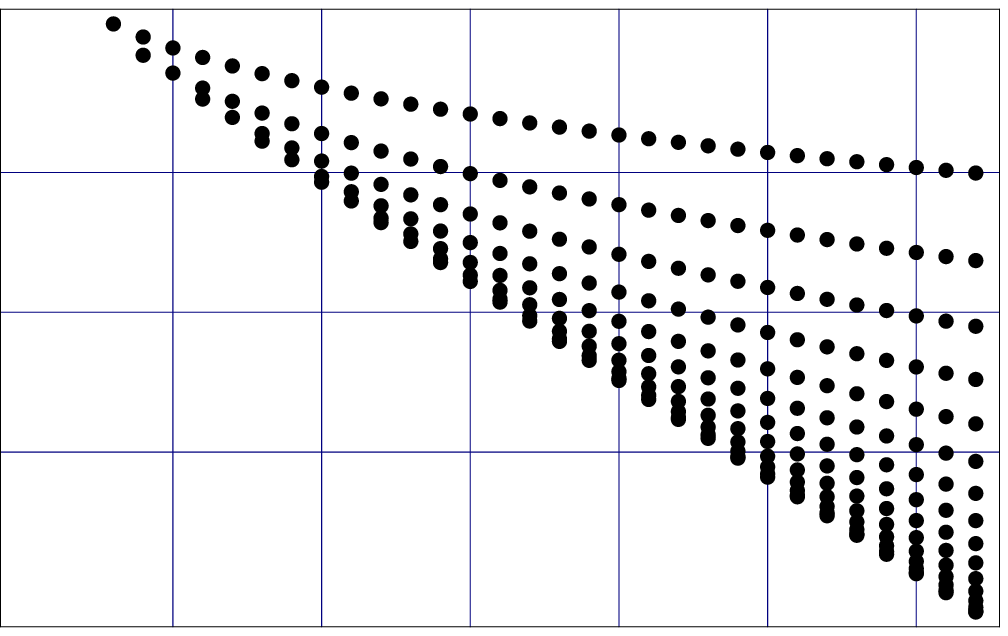}}}
\put(10,7){\scalebox{.4}[.4]{\includegraphics*{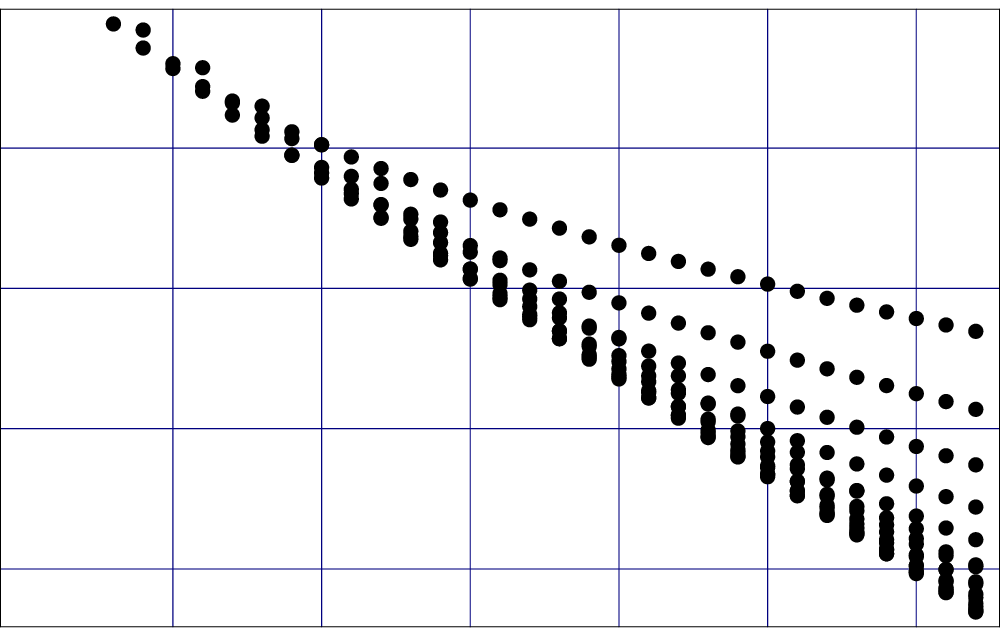}}}
\put(2.,7.7){\tsz$g=-.01$}
\put(11.5,7.7){\tsz$g=-.1$}
\put(2.,1.4){\tsz$g=-1$}
\put(11.41,1.4){\tsz$g=-10$}
\put(2.95,6.7){\tsz$10$}
\put(5.35,6.7){\tsz$20$}
\put(7.75,6.7){\tsz$30$}
\put(8.55,6.7){\tsz$L$}
\put(2.95,.2){\tsz$10$}
\put(5.35,.2){\tsz$20$}
\put(7.75,.2){\tsz$30$}
\put(8.55,.2){\tsz$L$}
\put(12.45,6.7){\tsz$10$}
\put(14.85,6.7){\tsz$20$}
\put(17.25,6.7){\tsz$30$}
\put(18.05,6.7){\tsz$L$}
\put(12.45,.2){\tsz$10$}
\put(14.85,.2){\tsz$20$}
\put(17.25,.2){\tsz$30$}
\put(18.05,.2){\tsz$L$}
\put(-.1,10.85){\tsz$-20$}
\put(-.1,9.1){\tsz$-30$}
\put(-.1,7.4){\tsz$-40$}
\put(9.4,10.87){\tsz$-10$}
\put(9.4,8.54){\tsz$-20$}
\put(-.1,4.17){\tsz$-10$}
\put(-.1,1.76){\tsz$-20$}
\put(9.4,4.15){\tsz$-10$}
\put(9.4,1.88){\tsz$-20$}
\end{picture}
\caption[NB-I]{Logarithm of deviation of the variational results
from  exact values $\log{|Z_{\rm var}^{(L)}-Z_{\rm exact}|}$
 plotted against the  order $L$ for different $g<0$ in the
non-Borel-summable region. All complex optimal $ \Omega $'s have been used.}
\label{NB-I}
\end{center}
\end{figure}
\begin{figure}[htp!]
\begin{center}
\setlength{\unitlength}{.5cm}
\begin{picture}(15,9)
\put(0,.7){\scalebox{.625}[.625]{\includegraphics*{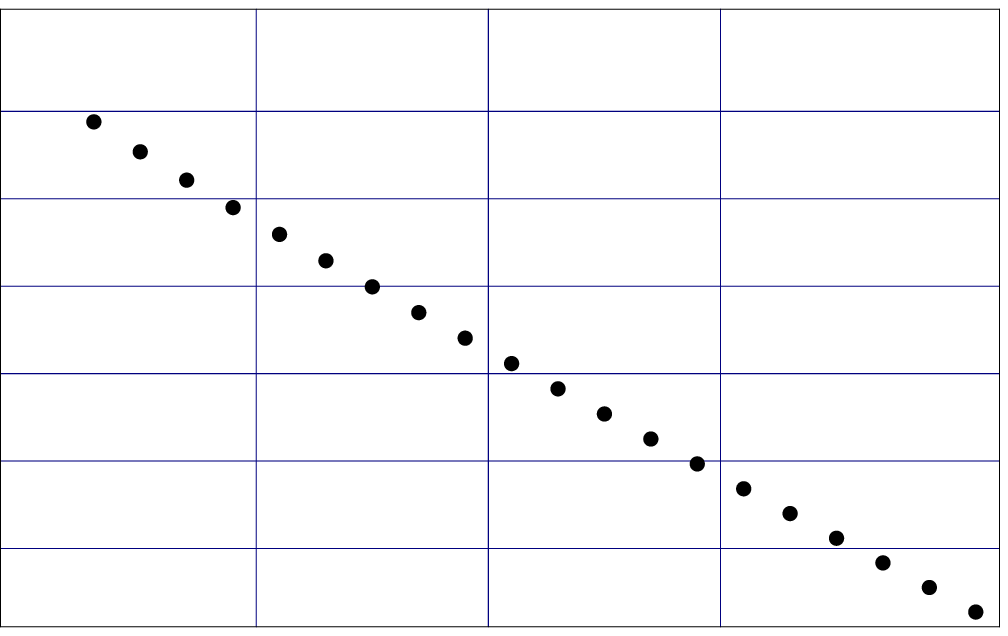}}}
\put(7.5,6.7){\tsz$ \Delta (L)$}
\put(12.5,0.3){\tsz$L$}
\put(3.05,0.3){\tsz$10$}
\put(6.02,0.3){\tsz$20$}
\put(8.98,0.3){\tsz$30$}
\put(-.6,6.08){\tsz$-10$}
\put(-.6,3.85){\tsz$-20$}
\put(-.6,1.62){\tsz$-30$}
\end{picture}
\caption[NB-II]{Logarithm of deviation of
 variational results from
exactly known value $\Delta(L)=\log{|Z_{\rm var}^{(L)}-Z_{exact}|}$,
 plotted against the  order $L$ for  $g=10$ in Borel-summable region.
The real positive
 optimal $ \Omega $
 have been used.
There is only one real zero of the first derivative
in every odd order $L$
and none for even orders.  There is
excellent convergence $\Delta(L)\simeq 0.02\exp{(-0.73L)}$
for $L\to\infty$.
}
\label{NB-II}
\end{center}
\end{figure}
\begin{figure}[htp!]
\begin{center}
\setlength{\unitlength}{.5cm}
\begin{picture}(19,6.5)
\put(0.5,0.5){\scalebox{.4}[.4]{\includegraphics*{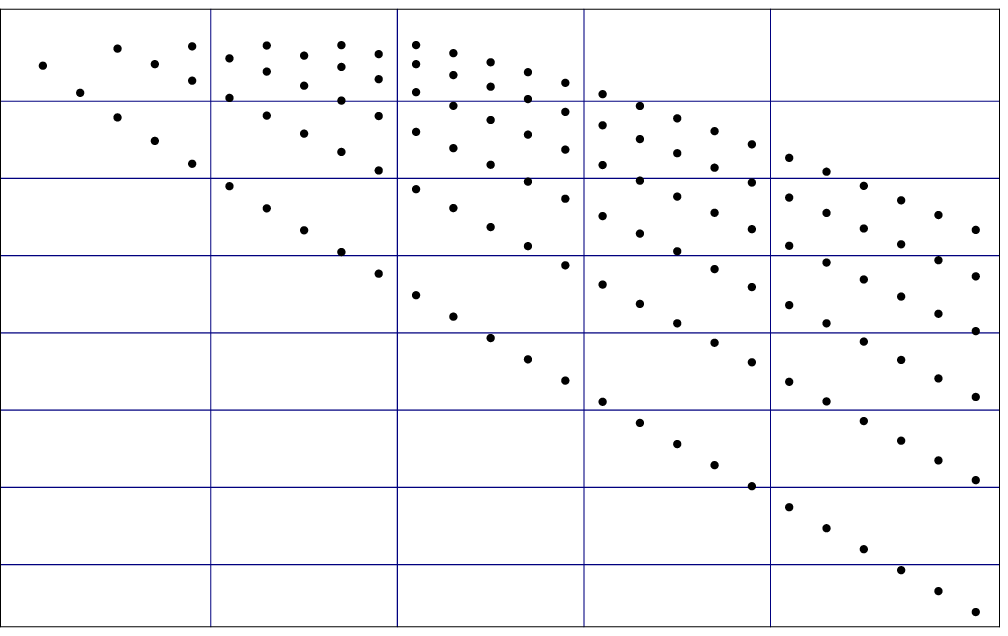}}}
\put(10,0.5){\scalebox{.4}[.4]{\includegraphics*{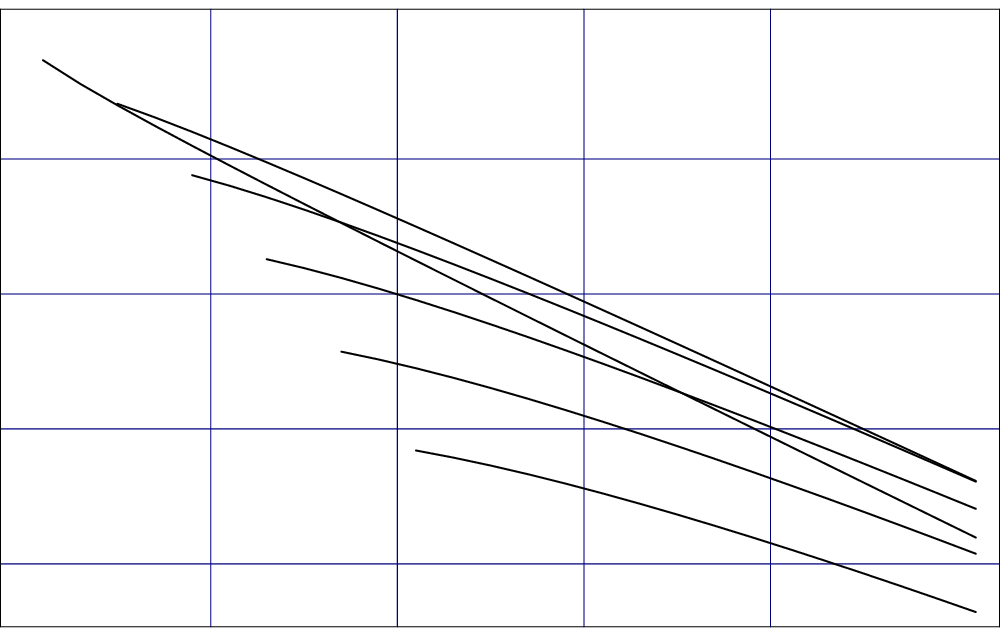}}}
\put(8.5,0.2){\tsz$L$}
\put(2.03,0.2){\tsz$10$}
\put(3.55,0.2){\tsz$20$}
\put(5.07,0.2){\tsz$30$}
\put(6.59,0.2){\tsz$40$}
\put(18,0.2){\tsz$L$}
\put(11.53,0.2){\tsz$10$}
\put(13.05,0.2){\tsz$20$}
\put(14.57,0.2){\tsz$30$}
\put(16.09,0.2){\tsz$40$}
\put(-.09,4.08){\tsz$-10$}
\put(-.09,2.82){\tsz$-20$}
\put(-.09,1.58){\tsz$-30$}
\put(9.4,4.28){\tsz$-10$}
\put(9.4,3.18){\tsz$-20$}
\put(9.4,2.08){\tsz$-30$}
\put(9.4,.98){\tsz$-40$}
\put(2.6,1.2){\tsz$\Delta_r$}
\put(12.1,1.2){\tsz$\Delta_a$}
\put(5.8,2){\tsz$0$}
\put(6.41,2.68){\tsz$4$}
\put(7,3){\tsz$8$}
\put(7.21,3.3){\tsz$12$}
\put(7.5,3.65){\tsz$16$}
\put(7.86,3.9){\tsz$20$}
\put(10.15,5.08){\tsz$0$}
\put(11.4,4.63){\tsz$4$}
\put(11.37,4.1){\tsz$8$}
\put(11.8,3.44){\tsz$12$}
\put(12.4,2.7){\tsz$16$}
\put(12.89,1.9){\tsz$20$}
\end{picture}
\caption[COEFF]{Relative logarithmic error $\Delta_r=\log{|1-b_l^{(L)}/b_l^{\rm (exact)}|}$ on the left,
and the absolute logarithmic error
$\Delta_a=\log{|b_l^{(L)}-b_l^{\rm (exact)}|}$ on the right,
plotted for some strong-coupling coefficients
$b_l$ with $l=0,4,8,12,16,20$
 against the order $L$.}
\label{COEFF}
\end{center}
\end{figure}
We wish to emphasize, that for the Borel-summable
domain with $g>0$,
 variational perturbation theory has the usual
fast convergence in this model.
In fact, for $g=10$, probing deeply into the
strong-coupling domain, we find rapid convergence like $\Delta(L)\simeq 0.02\exp{(-0.73L)}$ for $L\to\infty$, where $\Delta(L)=\log{|Z_{\rm var}^{(L)}-Z_{\rm exact}|}$ is the logarithmic error as a function of the order $L$. This
is shown in Fig.~\ref{NB-II}.
Furthermore, the strong-coupling coefficients $b_l$
of Eq.~(\ref{FP-STRONG}) are reproduced quite satisfactorily.
Having solved $P^{(L)}(\sigma)=0$ for $\sigma$,
we obtain $\Omega^{(L)}(g)$ by solving Eq.~(\ref{FP-s}). Inserting this
and (\ref{FP-EPS}) into (\ref{FP-VAR}), we bring $g^{1/4}\ Z_{\rm var}^{(L)}(g)$ into a form suitable for expansion in powers of $g^{-1/2}$. The expansion coefficients are the strong-coupling coefficients $b_l^{(L)}$ to order $L$. In Fig.~\ref{COEFF} we have plotted the logarithms of their absolute
and relative errors over the order $L$,
and find very good convergence, showing that
variational perturbation theory
works well for
our test-model $Z(g)$.

\begin{figure}[htp!]
\begin{center}
\setlength{\unitlength}{.5cm}~\\[1em]
\begin{picture}(12,8)
\put(0,.7){\scalebox{.6}[.6]{\includegraphics*{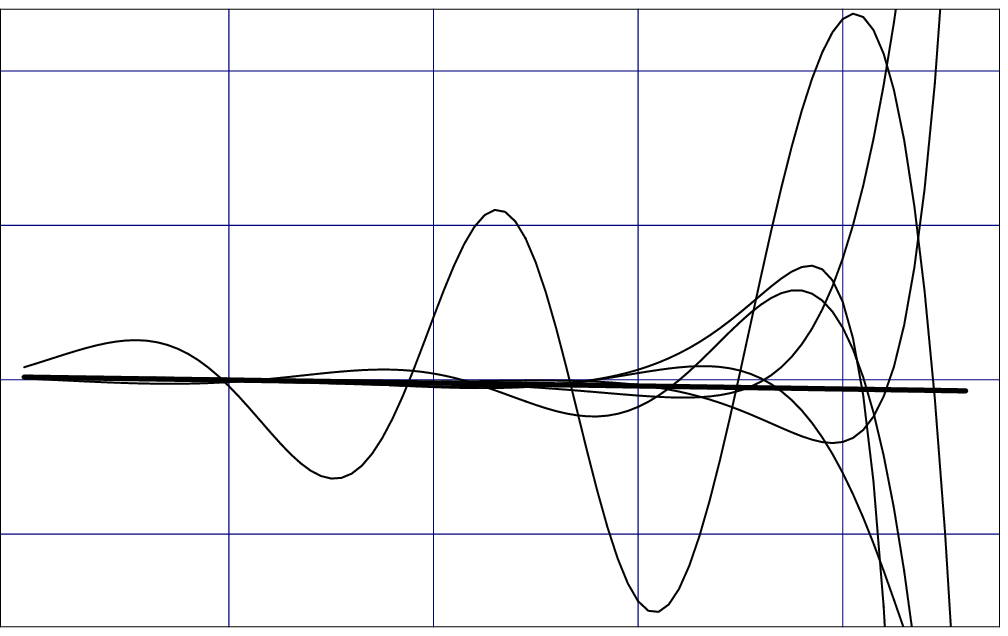}}}
\put(12.1,.3){\tsz$g$}
\put(2.2,.3){\tsz$-.014$}
\put(4.7,.3){\tsz$-.012$}
\put(7.2,.3){\tsz$-.01$}
\put(9.7,.3){\tsz$-.008$}
\put(-.7,1.8){\tsz$-.75$}
\put(-.6,3.67){\tsz$-.7$}
\put(-.7,5.54){\tsz$-.65$}
\put(-.6,7.4){\tsz$-.6$}
\put(10.9,7.9){\ssz A}
\put(11.45,7.9){\ssz B}
\put(10.26,1.9){\ssz C}
\put(10.35,2.6){\ssz D}
\put(11.1,.9){\ssz E}
\put(11.6,.9){\ssz F}
\end{picture}
\caption[NB-III]{Normalized imaginary part Im$[Z_{\rm var}^{(16)}(g)\exp{(-1/4g)}]$
 as a function of $g$ based on six different complex zeros (thin curves). The
fat curve represents the exact value, which is $Z_{\rm exact}(g)\simeq -0.7071+.524g-1.78g^2$. Oscillations of varying strength can be observed near $g=0$. Curves A
and C carry most smoothly near up to the origin.
Evaluation  based
on
 either of them yields equally good results.
 We have selected the zero belonging to curve C as our best choice to this order $L=16$.}
\label{NB-III}
\end{center}
\end{figure}

\begin{figure}[htp!]
\begin{center}
\setlength{\unitlength}{.5cm}
\begin{picture}(19,5.5)
\put(0.5,0.5){\scalebox{.4}[.4]{\includegraphics*{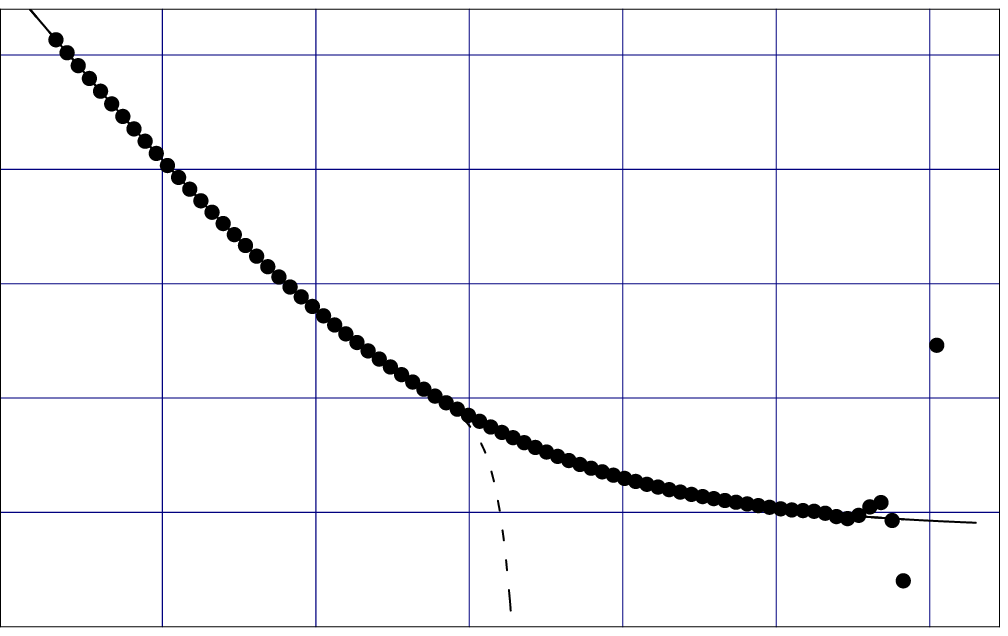}}}
\put(10,0.5){\scalebox{.4}[.4]{\includegraphics*{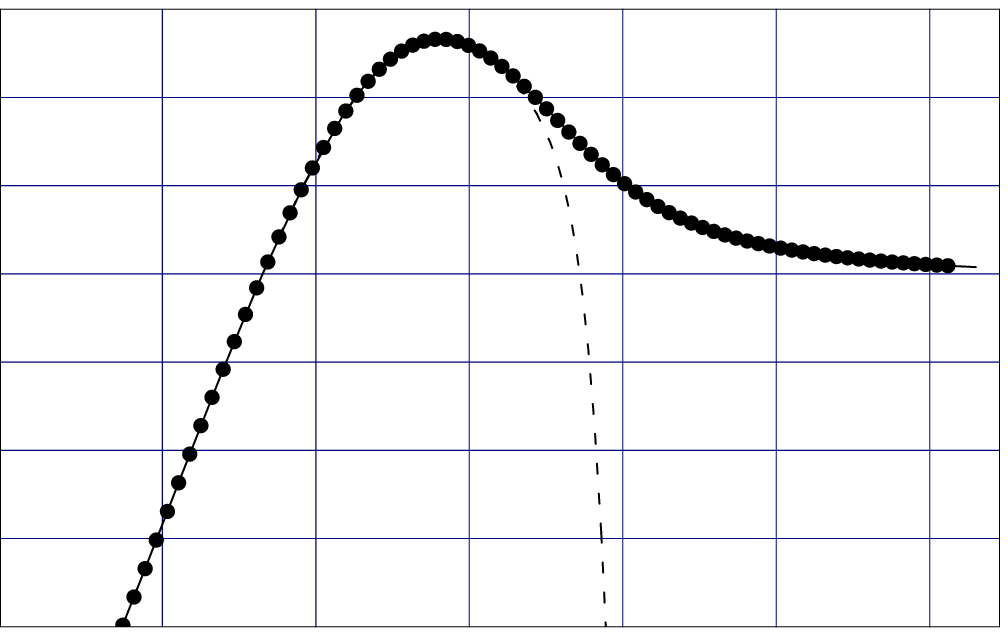}}}
\put(7.7,0.2){\tsz$\log{(-g)}$}
\put(1.72,0.2){\tsz$0$}
\put(4.,.2){\tsz$-2$}
\put(6.55,.2){\tsz$-4$}
\put(17.2,0.2){\tsz$\log{(-g)}$}
\put(11.22,0.2){\tsz$0$}
\put(13.55,.2){\tsz$-2$}
\put(16.05,.2){\tsz$-4$}
\put(9.55,1.9){\tsz$.9$}
\put(9.55,3.3){\tsz$1.0$}
\put(9.55,4.7){\tsz$1.1$}
\put(-.02,1.4){\tsz$-.7$}
\put(-.02,3.25){\tsz$-.6$}
\put(-.02,5.1){\tsz$-.5$}
\end{picture}
\caption[NB-IV]{Normalized imaginary part
Im$[Z_{\rm var}^{(16)}(g)\exp{(-1/4g)}]$ to the left
and the real part Re$[Z_{\rm var}^{(16)}(g)]$ to the right, based on the best zero C from Fig.~\ref{NB-III}, are plotted against $\log{|g|}$ as dots. The solid curve represents the exact function. The dashed curve is the 16th order of the strong-coupling expansion $Z^{(L)}_{\rm strong}(g)$ of equation (\ref{FP-STRONG}).}
\label{NB-IV}
\end{center}
\end{figure}

A better selection of the optimal $ \Omega $ values
comes from the following observation.
The
imaginary parts
of the approximations
near the singularity at $g=0$
show tiny
 oscillations.
The exact imaginary part is
known to decrease extremely fast,  like $\exp{(1/4g)}$, for $g \to 0-$,
practically without oscillations.
We can make  the tiny oscillations
more visible by taking
this exponential  factor out
of the imaginary part.
This is done in Fig.~\ref{NB-III}.
The oscillations differ
strongly  for
different choices of $ \Omega ^{(L)}$
 from the central region of the cluster. To each order $L$
we see that
one of them is smoothest in the sense that
the approximation
 approaches the singularity most closely before oscillations begin.
If this
$ \Omega ^{(L)}$
is chosen as the optimal one, we obtain excellent results
for the
entire
non-Borel-summable region $g<0$.
As an example, we pick the best zero for the $L=16$th order. Fig.~\ref{NB-III} shows the normalized imaginary part calculated to this order, but based on different zeros from the central cluster. Curve C appears optimal. Therefore we select the underlying zero as our best choice at order $L=16$
and calculate with it real
and imaginary part for the
non-Borel-summable region $-2<g<-.008$,
to be compared with the exact values.
Both are shown in Fig.~\ref{NB-IV}, where
we have again renormalized the imaginary part
by the exponential factor $\exp{(-1/4g)}$.
The agreement with the exact result (solid curve) is excellent
as was to be expected because of the fast convergence
observed in Fig.~\ref{NB-I}. It is
indeed much better than the strong-coupling expansion
to the same order, shown as a  dashed curve.
This is the essential improvement of our present theory as compared
to previously known methods probing
 into the tunneling regime \cite{Tunn}.

This non-Borel-summable regime will now be
investigated for the
quantum-mechanical
 anharmonic oscillator.
\subsection{
Tunneling Regime
of Quantum-Mechanical
 Anharmonic Oscillator}
\label{@sec3}
\begin{figure}[htp!]
\begin{center}
\setlength{\unitlength}{.7cm}
\begin{picture}(10,6)
\put(.5,.4){\scalebox{.56}[.56]{\includegraphics*{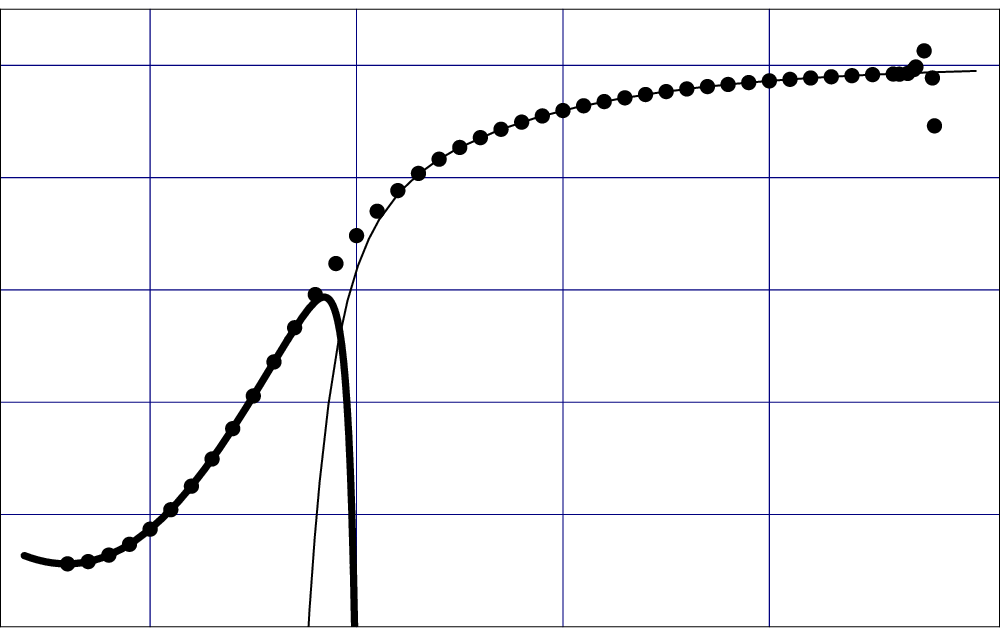}}}
\put(-.1,1.3){\tsz$\small {-.8}$}
\put(-.1,3.1){\tsz$\small {-.4}$}
\put(4.5,3.5){\tsz$l(g)$}
\put(.15,4.9){\tsz$\small {0}$}
\put(1.5,.1){\tsz$\small {-2}$}
\put(3.15,.1){\tsz$\small {-3}$}
\put(4.85,.1){\tsz$\small {-4}$}
\put(6.55,.1){\tsz$\small {-5}$}
\put(8.,.1){\tsz$\log{(-g)}$}
\end{picture}
\caption[I]{Logarithm of
 the imaginary part of
 the ground state energy of
 the anharmonic oscillator with the essential singularity
factored out for better visualization,
$
l(g)=\log\left[ {\sqrt{-\pi g/2}~E_{0,\rm var}^{(64)}(g)}\right] -1/3g$,
plotted against
 small negative values of
 the coupling constant $-0.2<g<-.006$
where the series is
 non-Borel-summable.
The thin curve represents
the divergent expansion around a
critical bubble
of Ref.~\cite{ZINNJ}.
The fat curve is the
 $22$nd order approximation of the
 strong-coupling expansion,
 analytically continued
to negative $g$ in the sliding regime calculated
 in
Chapter 17 of the textbook \cite{PI}.
 }
\label{I}
\end{center}
\end{figure}

The divergent weak-coupling perturbation expansion for the ground state energy of the anharmonic oscillator in the potential $V(x)=x^2/2+g\,x^4$ to order $L$
\begin{align}
\label{WEAK}
E_{0,\rm weak}^{(L)}(g)
 = \sum_{l=0}^L\; a_l\;g^l\,,
\end{align}
where $a_l=(1/2,\ 3/4,\ -21/8,\ 333/16,\ -30885/128,\ \dots)$, is
non-Borel-summable for $g<0$. It may be treated in the same way
as $Z(g)$ of the previous  model, making use as before
of Eqs.~(\ref{FP-VAR})--(\ref{FP-DERIV}), provided we set $p=1$
and $\omega=2/3$, so that $q=3$, accounting for the correct
power behavior $E_0(g)\propto g^{1/3}$ for $g \to \infty$. According to the principle of minimal dependence
and oscillations, we pick a best zero for the order $L=64$ from the
cluster of zeros of $P_L( \sigma )$,
and use it to calculate the logarithm of the normalized imaginary part:
\begin{align}
\label{IM64}
f(g):= \log{\left[ \sqrt{-\pi g/2}\ E_{0,\rm var}^{(64)}(g)\right] }-1/3g\,.
\end{align}
This quantity is plotted
in Fig.~\ref{I}
against $\log (-g)$
close to the tip of the left-hand cut
for $-.2<g<-.006$.
\noindent{}
Comparing our result to older values from
semi-classical calculations \cite{ZINNJ}
\begin{align}
\label{ZJ}
f(g)&=b_1 g-b_2 g^2+b_3 g^3-b_4 g^4+\dots ,
\end{align}
with
\begin{align}
\label{ZJ1}
 b_1 &=3.95833 \quad b_2=19.344 \quad b_3=174.21 \quad b_4=2177\ ,
\end{align}
 shown in Fig.~\ref{I} as
a thin curve, we find very good agreement.
This
expansion
contains the information
on the fluctuations
around the critical bubble.
It is divergent and
non-Borel-summable for $g<0$.
In Appendix B we have rederived
it in a novel way
which allowed us to extend and improve
it considerably.

Remarkably, our theory allows us
 to retrieve
the first three
terms of this expansion
from the
perturbation expansion.
Since our result
provides us with  a regular approximation
to
the essential singularity,
  the fitting procedure
 depends somewhat on
 the
 interval over which we fit
our curve by a power series.
A compromise between a sufficiently long
 interval and
the runaway of
 the divergent
 critical-bubble expansion is obtained for
a lower limit
 $g>-.0229\pm .0003$
and an
 upper limit $g=-0.006$. Fitting a polynomial to
 the data, we extract
 the following
 first three coefficients:
\begin{align}
\label{INST}
b_1 &=3.9586\pm .0003 \quad b_2=19.4\pm .12 \quad b_3=135\pm 18\ .
\end{align}
The agreement
of these numbers
 with those in
(\ref{ZJ})
demonstrates
 that
our method is capable of probing
 deeply into
 the
 critical-bubble region of the coupling constant.
\\

Further evidence
for the quality
of our theory
comes from a comparison
with  the analytically continued strong-coupling result
plotted to order $L=22$ as a
fat curve in Fig.~\ref{I}.
This expansion was derived by a
procedure
of summing non-Borel-summable series  developed
in Chapter 17
of the textbook \cite{PI}.
It was based on a two-step process:
the derivation of a strong-coupling
expansion of the type
(\ref{FP-STRONG}) from the divergent weak-coupling expansion,
and an
analytic
continuation of the strong-coupling expansion to negative $g$.
This method was applicable
only for large enough
coupling strength
where the
strong-coupling expansion converges,
the so-called {\em sliding regime\/}. It could
  not invade into the
 tunneling regime
at
small $g$
governed
by critical bubbles, which
 was treated
in \cite{PI}
 by a separate variational procedure.
The present
work fills the missing gap
by
extending variational perturbation theory
to {\em all\/} $g$ arbitrarily close to zero, without the need
 for a separate treatment of the tunneling regime.
\begin{figure}[tb]
\begin{center}
\setlength{\unitlength}{.5cm}
\begin{picture}(12,8)
\put(0,.7){\scalebox{.6}[.6]{\includegraphics*{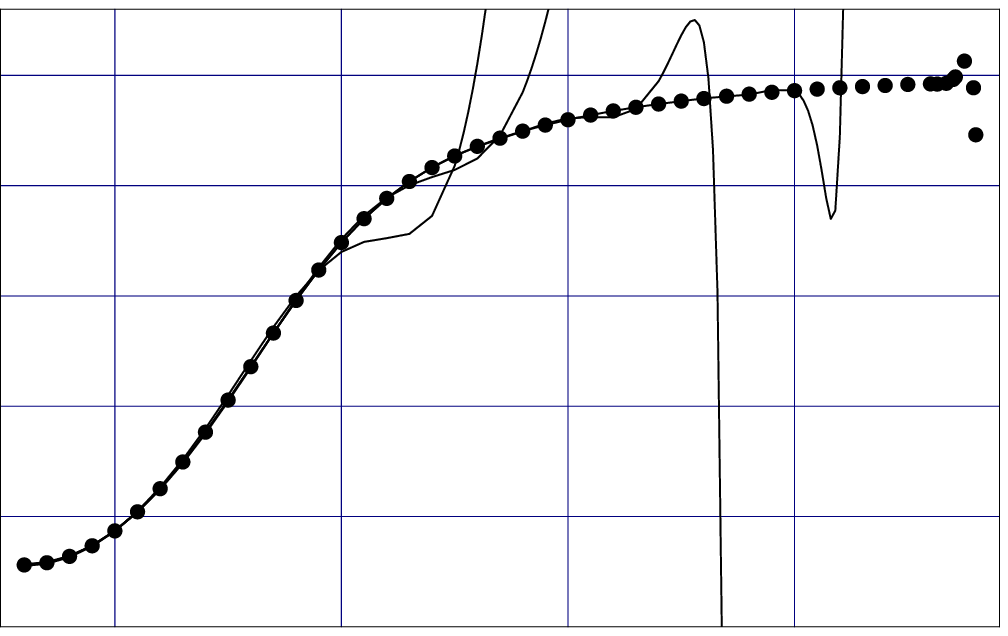}}}
\put(11.5,.3){\tsz$\log (-g) $}
\put(1.1,.3){\tsz$-2$}
\put(3.85,.3){\tsz$-3$}
\put(6.6,.3){\tsz$-4$}
\put(9.4,.3){\tsz$-5$}
\put(-.6,2){\tsz$-.8$}
\put(-.6,3.3){\tsz$-.6$}
\put(-.6,4.65){\tsz$-.4$}
\put(-.6,6){\tsz$-.2$}
\put(-.3,7.37){\tsz$0$}
\put(6.69,7.9){\tsz$8$}
\put(10.35,7.9){\tsz$32$}
\put(5.93,7.9){\tsz$4$}
\put(8.62,7.9){\tsz$16$}
\end{picture}
\caption[II]{Logarithm of the normalized imaginary part of the
ground state energy $\log{(\sqrt{-\pi g/2}\ E_{0,\rm var}^{L)}(g))}-1/3g$,
 plotted against $\log{(-g)}$ for orders $L=4,\ 8,\ 16,\ 32$ (curves). It is compared with the corresponding results for $L=64$ (points). This is shown for small negative values of the coupling constant $-0.2<g<-.006$, i.e. in the
non-Borel-summable critical-bubble region. Fast convergence is easily recognized. Lower orders oscillate more heavily. Increasing orders allow closer approach to the singularity at $g=0-$. }
\label{II}
\end{center}
\end{figure}

It is interesting to see, how the correct limit is approached as the order $L$ increases.
This is shown in Fig.~\ref{II}, based on the
optimal zero in each order. For large negative $g$,
 even the small orders give excellent results.
Close to the singularity the scaling factor $\exp{(-1/3g)}$
will always win over
 the perturbation results.
It is surprising, however, how
fantastically close to the singularity we can go.

\subsection{Dynamic Approach to the Critical-Bubble Regime}
\label{@sec4}
Regarding the computational challenges connected with the
critical-bubble regime of small $g<0$, it is
 worth to develop an independent
 method to calculate imaginary parts
in the tunneling regime.
For a quantum-mechanical system
with an interaction potential
$g\,V(x)$,
such as a
the harmonic oscillator,
we may study the
effect of an  infinitesimal increase in $g$
upon the system.
It  induces an infinitesimal unitary transformation of the
 Hilbert space. The new Hilbert space can be made
 the starting point for the next infinitesimal increase in $g$.
In this way we derive an infinite set of first order ordinary
differential equations for the change of the energy levels
and matrix elements (for details see Appendix C):
\begin{figure}[tb]
\begin{center}
\setlength{\unitlength}{.5cm}
\begin{picture}(12,8.5)
\put(0,.7){\scalebox{.6}[.6]{\includegraphics*{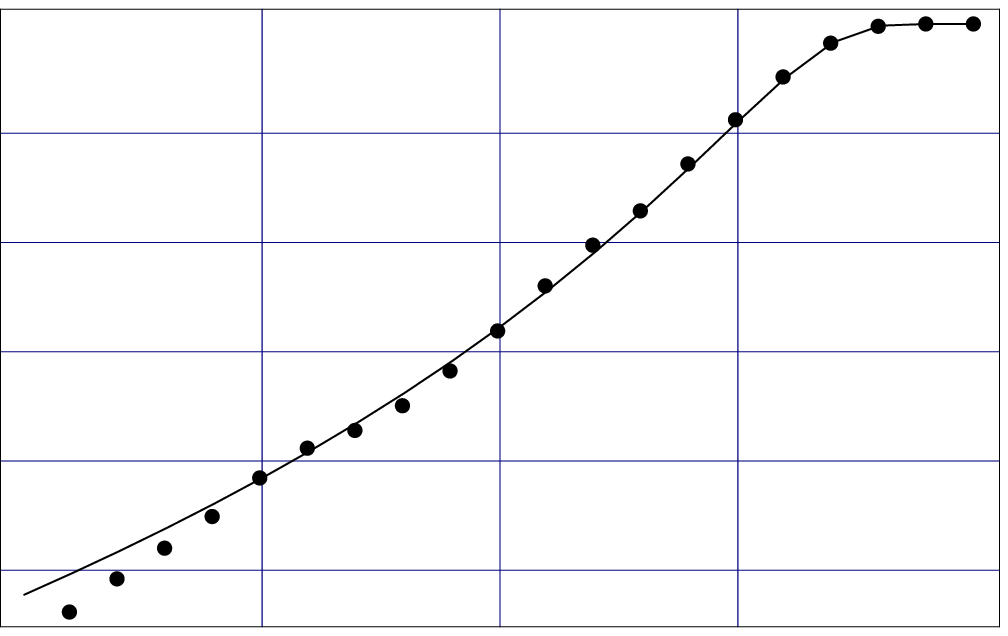}}}
\put(12,.3){\tsz$g$}
\put(2.85,.3){\tsz$-.3$}
\put(5.75,.3){\tsz$-.2$}
\put(8.6,.3){\tsz$-.1$}
\put(-.6,2.7){\tsz$-.2$}
\put(-.6,5.35){\tsz$-.1$}
\end{picture}
\caption[VIx]{inary part of the ground state energy of
the anharmonic oscillator as solution of the coupled set of
differential equations (\ref{DYN}), truncated at the energy level of $n=64$ (points), compared with the corresponding quantity from the $L=64$th order of
non-Borel-summable  variational perturbation theory (curve), both shown as functions of the coupling constant $g$.}
\label{VIx}
\end{center}
\end{figure}
\begin{align}
\label{DYN}
E'_n(g) =& V_{nn}(g), \\
V'_{mn}(g) =& \sum_{k\neq n}\frac{V_{mk}(g)V_{kn}(g)}{E_m(g)-E_k(g)}+ \sum_{k\neq m}\frac{V_{mk}(g)V_{kn}(g)}{E_n(g)-E_k(g)}.
\end{align}
This system of equations
holds for  any one-dimensional Schroedinger problem.
Individual differences  come from  the
initial conditions,
which are the energy levels $E_n(0)$ of the unperturbed system
and the matrix elements $V_{nm}(0)$ of the interaction $V(x)$ in the unperturbed basis. For a numerical integration of the system a truncation is necessary. The obvious way is to restrict the Hilbert
space to the manifold spanned by the lowest $N$ eigenvectors of the unperturbed system. For cases like the anharmonic oscillator, which are even, with even perturbation
and with only an even state to be investigated, we may span the Hilbert space by even basis vectors only. Our initial conditions are thus for $n=0,\ 1,\ 2,\ \dots,\ N/2$:
\begin{align}
\label{INI}
E_{2n}(0) =& 2n+1/2 \\
V_{2n,2m} =& 0 \quad \mbox{if } m<0 \; \mbox{or } m>N/2 \\
V_{2n,2n}(0) =& 3(8n^2 + 4n + 1)/4\\
V_{2n,2n\pm 2}(0) =& (4n+3)\sqrt{(2n+1)(2n+2)}/2 \\
V_{2n,2n\pm 4}(0) =& \sqrt{(2n+1)(2n+2)(2n+3)(2n+4)}/4 \\
\end{align}
For the anharmonic oscillator with a $V(x)=x^4$ potential,
 all sums in equation (\ref{DYN}) are finite with at most four terms
due to the near-diagonal structure of the perturbation.

In order to find a solution for some $g<0$, we first integrate
the system from $0$ to $|g|$,
then around a semi-circle $g=|g|\exp{(i\varphi)}$
 from $\varphi=0$ to $\varphi=\pi$.
The imaginary part of $E_0(g)$ obtained
from a calculation with $N=64$ is shown in
Fig.~\ref{VIx}, where it is
 compared with the variational result for $L=64$. The
agreement is excellent.
 It must be noted, however,
that the necessary truncation of the system of differential
equations introduces an error,
which cannot be made arbitrarily small by
increasing the truncation limit $N$. The approximations
are
asymptotic
sharing this property with the original weak-coupling series.
Its divergence is, however,
reduced considerably,
which  is the reason
why we obtain
accurate results
for the
critical-bubble regime, where
 the weak-coupling series
fails completely to
reproduce
the imaginary part.

\section{Hydrogen Atom in Strong Magnetic Field}

A point particle in $D$ dimensions with a potential $V({\bf x})$ and a vector
potential ${\bf A}({\bf x})$ is described by
a Hamiltonian
\begin{equation}
\label{ham00}
H({\bf p},{\bf x})=\frac{1}{2M}\left[{\bf p}-\frac{e}{c}{\bf A}({\bf x})\right]^2-\frac{e^2}{4\pi |{\bf x}|}.
\end{equation}
The quantum statistical partition function is given by the euclidean
phase space path integral
\begin{equation}
\label{ham01}
Z=\oint {\cal D'}^Dx {\cal D}^Dp\,e^{-{\cal A}[{\bf p},{\bf x}]/\hbar}
\end{equation}
with an action
\begin{equation}
\label{ham02}
{\cal A}[{\bf p},{\bf x}]=\int_0^{\hbar\beta}d\tau\left[-i{\bf p}(\tau)\cdot\dot{\bf x}(\tau)
+H({\bf p}(\tau),{\bf x}(\tau))\right],
\end{equation}
and the path measure
\begin{equation}
\label{ham03}
\oint{\cal D'}^Dx {\cal D}^Dp=\lim_{N\to\infty}\prod\limits_{n=1}^{N+1}
\left[\int\frac{d^Dx_nd^Dp_n}{(2\pi\hbar)^D} \right].
\end{equation}
The parameter $\beta=1/k_BT$ denotes the usual inverse thermal energy at
temperature $T$, where $k_B$ is the Boltzmann constant. From $Z$ we obtain the free energy
of the system:
\begin{equation}
\label{ham04}
F=-\frac{1}{\beta}{\rm ln}\,Z.
\end{equation}

Applying variational perturbation theory
to the path integral
(\ref{ham01})
leads to a variational binding
energy \cite{BKP}
defined by $ \varepsilon (B)\equiv
B/2-E(B)$
in atomic natural
with $\hbar =1$, $M=1$, $e=1$, energies in units
of 2\,Ryd$=e^4M^2/\hbar ^3$.
\begin{equation}
\label{we00}
\varepsilon^{(1)}_{\eta,\Omega}(B)=
\frac{B}{2}-\frac{\Omega}{4}\left(1+\frac{\eta}{2}\right)-\frac{B^2}{4\Omega}
-\sqrt{\frac{\eta\Omega}{2\pi}}h(\eta)
\end{equation}
with
\begin{equation}
\label{we01}
h(\eta)=\frac{1}{\sqrt{1-\eta}}\,{\rm ln}\,\frac{1-\sqrt{1-\eta}}{1+\sqrt{1-\eta}}.
\end{equation}
Here we have introduced variational parameters
\begin{equation}
\label{we02}
\eta\equiv\frac{2\op}{\osb}\leq1,\qquad \Omega\equiv\osb.
\end{equation}
Extremizing
the energy
with respect to
 these
yields the conditions
\begin{eqnarray}
\label{we03}
\frac{\Omega}{8}+\sqrt{\frac{\Omega}{2\pi\eta}}\frac{1}{1-\eta}\left(1+\frac{1}{2}
\frac{1}{\sqrt{1-\eta}}{\rm ln}\,\frac{1-\sqrt{1-\eta}}{1+\sqrt{1-\eta}}\right)&\stackrel{!}{=}&0,
\nonumber\\
\label{we04}
\frac{1}{4}+\frac{\eta}{8}-\frac{B^2}{4\Omega^2}+\frac{1}{2}\sqrt{\frac{\eta}{2\pi\Omega}}
\frac{1}{\sqrt{1-\eta}}{\rm ln}\,\frac{1-\sqrt{1-\eta}}{1+\sqrt{1-\eta}}&\stackrel{!}{=}&0.
\end{eqnarray}
Expanding the variational parameters into perturbation series of the square magnetic field
$B^2$,
\begin{equation}
\label{we05}
\eta(B)=\sum\limits_{n=0}^{\infty}\,\eta_n B^{2n},\qquad
\Omega(B)=\sum\limits_{n=0}^{\infty}\,\Omega_n B^{2n}
\end{equation}
and inserting these expansions into the self-consistency conditions (\ref{we03}) and
(\ref{we04}) we obtain order by order the coefficients given in Table~\ref{tab1}.
Inserting these values into the expression for the binding energy (\ref{we00}) and expand
with respect to $B^2$, we obtain the perturbation series
\begin{equation}
\label{we06}
\varepsilon^{(1)}(B)=\frac{B}{2}-\sum\limits_{n=0}^\infty\,\varepsilon_n B^{2n}.
\end{equation}
The first coefficients are also given in Table~\ref{tab1}. We find thus the important
result that the first-order variational perturbation solution possesses a perturbative
behavior with respect to the square magnetic field strength $B^2$ in the weak-field limit thus
yielding the correct asymptotic. The coefficients differ in higher order from the
exact ones
but are improved by
 variational perturbation
theory~\cite{PI}.
\begin{table}
\caption[]{\label{tab1} Perturbation coefficients up to order $B^6$ for the weak-field
expansions of the variational parameters and the binding energy in comparison to the
exact ones of Ref.~\cite{cizek1}.}
\centerline{\phantom{XXx}
\begin{tabular}{c|cccc}
$n$ & 0 & 1 & 2 & 3\\ \hline\hline
$\eta_n$ & $1.0$ & \fsz$-\frac{405\pi^2}{7168}\approx -0.5576$ &
\fsz$\frac{16828965\pi^4}{1258815488}\approx 1.3023$ &
\fsz$-\frac{3886999332075\pi^6}{884272562962432}\approx -4.2260$\\[1mm] \hline
$\Omega_n$ & \fsz$\frac{32}{9\pi}\approx 1.1318$ & \fsz$\frac{99\pi}{224}\approx 1.3885$ &
\fsz$ -\frac{1293975\pi^3}{19668992}\approx -2.03982$ &
\fsz$\frac{524431667187\pi^5}{27633517592576}\approx 5.8077$
\\[1mm] \hline
$\varepsilon_n$ & \fsz$ -\frac{4}{3\pi}\approx -0.4244$ & \fsz$\frac{9\pi}{128}\approx 0.2209$ &
\fsz$ -\frac{8019\pi^3}{1835008}\approx -0.1355$ & \fsz$\frac{256449807\pi^5}{322256764928}
\approx 0.2435$\\[1mm] \hline
$\varepsilon_n$~\cite{cizek1} & $-0.5$ & $0.25$ & \fsz$ -\frac{53}{192}\approx -0.2760$ &
\fsz$\frac{5581}{4608}\approx 1.2112$
\end{tabular}
}\end{table}

In a strong magnetic field one has
\begin{equation}
\label{st00}
\os \gg 2\op,\qquad \op \ll B
\end{equation}
and the variational expression   simplifies to
\begin{equation}
\label{st01}
\varepsilon_{\os,\op}^{(1)}=\frac{B}{2}-\left(\frac{\os}{4}+\frac{B^2}{4\os}+\frac{\op}{4}
+\sqrt{\frac{\op}{\pi}}\,{\rm ln}\,\frac{\op}{2\os}\right),
\end{equation}
which is minimal at
\begin{eqnarray}
\label{st03}
\sqrt{\op}&=&-\frac{2}{\sqrt{\pi}}\left({\rm ln}\,\op-{\rm ln}\,\os+2-{\rm ln}\,2 \right),\\
\label{st04}
\os&=&2\sqrt{\frac{\op}{\pi}}+B\sqrt{1+4\frac{\op}{\pi B^2}}.
\end{eqnarray}
Expanding
the second conditions as
\begin{equation}
\label{st05}
\os=B+2\sqrt{\frac{\op}{\pi}}+2 \frac{\op}{\pi B}-4\frac{\op^2}{\pi^2 B^3}+\ldots,
\end{equation}
and inserting only the first two terms
 into the first
 condition (\ref{st03}), we
 neglect terms of order $1/B$, and find
\begin{equation}
\label{st07}
\sqrt{\op}\approx\frac{2}{\sqrt{\pi}}\left({\rm ln}\,B-{\rm ln}\,\op^{(1)}+{\rm ln}\,2-2 \right).
\end{equation}
To obtain a tractable approximation for $\op$, we perform some iterations starting from
\begin{equation}
\label{st08}
\sqrt{\op^{(1)}}=\frac{2}{\sqrt{\pi}}{\rm ln}\,2Be^{-2}
\end{equation}
Reinserting this on the right-hand side of Eq.~(\ref{st07}), one obtains the second iteration
$\sqrt{\op^{(2)}}$. We stop this procedure after an additional reinsertion which yields
\begin{equation}
\label{st09}
\sqrt{\op^{(3)}}=\frac{2}{\sqrt{\pi}}\left({\rm ln}\,2Be^{-2}-2{\rm ln}\left[\frac{2}{\sqrt{\pi}}
\left\{{\rm ln}\,2Be^{-2}-2{\rm ln}\,\left(\frac{2}{\sqrt{\pi}}{\rm ln}\,2Be^{-2}\right)
\right\}\right] \right).
\end{equation}
The reader may convince himself that this iteration procedure indeed converges.
For a subsequent systematical extraction of terms essentially contributing to the binding
energy, the expression (\ref{st09}) is not satisfactory. Therefore it is better to separate
the leading term in the curly brackets and expand the logarithm of the remainder. Then this
procedure
is applied to the expression in the square
 brackets and so on. Neglecting terms
of order ${\rm ln}^{-3} B$, we obtain
\begin{equation}
\label{st10}
\sqrt{\op^{(3)}}\approx \frac{2}{\sqrt{\pi}}\left({\rm ln}\,2Be^{-2}+{\rm ln}\frac{\pi}{4}
-2{\rm ln}{\rm ln}\,2Be^{-2}\right).
\end{equation}
The double-logarithmic term can be expanded in a similar way as described above:
\begin{equation}
\label{st11}
{\rm ln}{\rm ln}\,2Be^{-2}={\rm ln}\left[{\rm ln}\,B\left(1+\frac{{\rm ln}\,2-2}{{\rm ln}\,B}
\right) \right]= {\rm ln}{\rm ln}\,B+\frac{{\rm ln}\,2-2}{{\rm ln}\,B}-\frac{1}{2}
\frac{({\rm ln}\,2-2)^2}{{\rm ln}^2B}+{\cal O}({\rm ln}^{-3} B).
\end{equation}
Thus the expression (\ref{st10}) may be rewritten as
\begin{equation}
\label{st12}
\sqrt{\op^{(3)}}= \frac{2}{\sqrt{\pi}}\left( {\rm ln}\,B -2{\rm ln}{\rm ln}\,B+
\frac{2a}{{\rm ln}\,B}+\frac{a^2}{{\rm ln}^2B}+b \right) +{\cal O}({\rm ln}^{-3} B)
\end{equation}
with abbreviations
\begin{equation}
\label{st13}
a=2-{\rm ln}\,2 \approx 1.307,\qquad b= {\rm ln}\frac{\pi}{2}-2\approx -1.548.
\end{equation}
The first observation is that the variational parameter $\op$ is always
much smaller
than $\os$ in the high $B$-field limit. Thus we can further simplify
the approximation
(\ref{st05}) by replacing
\begin{equation}
\label{st14}
\os\approx B\left(1+\frac{2}{B}\sqrt{\frac{\op}{\pi}}\right)\longrightarrow B
\end{equation}
without affecting the following expression for the binding energy. Inserting the solutions
(\ref{st12}) and (\ref{st14}) into the equation for the binding energy (\ref{st01}) and
expanding the logarithmic term once more as described, we find up to the
order ${\rm ln}^{-2}B$:
\begin{eqnarray}
\label{st15}
\varepsilon^{(1)}(B)&=&\frac{1}{\pi}\left({\rm ln}^2B -4\,{\rm ln}\,B\; {\rm ln}{\rm ln}\,B
+4\,{\rm ln}^2{\rm ln}\,B -4b\,{\rm ln}{\rm ln}\,B
+2(b+2)\,{\rm ln}\,B
+b^2
-\frac{1}{{\rm ln}\,B}\left[8\,{\rm ln}^2{\rm ln}\,B-8b\,{\rm ln}{\rm ln}\,B+2b^2
\right]\right)\nonumber\\
&&+{\cal O}({\rm ln}^{-2}B)
\end{eqnarray}
Note that the prefactor $1/\pi$ of the leading ${\rm ln}^2B$-term
differs from a value $1/2$ obtained by
Landau and Lifschitz~\cite{landau}.
Our different value is a consequence of using a harmonic trial system.
The calculation of higher orders
in variational perturbation theory would improve the value of the prefactor.

At a magnetic field strength $B=10^5 B_0$, which corresponds to $2.35\times 10^{10}\,{\rm T}=
2.35\times10^{14}\,{\rm G}$,
the contribution from the first six terms is $22.87\,[2\,{\rm Ryd}]$.
The next three terms suppressed
by a factor ${\rm ln}^{-1} B$ contribute $-2.29\,[2\,{\rm Ryd}]$,
while an estimate for the ${\rm ln}^{-2} B$-terms
yields nearly $-0.3\,[2\,{\rm Ryd}]$. Thus we find
\begin{equation}
\label{st16}
\varepsilon^{(1)}(10^5)=20.58\pm 0.3\,[2\,{\rm Ryd}].
\end{equation}
This is in very good agreement with the value $20.60\,[2\,{\rm Ryd}]$
obtained from an accurate numerical treatment
\cite{wunner}.
\begin{table}
\caption{\label{asymp} Example for the competing leading six terms in Eq.~(\ref{st15})
at $B=10^5B_0\approx 2.35\times 10^{14}\,{\rm G}$.}
\centerline{\phantom{xxxxxxxx}\begin{tabular}{cccccc}
\fsz $(1/\pi){\rm ln}^2B$ & \fsz $-(4/\pi){\rm ln}\,B\;{\rm ln}{\rm ln}\,B$ &
\fsz $(4/\pi)\,{\rm ln}^2{\rm ln}\,B$ &
\fsz $-(4b/\pi)\,{\rm ln}{\rm ln}\,B$ & \fsz $[2(b+2)/\pi]\,{\rm ln}\,B$ & \fsz $b^2/\pi$\\ \hline
\fsz $42.1912$ & \fsz $-35.8181$ & \fsz $7.6019$ & \fsz $4.8173$ & \fsz $3.3098$ & \fsz $0.7632$
\end{tabular}}
\end{table}

\begin{figure}[thb]
\centerline{
\epsfxsize=12cm \epsfbox{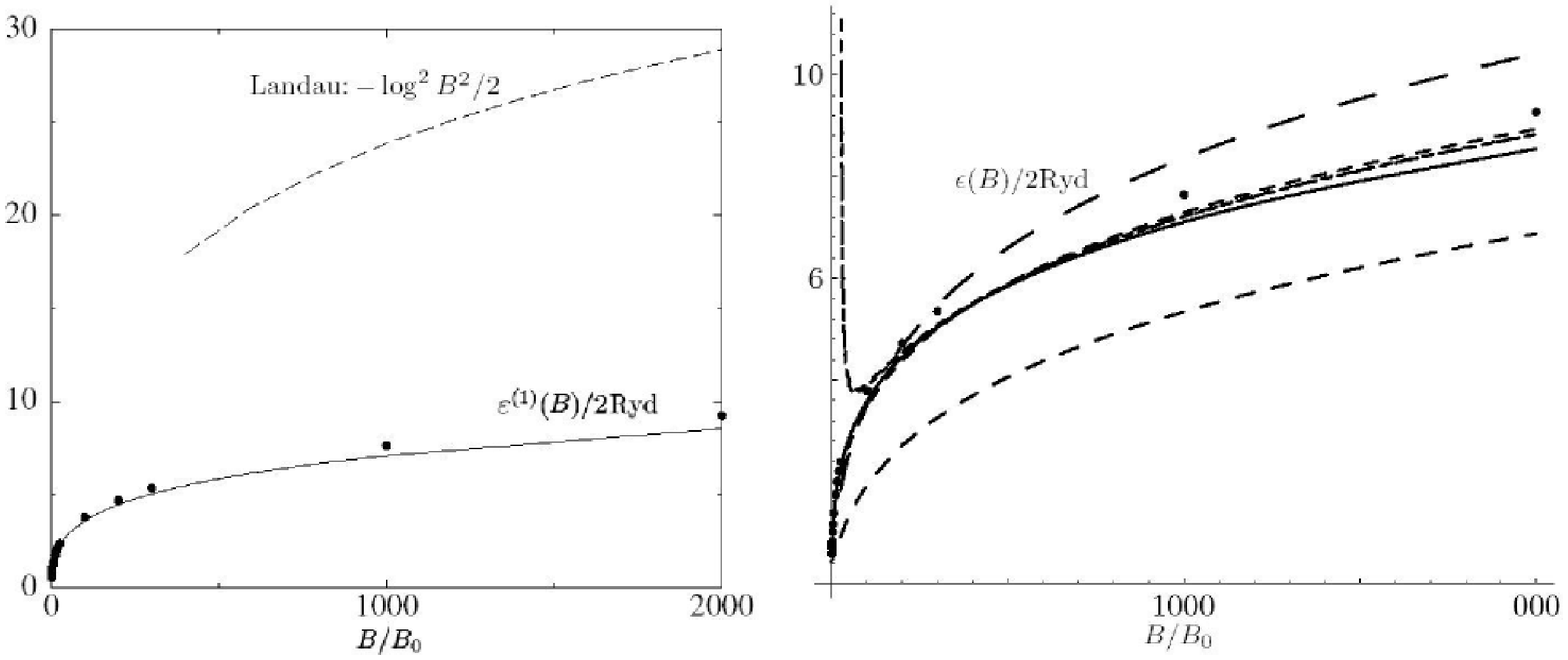}
}   \vspace{-.5cm}
\caption{
Ground state energy $E(B)$ of
hydrogen in a strong magnetic field
The dotted figure on the left is Landau's
old upper limit.
On the right-hand side
our curve is compared with the accurate values (dots \cite{AVR,wunner}). It also shows
various lower-order approximations within our procedure.
The quantity $ \varepsilon (B)$ is the binding energy
defined by $ \varepsilon (B)\equiv
B/2-E(B)$.
All quantities are in atomic natural units
$\hbar =1$, $M=1$, $e=1$, energies in units of 2\,Ryd$=e^4M^2/\hbar ^3$. }
 \label{@figB}
\end{figure}

Table~\ref{asymp} lists the values of the first six terms of Eq.~(\ref{st15}).
This shows in particular the significance of the second-leading term
$-(4/\pi){\rm ln}\,B\;{\rm ln}{\rm ln}\,B$, which is of the same order of the leading term
$(1/\pi){\rm ln}^2B$ but with an opposite sign. In Fig.~\ref{@figB},
we have plotted the expression
\begin{equation}
\label{st17}
\varepsilon_L(B)=\frac{1}{2}\,{\rm ln}^2B
\end{equation}
from Landau and Lifschitz~\cite{landau} to illustrate that it gives far too large
binding energies even at very large magnetic fields, e.g. at $2000 B_0\propto 10^{12}\,{\rm G}$.

This strength of magnetic field appears on surfaces of neutron stars
($10^{10}-10^{12}\,{\rm G}$). A recently discovered new type of neutron star is the
so-called magnetar. In these, charged particles such as protons and electrons produced by
decaying neutrons give rise to the giant magnetic field of $10^{15}\,{\rm G}$.
Magnetic fields of white dwarfs reach only up to $10^6-10^8\,{\rm G}$.
All these magnetic field strengths are far from realization in
experiments. The strongest magnetic fields ever produced in a laboratory were only of the order
$10^5\,{\rm G}$, an order of magnitude larger than the fields in sun spots which reach about
$0.4\times10^4\,{\rm G}$. Recall, for comparison, that the earth's
magnetic field has the small value of $0.6\,{\rm G}$.

The nonleading terms in Eq.~(\ref{st15})
give important contributions to the asymptotic behavior even at such large magnetic fields,
as we can see in Fig.~\ref{@figB}.
It is an unusual property of the asymptotic behavior that the absolute value of
the difference between the Landau-expression (\ref{st17}) and our approximation (\ref{st15})
diverges with increasing magnetic field strengths $B$, only the relative difference
decreases.

\comment{
variational perturbation theory
permits to find an accurate approximation
to the ground state energy as a function of the field strength
shown in Fig.~\ref{@figB}.
}

\comment{
\medskip
\noindent{\bf Remark:} We note that similar results can be proven
for the zeros of the Dual Hahn, Continuous Hahn and Continuous Dual
Hahn polynomials using the contiguous relations for generalized
hypergeometric polynomials (cf \cite[eq, 14, 15, p. 82]{Rainville})
as above.
}




%
\section{Appendix A: Modification of Principle of Minimal Sensitivity}
\label{APPA}
The naive quantum mechanical variational perturbation theory
has been used by many authors under the name $ \delta $-expansion.
This name stems from the fact that
one may
write the Hamiltonian
of an anharmonic oscillator
\begin{eqnarray}
H=\frac{p^2}{2M}+ \frac{M}2\omega ^2x^2 +\frac{g}{4}x^4
\label{@}\end{eqnarray}
alternatively as
\begin{eqnarray}
H=\frac{p^2}{2M}+M \frac{\Omega ^2}2x^2 +
 \delta \left[ \frac{M}{2}\
\left( \omega ^2- \Omega ^2\right)+\frac{g}{4}x^4\right],
\label{@}\end{eqnarray}
and expand the eigenvalues
systematically
in powers of $ \delta $.
Each partial sum of order $L$ is evaluated at $ \delta =1$
and extremized in $ \Omega $.  It is obvious that this procedure
is equivalent the
re-expansion method in Section \ref{QME}.

As mentioned in the text and pointed out in \cite{HK1},
such an analysis is inapplicable
in quantum field theory, where
the Wegner exponent
$ \omega $
is anomalous and
must be determined dynamically.
Most recently, the false treatment
was given
to the
shift of the
critical temperature
 in a Bose-Einstein condensate caused by a small interaction
\cite{Braaten,ramospr,prb}.
We have seen in Section
\ref{BECa} that
the
perturbation expansion for this quantity
is a function of $g/\mu$ where $\mu$ is the
chemical potential
which goes to zero
at the critical point,
we are faced with a
typical strong-coupling problem
of critical phenomena.
In order to justify the
application of the
$ \delta $-expansion
to this problem,
 BR \cite{Braaten}
studied the
convergence
properties
of the
 method
by applying it to
a certain  amplitude $ \Delta (g)$ of an
$O(N)$-symmetric $\phi^4$-field theory
in the
limit of large $N$, where
the model is exactly solvable.

Their procedure must be criticized in two ways.
First, the amplitude $ \Delta (g)$ they considered
 is not a good candidate for
a  resummation  by a $ \delta $-expansion
since
 it does not possess the
characteristic
strong-coupling power structure
 \cite{HKSC}
 of quantum mechanics and
 field theory,
which the
final
resummed expression will always have
by construction.
The power structure is disturbed
by additional logarithmic terms.
Second, the
 $ \delta $-expansion
is, in the example,
equivalent to  choosing, on dimensional grounds, the
 exponent
 $\omega=2$ in
 \cite{HKSC}, which is far from the
correct value $\approx 0.843$
to be derived below.
Thus the $ \delta $-expansion is
inapplicable, and this
 explains  the
 problems into which BR run in
their resummation attempt.
Most importantly, they
do
 not
 find
a well-shaped plateau of the variational expressions
 $ \Delta^{(L)} (g,z)$ as a function of $z$ which would be
 necessary for invoking the
principle of minimal sensitivity.
Instead, they observe
that the
zeros of the
first derivatives
$\partial _z \Delta^{(L)} (g,z)$
 run  away
far into the
complex plain.
Choosing the complex
 solutions
to determine their final resummed value misses the
correct one by 3\%
up to the 35th order.

One may improve the situation
by
 trying out
various different  $\omega$-values and
 choosing
the
best of them yielding
an acceptable plateau in $ \Delta (g,z)$.
This happens for $\omega \approx 0.843$.
However, even
for this optimal value, the
resummation result
never converges to the
correct limit.
For $ \Delta (g)$
the error happens to be numerically small, only
 0.1\%,
but it will be uncontrolled
in physical problems
where the result is unknown.

Let us explain these points in more detail.
BR
consider
the
weak-coupling series
with the reexpansion parameter $ \delta $:
\begin{align}
\label{BRAA1}
\Delta(\delta, g) &=-\sum_{l=1}^\infty
\Big(-\frac{\delta \,g}{\sqrt{1-\delta}}\Big)^l a_l~,  &\!\!\!\!\!\!
{\rm where}~~~
a_l &\equiv  \int_0^\infty K(x)f^l(x)\,dx\,,
\end{align}
with
\begin{align}
\label{BRAA1.1}
K(x) &\equiv  \frac{4 x^2}{\pi (1+x^2)^2}~,  &
f(x) &\equiv \frac{2}{x}\arctan \frac{x}{2}\,.
\end{align}
The geometric series in (\ref{BRAA1}) can be summed exactly,
and
 the
result may formally be reexpanded into
a strong-coupling series in $h\equiv
{\sqrt{1-\delta}}/({\delta \,g})$:
\begin{align}
\label{BRAA2}
\Delta(\delta, g) &= \int_0^\infty K(x) \frac{\delta g f(x)}{\sqrt{1-\delta}+\delta g f(x)}\,dx
=\sum_{m=0}^\infty b_m
\left(-h\right)^m,
 &  \mbox{where~~  }
b_m= \int_0^\infty K(x)f^{-m}(x)\,dx\,.
\end{align}
The strong-coupling limit  is found for $h\rightarrow 0$
where
$ \Delta \rightarrow b_0
=\int_0^\infty dx\, K(x)=1$.
The approach to this limit is, however,
{\em not\/}  given by a
strong-coupling expansion of the
form
 (\ref{BRAA2}). This would only happen
if all the
integrals $b_m$ were to exist which,
unfortunately, is not the
case
since
all
 integrals for $b_m$
with
  $m>0$
diverge at the
upper limit, where
\begin{align}
\label{BRAA1.2}
f(x) &= \frac{2}{x}\arctan \frac{x}{2} \sim \frac{\pi}{x}\,.
\end{align}
The exact behavior of $\Delta$
in the strong-coupling limit
 $h \to 0$
is found
 by studying the
effect of the
asymptotic  $\pi /x$-contribution of $f(x)$
to the
integral in (\ref{BRAA2}). For $f(x)=\pi /x$ we obtain
\begin{align}
\label{BRAA3}
\int_0^\infty K(x) \frac{1}{1+h / f(x)}\,dx = \frac{\pi^4+2\pi²h-\pi²h²+2h³+4\pi²h \log{h/\pi}}{(\pi²+h²)²}\,.
\end{align}
The logarithm of $h$
shows a mismatch with the
general
asymptotic form of the result
 \cite{HKSC}, which
and prevents
 the
expansion
(\ref{BRAA1}) to be a candidate
for variational perturbation theory.

We now explain the
second criticism.
Suppose we
 ignore the
just-demonstrated fundamental
 obstacle and
follow the
rules of the
$\delta$-expansion, defining
 the
$L$th order
approximant  $\Delta( \delta ,\infty)$
by expanding (\ref{BRAA1})
 in powers of $\delta$ up to order $ \delta ^L$,
 setting $\delta=1$, and
 defining $z\equiv g$.
Then we
 obtain the
$L$th variational expression for $b_0$:
\begin{align}
\label{BRAAST}
b_0^{(L)}( \omega ,z)=\sum_{l=1}^L a_l z^l \binom{L-l+l/\omega}{L-l}~,
\end{align}
with $ \omega =2$, to be optimized in $z$.
This $ \omega $-value
would only be adequate
if the
approach to the
strong-coupling limit
 behaved like $A + B/h^2+\dots$, rather
than
 (\ref{BRAA3}).
This is the
reason why BR  find
no real regime of minimal sensitivity on $z$.
\begin{figure}[htp]
\begin{center}
\setlength{\unitlength}{1cm}
\centerline
{\phantom{}
\scalebox{.65}[.65]{\includegraphics*{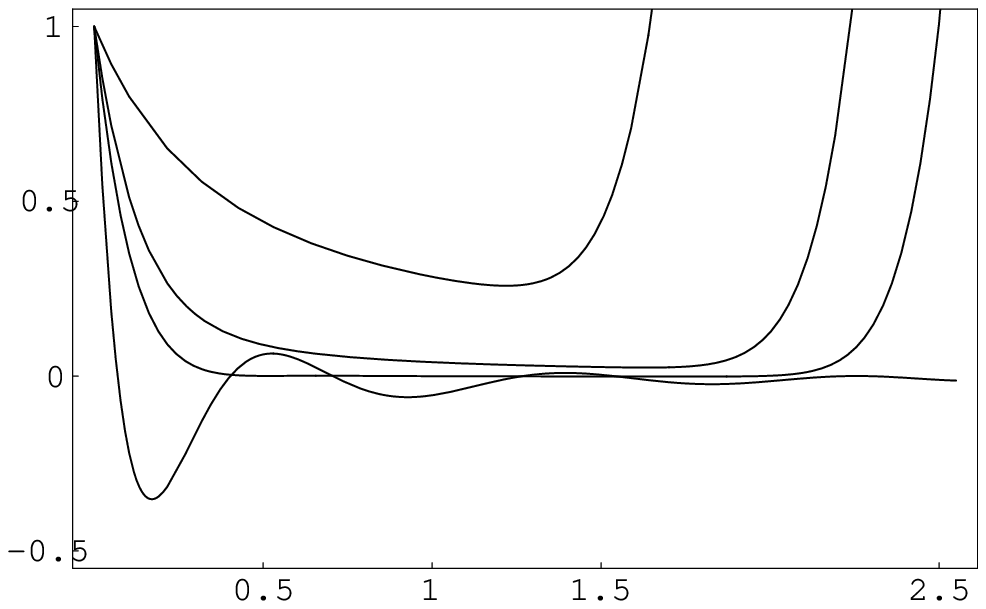}}
}\caption[BRI]{Plot of $1-b_0^{(L)}(\omega,z)$
versus $z$ for $L=10$ and
$\omega=0.6,~0.843,~ 1,~2~$.
The
curve
 with $\omega=0.6$
 shows oscillations. They decrease with
increasing $\omega$
and
 becomes flat
 at about
$\omega=0.843$. Further increase
of $ \omega $  tilts the
plateau and
 shows no regime of
minimal sensitivity.
At the
same time, the
minimum of the
curve rises
 rapidly above the
correct value of $1-b_0=0$,
as can be seen from the
upper two curves for $\omega=1$
and
  $\omega=2$, respectively. }
\label{BRI}
\end{center}
\end{figure}%

Let us attempt to
improve the situation
by determining $\omega$ dynamically
by making the plateau
in the plots of $\Delta^{(L)}(\omega,h)$
versus $h$
 horizontal
 for several different $\omega$-values.
The result is
 $\omega \approx 0.843$, quite far from the
naive value $2$.
This value can also  be
estimated by inspecting plots of $\Delta^{(L)}(\omega,h)$
versus $h$ for several different $\omega$-values in
Fig.~\ref{BRI}, and
 selecting the
one producing minimal sensitivity.
\begin{figure}[htp!]
\begin{center}
\setlength{\unitlength}{1cm}
\begin{picture}(19,6)
\phantom{XxxxxXXXX}
\scalebox{1.2}[1.2]{\includegraphics*{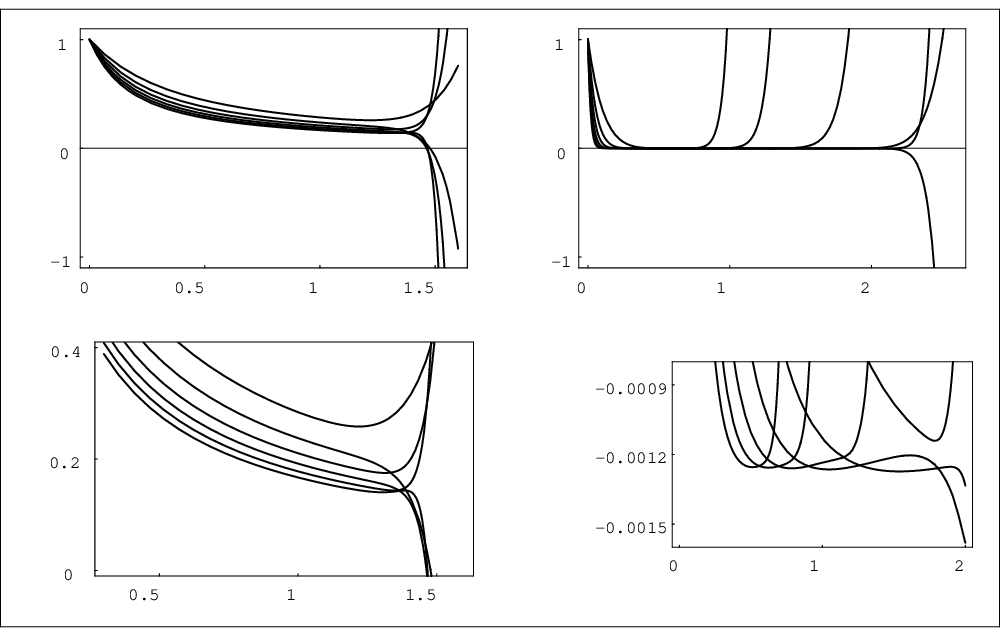}}
\end{picture}
\caption[BRII]{
Left-hand
 column shows plots
of
$1-b_0^{(L)}(\omega,z)$ for $L=10, ~17, ~24, ~31, ~38,~45$
with $\omega=~2$ of $ \delta $-expansion
of BR,
right-hand
 column with optimal $\omega=0.843$.
The lower row  enlarges the
interesting plateau regions
of the
plots above.
Only the
right-hand
 side
shows minimal sensitivity, and
 the
associated
plateau lies closer
to the
correct value  $1-b_0 = 0$
than
 the
minima in
  the
left column by
 two orders of magnitude.
Still the
right-hand
 curves do not approach the
exact limit
for $L\rightarrow \infty$
due to the
wrong strong-coupling behavior of the
initial function.}
\label{BRII}
\end{center}
\end{figure}%
It produces reasonable results also in higher orders,
as is seen in Fig. \ref{BRII}.
\begin{figure}[htp!]
\begin{center}
\setlength{\unitlength}{1cm}
\begin{picture}(19,2)
\centerline{\scalebox{.5}[.5]{\includegraphics*{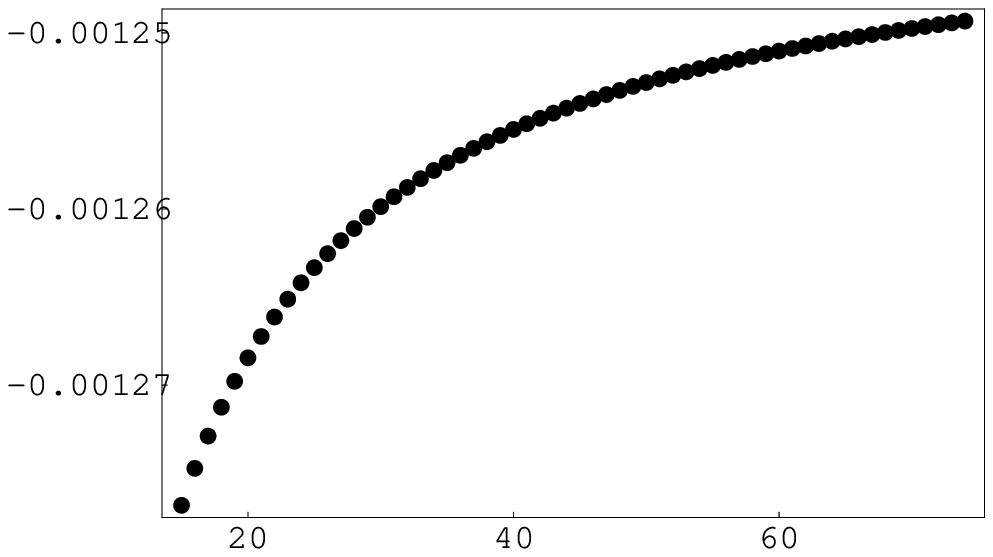}}}
\end{picture}
\caption[BRIII]{Deviation of $1-b^{(L)}_{0,{\rm plateau}}(\omega=0.843)$
from zero
 as a function of the
order $L$. Asymptotically the
value
$-.001136$ is reached, missing the
correct number by about $0.1\%$.}
\label{BRIII}
\end{center}
\end{figure}%
The
approximations appear
to converge rapidly. But
 the
limit
does not coincide with the
known exact value, although it happens to lie numerically quite close.
 Extrapolating the successive approximations
by an extremely accurate fit to
the
analytically known large-order
behavior \cite{HKSC} with a function
$b^{(L)}_{0,{\rm plateau}}(\omega=0.843)=A+B\,L^{-\kappa}$,
we find  convergence to $A=1-0.001136$,
which misses the
correct limit $A=1$.
The other two parameters are fitted best by $B=-0.002495$ and
$\kappa=0.922347$ (see
Fig.~\ref{BRIII}).

We may easily convince ourselves by numerical analysis that the
error
in the
limiting value is
indeed linked to the
failure of the
strong-coupling behavior
(\ref{BRAA3})
to have the
power structure
of
\cite{HKSC}.
For this purpose
we change
the
function $f(x)$ in equation (\ref{BRAA1.1}) slightly
into $f(x) \to \tilde f(x)=f(x)+1$, which makes the
integrals for $\tilde b_m$ in (\ref{BRAA2}) convergent.
The exact limiting value $1$ of $\tilde \Delta$ remaines unchanged,
but $\bar b_0^{(L)}$
acquires
now the
correct strong-coupling power structure
of \cite{HKSC}.
For this reason,
 we can easily verify
that the
application of
variational theory with a dynamical determination of $ \omega $
yields
the
correct strong-coupling limit $1$
with the
exponentially  fast convergence of the successive approximations
for $L\rightarrow \infty$ like
  $\bar b_0^{(L)} \approx 1-\exp{(-1.909-1.168~L)}$.

It is worthwhile emphasizing
that an escape
to complex zeros
which BR propose to
remedy the problems
of the $ \delta $-expansion is really of no help.
%
%
%
%
It has been claimed
\cite{BelGarNev} and
repeatedly cited
\cite{RULES}, that the
study of the
anharmonic oscillator in quantum mechanics suggests
the
use of complex extrema to optimize the
$ \delta $-expansion.
In particular, the
use
 of so-called {\em families\/}
of optimal candidates for the
variational parameter
$z$  has been suggested.
We are now going to show, that following these suggestions
one obtains bad  resummation results for the
anharmonic oscillator.
Thus we expect such procedures to lead to even worse results
in field-theoretic applications.

In quantum mechanical applications there
are
 no
 anomalous dimensions
in the
strong-coupling behavior of
the energy eigenvalues.
The
growth  parameters $\alpha$ and
 $\omega$ can be directly read off from
the Schr\"odinger equation;
they are
 $\alpha=1/3$ and
 $\omega=2/3$  for the
anharmonic oscillator
(see
Appendix A).
The
variational perturbation theory is applicable
for all couplings strengths
$g$
as long as $b_0^{(L)}(z)$ becomes stationary for a certain value of $z$.
For higher orders $L$
 it must
exhibit a well-developed
plateau. Within the
range of the
plateau, various derivatives of $b_0^{(L)}(z)$ with respect to $z$ will vanish. In addition there will be complex zeros with
small imaginary parts clustering around the
plateau. They are, however, of limited
use for designing an automatized
computer program
for localizing
 the
position of the
plateau. The study of several examples shows that plotting $b_0^{(L)}(z)$ for various values of
$\alpha$
and
 $\omega$ and
 judging visually  the
plateau is by far the
safest method, showing immediately
  which values of $\alpha$
and
 $\omega$
lead to  a    well-shaped
plateau.

\begin{figure}[htp!]
\begin{center}
\setlength{\unitlength}{1cm}
\begin{picture}(19,4.1)
\centerline{\scalebox{.65}[.65]{\includegraphics*{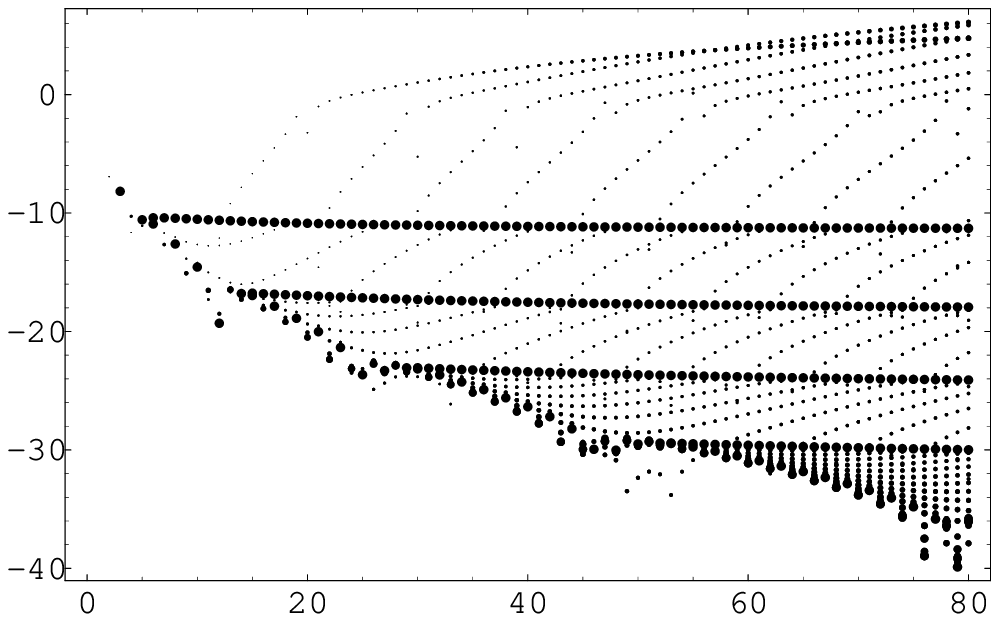}}}
\end{picture}
\caption[FigI]{Logarithmic error of the
leading strong-coupling coefficient $b_0^{(L)}$ of the
ground state energy of the
anharmonic oscillator with $x^4$ potential. The errors are plotted over the
order $L$ of the
variational perturbation expansion. At each order, all zeros of the
first derivative have been exploited. Only the
real parts of the
complex roots have been used
to evaluate $b_0^{(L)}$.
The fat points show
results from real zeros, the
smaller points
those
 from complex zeros,
 size is decreasing
with
distance from real axis.
}
\label{FigI}
\end{center}
\end{figure}

Let us review briefly the
properties
 of the
results obtained from
real and
 complex zeros of $\partial_z b_0^{(L)}(z)$ for the
anharmonic oscillator.
In Fig.~\ref{FigI}, the
logarithmic error of $b_0^{(L)}$ is plotted
versus the
order $L$. At each order, all zeros of the
first derivative
are
exploited. To test the rule
suggested in
 \cite{BelGarNev}, only the
real parts of the
complex roots have been used to evaluate $b_0^{(L)}$. The
fat points represent the
results of real zeros, the
thin points stem from the
real parts of complex zeros. It is readily seen that
the
real zeros give the
better result. Only by chance may a complex zero yield a
smaller error.
Unfortunately, there
is no rule to detect these accidental events.
Most complex zeros produce  large errors.

\begin{figure}[bt]
\begin{center}
\setlength{\unitlength}{1cm}
\begin{picture}(19,8)
\centerline{\scalebox{.75}[.75]{\includegraphics*{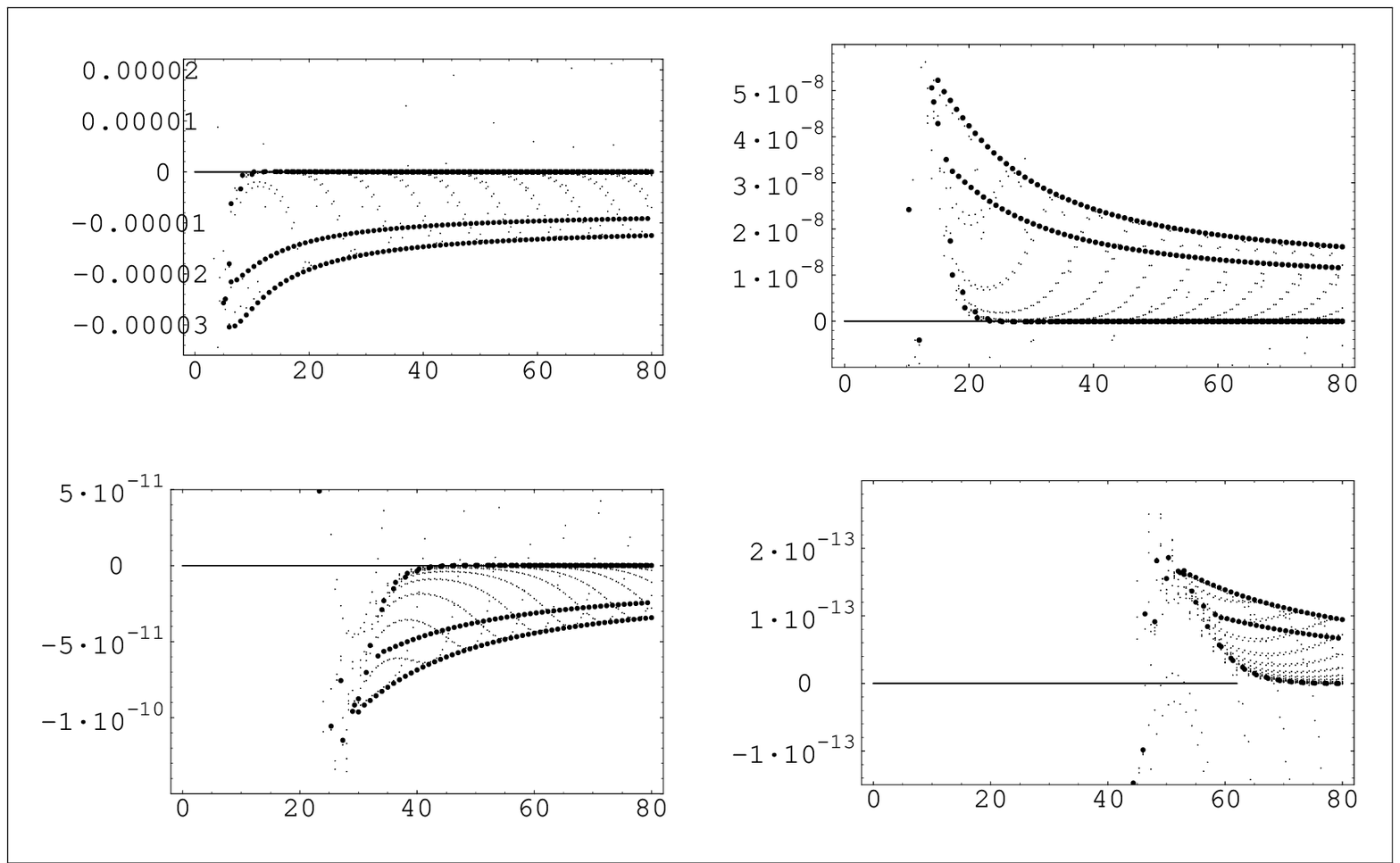}}}
\end{picture}
\caption[FgII]{Deviation of the
coefficient $b_0^{(L)}$ from the
exact value is shown as a function of perturbative order $L$ on a linear scale. As before, fat
 dots represent real zeros. In addition to Fig.~\ref{FigI}, the
results obtained from zeros of the
second derivative of $b_0^{(L)}$ are shown.
They give rise to own families with smaller errors by about 30\%.
At $N=6$, the upper left plot shows
 the start of two families
belonging to the first and second
derivative of $b_0^{(L)}$, respectively.
The deviations of both families
are negative.
 On the upper
right-hand figure, an enlargement visualizes the
next two families starting at $N=15$. Their deviations are
positive. The
bottom row shows two more enlargements
of
families starting at $N=30$ and
 $N=53$, respectively. The deviations  alternate again
in sign.
}
\label{FgII}
\end{center}
\end{figure}

We observe the
existence of families described in detail in the
textbook \cite{PI} and
 rediscovered in Ref.~\cite{BelGarNev}. These families start at about $N=6,~15,~30,~53$,
 respectively. But each family fails to converge to the
correct result. Only a
sequence of selected members
in each family
leads to an exponential convergence.
 Consecutive families alternate
around the
correct result, as can be seen more clearly in a plot of the
deviations of $b_0^{(L)}$
 from their $L\rightarrow  \infty$ -limit in Fig.~\ref{FgII}, where values derived from the
zeros of the
second derivative of $b_0^{(L)}$ have been included.
These give rise to accompanying families of similar behavior,
deviating with the same sign pattern from the exact result,
but lying closer to the
correct result by about 30\%.
~\\[.6cm]
\comment{
\section{Appendix B: }
In order to gain further insight into the
working of the variational resummation procedure, we apply it to the simple
test
function
\begin{align}
\label{E1}
f(x)=(1+x)^\alpha=x^\alpha \Big(1+\frac{1}{x}\Big)^\alpha
\end{align}
with
 weak coupling coefficients $a_n=\binom{\alpha}{n}$
and
a leading strong-coupling behavior $f\sim x^\alpha(1+{\alpha}/{x}+\dots)$,
so that
 $b_0=1$.
Inserting this information into
From the variational perturbation treatment of the weak-coupling expansion
of $f(x)$, we find
the
$L$th-order approximation to the
leading coefficient $b_0$ \cite{PI}:
\begin{align}
\label{B0}
b_0^{(L)}(z) = z^{-\alpha}\;\sum_{l=0}^L\; a_l\;z^l
 \binom{L-l+(l-\alpha)/\omega}{L-l} \,,
\end{align}
where
the
$z\equiv g/ \Omega ^{1/\alpha}$ is
the variational parameter to be optimized
for minimal
sensitivity on $z$.
Inserting the expansion coefficients of $f(x)$,
 we obtain
the variational
leading coefficient to $L$th order:
\begin{align}
\label{E2}
b_0^{(L)}(z)=\sum_{l=0}^L\binom{\alpha}{l}\binom{L-\alpha}{L-l}\,z^{l-\alpha},
\end{align}
which is easily transformed into the
expression:
\begin{align}
\label{E3}
b_0^{(L)}(z)=\binom{\alpha}{L}\sum_{l=0}^L\binom{L}{l}\,\frac{L-\alpha}{l-\alpha}\,(-1)^{L+l}z^{l-\alpha}
\end{align}
Determining the
variational parameter $z$ according to the
principle of minimal sensitivity requires a well developed
plateau of $b_0^{(L)}$ as a function of $z$.
For the
simple test function,
 the derivative $\partial_z b_0^{(L)}(z)$ can be obtained in the closed form:
\begin{align}
\label{E4}
\frac{d}{dz}b_0^{(L)}(z) &=(-1)^{L+1}\frac{L-\alpha}{z^{\alpha+1}}\binom{\alpha}{L}\sum_{l=0}^L(-z)^l\binom{L}{l}\\
 &=(-1)^{L+1}(L-\alpha)\binom{\alpha}{L}\frac{(1-z)^L}{z^{\alpha+1}}~.
\end{align}
This exhibits a flat plateau around
$z=1$ if the order $L$ is much larger than $\alpha$. An equally flat
 plateau is found for $b_0^{(L)}(z)$.
The
value of the
leading strong coupling coefficient $b_0^{(L)}$ at the
plateau is
\begin{align}
\label{E5}
b_0^{(L)}(1)=\binom{\alpha}{L}\sum_{l=0}^L\binom{L}{l}\,\frac{L-\alpha}{l-\alpha}\,(-1)^{L+l}=1,
\end{align}
in perfect agreement with the exact result, thus
confirming the
applicability of the
resummation scheme for this class of problems.
}

\section{Appendix B:
Ground-State Energy from Imaginary Part}
We determine the ground state energy function $E_0(g)$ for the anharmonic oscillator on the cut, i.e. for $g<0$ in the bubble region, from the weak coupling coefficients $a_l$ of equation (\ref{WEAK}). The behavior of the $a_l$ for large $l$ can be cast into the form
\begin{align}
a_l/a_{l-1}=-\sum_{j=-1}^L \beta_j\ l^{-j}\ .
\label{B1}
\end{align}
The $\beta_j$ can be determined by a high precision fit to the data in the large $l$ region of $250<l<300$ to be
\begin{align}
\label{B2}
\beta_{-1,\ 0,\ 1,\ \dots}=&\left\{\ 3,\  -\frac{3}{2} ,\ \frac{95}{24},
  \ \frac{113}{6},\ \frac{391691}{3456},
  \ \frac{40783}{48},\ \frac{1915121357}{248832},
 \  \frac{10158832895}{124416},
 \  \frac{70884236139235}{71663616}, \right. \\& \nonumber
 \quad\  \frac{60128283463321}{4478976},
 \  \frac{286443690892}{1423},
 \  \frac{144343264152266}{43743},
 \  \frac{351954117229}{6},
 \  \frac{2627843837757582}{2339},\\& \nonumber \left.
 \quad  \frac{230619387597863}{10},
 \  \frac{12122186977970425}{24},
 \  \frac{41831507430222441029}{3550},\ \dots \right\}\ ,
\end{align}
where the rational numbers up to $j=6$ are found to be exact, whereas the higher ones are approximations.\\
Equation (\ref{B1}) can be read as recurrence relation for the coefficients $a_l$. Now we construct an ordinary differential equation for $E(g):=E_{\rm 0,weak}^{(L)}(g)$ from this recurrence relation and find:
\begin{align}
\left[\left(g \frac{d}{dg}\right)^L+g\sum_{j=0}^{L+1}\beta_{L-j} \left(g \frac{d}{dg}+1\right)^j \right]E(g)=0\ .
\label{B3}
\end{align}
All coefficients being real, real and imaginary part of $E(g)$ each have to satisfy this equation separately.
The point $g=0$, however,  is not a regular point. We are looking for a solution, which is finite when approaching it along the negative real axis. Asymptotically $E(g)$ has to satisfy $E(g)\simeq \exp{(1/g\beta_{-1})}= \exp{(1/3g)}$. Therefore we solve (\ref{B3}) with the ansatz
\begin{align}
E(g)=g^\alpha \ \exp{\left( \frac{1}{3g} -\sum_{k=1} b_k (-g)^k \right)}
\label{B4}
\end{align}
to obtain $\alpha = -1/2$ and
\begin{align}
\label{B5}
b_{1,2,3,\dots}=&\left\{\; \frac{95}{24},\ \frac{619}{32}\right.
     ,\ \frac{200689}{1152},
  \ \frac{2229541}{1024} ,
  \ \frac{104587909}{3072},
  \ \frac{7776055955}{12288}  ,
  \ \frac{9339313153349}{688128},
 \ \frac{172713593813181}{524288},\\ & \nonumber
     \quad\ \frac{1248602386820060039}
   {139886592},\ \frac{
       14531808399402704160316631}{
       54391637278720},
 \ \frac{12579836720279641736960567921}
   {1435939224158208},\\ & \nonumber
\left. \quad \frac{109051824717547897884794645746723}{348951880031797248},\ \frac{45574017678173074497482074500364087}{3780312033677803520}\ \dots
\right\}\ .
\end{align}
This is in agreement with equation (\ref{ZJ1}) and an improvement compared to the WKB results of \cite{ZINNJ}. Again, the first six rational numbers are exact, followed by approximate ones.
\section{Appendix C:
First-Order
Differential Equations for  $E_n(g)$
}
\noindent
Given a one-dimensional quantum system
\begin{align}
(H_0+g\ V)|n,g\rangle=E_n(g)|n,g\rangle
\label{A1}
\end{align}
with Hamiltonian $H=H_0+g\ V$, eigenvalues $E_n(g)$ and eigenstates $|n,g\rangle$ we consider an infinitesimal increase $dg$ in the coupling constant $g$. The eigenvectors will undergo a small change:
\begin{align}
|n,g+dg\rangle=|n,g\rangle+dg\ \sum_{k\ne n}u_{nk}|k,g\rangle
\label{A2}
\end{align}
so that
\begin{align}
\frac{d}{dg}|n,g\rangle=\sum_{k\ne n}u_{nk}|k,g\rangle\ .
\label{A3}
\end{align}
Given this, we take the derivative of (\ref{A1}) with respect to $g$ and multiply by $\langle m,g|$ from the left to obtain:
\begin{align}
\langle m,g|V-E'_n(g)|n,g\rangle=\sum_{k\ne n}u_{nk}\langle m,g|H_0+g\ V-E_n(g)|k,g\rangle\ .
\label{A4}
\end{align}
Setting now $m=n$ and $m \ne n$ in turn, we find:
\begin{align}
E'_n(g)=&V_{nn}(g)
\label{A5}\\
V_{mn}(g)=&u_{nm}\ \left(E_m(g)-E_n(g)\right)\ ,
\label{A6}
\end{align}
where $V_{mn}(g)=\langle m,g|V|n,g\rangle$.\\
Equation (\ref{A5}) governs the behavior of the eigenvalues as functions of the coupling constant $g$. In order to have a complete system of differential equations, we must also determine how the  $V_{mn}(g)$ change, when $g$ changes. With the help of equations (\ref{A3}) and (\ref{A6}), we obtain:
\begin{align}
V'_{mn}=&\sum_{k\ne m}u^*_{mk}\langle k,g|V|n,g\rangle + \sum_{k\ne n}u_{nk}\langle m,g|V|k,g\rangle
\label{A7} \\
V'_{mn}=& \sum_{k\ne m} \frac{V_{mk}V_{kn}}{E_m-E_k} +  \sum_{k\ne n} \frac{V_{mk}V_{kn}}{E_n-E_k} \ .
\label{A8}
\end{align}
Equations  (\ref{A5}) and (\ref{A8}) together describe a complete set of differential equations for the energy eigenvalues $E_n(g)$ and the matrix-elements $V_{nm}(g)$. The latter determine via (\ref{A6}) the expansion coefficients $u_{mn}(g)$. Initial conditions are given by the eigenvalues $E_n(0)$ and the matrix elements  $V_{nm}(0)$ of the unperturbed system.

\bibliographystyle{elsarticle-num}



\bibliographystyle{unsrt}

\bibliography{Weniger}

\begin{thebibliography}{9}\setlength{\itemsep}{0mm}





\bibitem{EFFAC}
J. Goldstone, A. Salam, and S. Weinberg, Phys. Rev. {\em 127\/}, 965 (1962).

\bibitem{CDED}
C. De Dominicis. J. Math. Phys. {\bf 3}, 983 (1962).

\bibitem{FEY1}
{R.P. Feynman}, {\it Statistical Mechanics\/},
Benjamin, Reading, 1972.


\bibitem{FK}
{R.P. Feynman} and H. Kleinert,
Phys. Rev. A {\bf 34}, 1986
({\tt http//physik.fu-berlin.de/\~{}kleinert/159/159.pdf}).

\bibitem{KLEX}
H. Kleinert
Phys. Lett. A {\bf 173}, 332 (1993)
({\tt
http://physik.fu-berlin.de/\~{}kleinert/213}).


\bibitem{PI}
H. Kleinert, \emph{Path Integrals in Quantum Mechanics,
     Statistics, Polymer Physics, and Financial Markets},
World Scientific, Singapore, 2009
({\tt http://www.phy\-sik.fu-ber\-lin.de/\~{}klei\-nert/b5}).
%

\bibitem{KBEY}
H. Kleinert,
Annals of Physics {\bf 266}, 135 (1998)
({\tt
http://physik.fu-berlin.de/\~{}kleinert/255/255.pdf}).


\bibitem{LIPA}
J.A. Lipa, D.R. Swanson, J.A. Nissen, T.C.P. Chui and
U.E. Israelsson, Phys. Rev. Lett. |bf 76, 944 (1996);
J.A. Lipa, D. R. Swanson,
J.A. Nissen, Z.K. Geng, P.R. Williamson, D.A. Stricker, T.C.P. Chui,
U.E.  Israelsson and M. Larson, Phys. Rev. Lett. |bf 84, 4894 (2000).

\bibitem{HK7L}
H. Kleinert,
Phys. Rev. D {\bf 60},  085001 (1999) (hep-th/9812197) (hep-th/9812197);
Phys.Lett. A {\bf 277},  205
(2000) (cond-mat/9906107).
\bibitem{STE}
{P.M.~Stevenson}, Phys.~Rev.~D {\bf 30}, 1712 (1985);
D {\bf 32}, 1389 (1985);
{P.M.~Stevenson} and
{R.~Tarrach}, Phys.~Lett.~B {\bf 176}, 436 (1986).


\bibitem{JK}
W. Janke, H. Kleinert,
Phys.\ Rev.\ Lett.\ {\bf 75}, 2787 (1995) (quant-ph/9502019).

\bibitem{EJW}
E.J. Weniger,
Phys. Rev. Lett. {\bf 77},
2859 (1996); see also
{F.~Vinette} and
{J.~\v{C}\'{\i}\v{z}ek},
J.~Math.~Phys.~{\em 32}, 3392 (1991).




\bibitem{KJ2}
H~Kleinert and W.~Janke,
   Phys.\ Lett.\ {\bf A 206}, 283 (1995) (quant-ph/9509005).

\bibitem{GKS}
R. Guida, K. Konishi, H. Suzuki, Ann. Phys. (N.Y.) {\bf 249}, 109 (1996).

\bibitem{HKSC}
H. Kleinert,
Phys. Rev. D {\bf 57}, 2264 (1998)
(cond-mat/9801167);
Phys. Rev. D {\bf 58}, 107702 (1998)
	    (cond-mat/9803268).



\bibitem{HK1}
B. Hamprecht and H. Kleinert,
Phys. Rev. D {\bf 68}, 065001 (2003)
(hep-th/0302116).


\bibitem{ODM}

R. Seznec and J. Zinn-Justin, J. Math. Phys. {\bf 20}, 1398 (1979).
This paper has developed
important techniques for understanding
the convergence
mechanism of variational perturbation theory.

\bibitem{KS}
H. Kleinert and V. Schulte-Frohlinde,
\emph{Critical Phenomena in $\Phi ^4$-Theory},
     World Scientific, Singapore, 2001
({\tt http://www.phy\-sik.fu-ber\-lin.de/\~{}klei\-nert/b8}).


\bibitem{Resumm}


 {J. Zinn-Justin}, {\it Quantum Field Theory and
Critical Phenomena\/},
             Clarendon, Oxford, 1989.


\bibitem{rem1}
See Section 19.4 in \cite{KS}.

\bibitem{Wegner}
{F.J. Wegner}, Phys. Rev. B {\bf 5}, 4529 (1972); B {\bf 6}, 1891 (1972).



\bibitem{NICK}

B.G. Nickel,
D.I. Meiron, and G.B. Baker, Univ. of Guelph preprint 1977
(unpublished). The preprint is readable on the WWW at
{\tt http://www.physik.fu-berlin.de/\~{}kleinert/nickel/guelph.pdf};\\
The results are cited and used in Chapters 19 and 20.
They were axtended to seven loops
by\\
D.B. Murray and B.G. Nickel,
Univ. of Guelph preprint 1991. \\
The additional $g^7$ coefficients
of the renormalization group functions are listed in
Section 20.4 of
the textbook \cite{KS}.


\bibitem{rem2}
See Section 20.2 in \cite{KS}.



\bibitem{HK7L}
H. Kleinert,
Phys. Rev. D {\bf 60}, 085001 (1999)
(hep-th/9812197).
See Fig. 9.


\bibitem{gordon} M. Holzmann, G. Baym, J.-P. Blaizot and F. Lalo\"e,
Phys. Rev. Lett. {\bf 87}, 120403 (2001).

\bibitem{second} P. Arnold, G. Moore and B. Tom\'{a}sik, cond-mat/0107124.

\bibitem{zj} J. Zinn-Justin, {\it Quantum Field Theory and Critical Phenomena}
(Oxford University Press, Oxford, England, 1996).

\bibitem{st}
M. Bijlsma and H. T. C. Stoof,
Phys. Rev. A {\bf 54}, 5085 (1996)




\bibitem{ramospr}
F.F. de Souza Cruz, M.B. Pinto and R.O. Ramos,
Phys. Rev. A {\bf 65},
 053613
 (2002)
(cond-mat/0112306).


\bibitem{braat}
E. Braaten, and E. Radescu,
(cond-mat/0206186).

\bibitem{braatexp}
These are the results of \cite{braat}.
They differ from those of
 Refs.~\cite{ramospr} which are
$a_3\! =\! 0.644519,
a_{41} \!=\! 0.87339,
a_{42}\! = \!3.15905,
a_{43} \!= \!1.70959,
a_{44} \!= \!4.4411,
a_{45} \!= \!2.37741.$
The coefficients of the series
(\ref{@fexpp})
to be resummed
differ mainly in the last term:
$f_{-1}=
-126.651\,10^{-4}
,f_0=0,
f_1\!=-4.04857 \, 10^{-4},
f_2\!=  2.40587\,{10}^{-4},
f_3\!=-2.06849\,{10}^{-4}.$


\bibitem{SC3}
H. Kleinert,
{\em Strong-Coupling Behavior of Phi$^4$-Theories and Critical Exponents\/},
Phys. Rev. D {\bf 57 },  2264 (1998);
Addendum: Phys. Rev. D {\bf 58 },  107702 (1998) (cond-mat/9803268);
 {\em Seven Loop Critical Exponents from Strong-Coupling
$\phi^4$-Theory in Three Dimensions\/},
Phys. Rev. D {\bf 60 },  085001 (1999) (hep-th/9812197);
{\em Theory and Satellite Experiment
on Critical Exponent alpha of Specific Heat in
Superfluid Helium\/}\\
Phys. Lett. A
  {\bf 277}, 205 (2000)
 (cond-mat/9906107).

\bibitem{SCE}
H. Kleinert,
{\em Strong-Coupling $\phi^4$-Theory in $4- \epsilon$ Dimensions,
and Critical
Exponent\/},
Phys. Lett. B {\bf  434 },  74 (1998) (cond-mat/9801167);
{\em Critical Exponents without beta-Function\/},
Phys. Lett. B {\bf463}, 69 (1999) (cond-mat/9906359).

\bibitem{ubi}
See
\cite{ramospr,braat,prb}
and references cited there.

\bibitem{rem}
With
standard  normalization conditions
used in the 3-dimensional $\phi^4$-theory, the approach to scaling is governed  by
Wegner's exponent
$ \omega $
 (see \cite{SC3}).
The present definition of $m$ differs from the
 inverse correlation length $m=\xi^{-1}$ by
a factor: $m=mZ_\phi^{-1}\propto m\,m^{-\eta/2}$ for $m\rightarrow 0$.
This changes the exponent of approach to $ \omega '= \omega /(1- \eta /2)$.
I thank B. Kastening for
noting this.



\bibitem{arnold1} P. Arnold and G. Moore, Phys. Rev. Lett. {\bf 87},
120401 (2001); Phys. Rev. E {\bf 64}, 066113 (2001).
The authors derive a $1/N$ correction factor $(1-0.527/N)$
to the leading $N\rightarrow \infty$ result.

\bibitem{russos} V.A.  Kashurnikov, N.V. Prokof'ev and B.V. Svistunov,
Phys. Rev. Lett. {\bf 87}, 120402 (2001).

\bibitem{baymprl}
G. Baym, J.-P. Blaizot M. Holzmann, F. Lalo\"e
and D. Vautherin, Phys. Rev. Lett. {\bf 83}, 1703 (1999).

\bibitem{baymN} G. Baym, J.-P. Blaizot and J. Zinn-Justin, Europhys.
Lett. {\bf 49}, 150 (2000).

\bibitem{arnold} P. Arnold and B. Tom\'{a}sik, Phys. Rev. {\bf A62},
063604 (2000).
This paper starts out  from the 3+1-dimensional
initial
theory and derives from it the
three-dimensional effective classical
field theory, the field-theoretic generalization of
the quantum-mechanical effective classical potential
of
 R.P Feynman and H. Kleinert,
     Phys.~Rev.~ A {\bf  34}, 5080 (1986).
 This reduction program was started
for the Bose-Einstein gas
 by
{A.M.J. Schakel}, Int. J. Mod. Phys. B {\bf 8}, 2021
(1994);
 J. Mod. Phys. B {\bf 8}, 2021 (1994);
{\em Boulevard of Broken Symmetries\/},
Habilitationsschrift, FU-Berlin, (cond-mat/9805152) (1998).
 Unfortunately, Schakel
did not go beyond the one-loop level
so that he was happy to have found
a positive shift
$ \Delta T_c/T_c$, and did see
the cancellation
at the two-loop level.
See his recent paper
in J. Phys. Stud. {\bf 7},
 140
 (2003)
(cond-mat/0301050).


\bibitem{prb}
F.F. de Souza Cruz, M.B. Pinto and R.O. Ramos,
Phys. Rev. B {\bf 64}, 014515 (2001).


\bibitem{Grue}
{P. Grueter}, {D. Ceperley}, {F. Laloe},
Phys. Rev. Lett. {\bf 79}, 3549 (1997)
(cond-mat/9707028).

\bibitem{omN}
See  Eq.~(20.23) in the textbook
\cite{KS} or
{S.E. Derkachov}, {J.A. Gracey}, and {A.N. Manashov},
Eur. Phys. J. C {\bf 2}, 569 (1998)
(hep-ph/9705268).




\bibitem{HKM}
  W. Janke and H. Kleinert,
    Phys.\ Lett.\ A {\bf  117}, 353 (1986)
{\tt http://www.phy\-sik.fu-ber\-lin.de/\~{}klei\-nert/133});
Phys. Rev. Lett. {\bf 58}, 144 (1986).
H. Kleinert, Phys. Lett. A 257, 269 (1999)
(cond-mat/9811308);
M. Bachmann, H. Kleinert, A. Pelster,
 Phys. Lett. A {\bf 261}, 127 (1999)
(cond-mat/9905397);
Physical Review E 63, 051709/1-10 (2001)
(cond-mat/0011281); see also
B. Kastening,
Phys.Rev. A {\bf 68},
 061601
(2003) (cond-mat/0303486);
 Phys.Rev. A {\bf 69}, 043613 (2004)
(cond-mat/0309060);
 Phys. Rev. E {\bf 73}, 011101 (2006)
(cond-mat/0508614);
 Phys. Rev. A 70, 043621 (2004)
(cond-mat/0406035).





\bibitem{Langer}

{J.S. Langer}, Ann. Phys. {\bf 41}, 108 (1967).



%
\comment{
\bibitem{refs}
{R.~Seznec} and
{J.~Zinn-Justin}, J.~Math.~Phys.~{\bf 20},
   1398 (1979);
{T.~Barnes} and
{G.I.~Ghandour}, Phys.~Rev.~D {\bf 22},
  924 (1980);
{B.S.~Shaverdyan} and
{A.G.~Usherveridze},
 Phys.~Lett.~B {\bf 123}, 316 (1983);
{K. Yamazaki},
J. Phys. A {\bf 17}, 345 (1984);
{H. Mitter} and
{K. Yamazaki}, J. Phys. A {\bf 17}, 1215 (1984);
{P.M.~Stevenson}, Phys.~Rev.~D {\bf 30}, 1712 (1985);
D {\bf 32}, 1389 (1985);
{P.M.~Stevenson} and
{R.~Tarrach}, Phys.~Lett.~B {\bf 176}, 436 (1986);
{A.~Okopinska}, Phys.~Rev. D {\bf 35}, 1835 (1987);
D {\bf 36}, 2415 (1987);
{W.~Namgung},
{P.M.~Stevenson}, and
{J.F.~Reed},
 Z.~Phys.~C {\bf 45}, 47 (1989);
{U.~Ritschel}, Phys.~Lett.~B {\bf 227}, 44 (1989);
    Z.~Phys.~C {\bf 51}, 469 (1991);
{M.H.~Thoma}, Z.~Phys.~C {\bf 44}, 343 (1991);
{I.~Stancu} and
{P.M.~Stevenson},
  Phys.~Rev.~D {\bf 42}, 2710 (1991);
{R.~Tarrach}, Phys.~Lett.~B {\bf 262}, 294 (1991);
{H.~Haugerud} and
{F.~Raunda}, Phys.~Rev.~D {\bf 43},
  2736 (1991);
{A.N.~Sissakian},
{I.L.~Solovtsov}, and
 {O.Y.~Shevchenko},
  Phys.~Lett.~B {\bf 313}, 367 (1993).
%
\bibitem{finiteg}
  {I.R.C.~Buckley},
 {A.~Duncan},
{H.F.~Jones}, Phys.~Rev.~{\bf D 47},
    2554 (1993);
{C.M.~Bender},
{A.~Duncan},
{H.F.~Jones},
 Phys.~Rev.~{\bf D 49},
    4219 (1994).
}
\comment{
\bibitem{KleinertJanke}
 { H.~Kleinert} and
{ W.~Janke},
      {\em Convergence behavior of variational perturbation
       expansion. \\A method for locating Bender-Wu singularities\/},
      Phys.~Lett.~A {\bf  206}, 283 (1995) (quant-ph/9509005).
\bibitem{Guida}
{R.~Guida}, {K. Konishi}, and
{H. Suzuki},
Annals Phys. {\bf 249}, 109 (1996)
(hep-th/9505084).
}
%
\comment{
}

%
%
\comment{
\bibitem{AntiBraaten}
{B. Hamprecht}
and {H. Kleinert},
{\em Dependence of Variational Perturbation Expansions
 on Strong-Coupling Behavior.
Inapplicability of $ \delta $-Expansion to Field Theory.\/}, (hep-th/0302116).
}
%

\comment{
\bibitem{JK}
 W.~Janke and H.~Kleinert,
    {\em Convergent Strong-Coupling Expansions from Divergent
      Weak-Coupling Perturbation Theory\/}\\
   Phys.\ Rev.\ Lett.\ {\bf 75}, 2787 (1995) (quant-ph/9502019)
}
%
%
 \comment{
\bibitem{strong}
H. Kleinert,
{\em Strong-Coupling Behavior of Phi$^4$-Theories and
Critical Exponents\/},
Phys. Rev. D {\bf 57},  2264 (1998);
Addendum: Phys. Rev. D {\bf 58},  107702 (1998) (cond-mat/9803268);
 {\em Seven Loop Critical Exponents from Strong-Coupling
$\phi^4$-Theory
 in Three Dimensions\/},
Phys. Rev. D {\bf 60},  085001 (1999) (hep-th/9812197);
H. Kleinert,
{\em Strong-Coupling $\phi^4$-Theory
 in $4- \epsilon$ Dimensions,
and
Critical
Exponent\/},
Phys. Lett. B {\bf  434},  74 (1998) (cond-mat/9801167).
}

\comment{
\bibitem{JKsig}
This was proved
in
 W.~Janke
and H.~Kleinert,
 {\em Scaling property of variational perturbation expansion
for general anharmonic oscillator with $x^P$-potential\/},
Phys.~Lett.~A {\bf 199},  287 (1995) (quant-ph/9502018)
 for $p/q=1$ (see also the textbook \cite{PI}, Appendix 5A),
but  can easily be generalized to hold for arbitrary $p$
and $q$.
}


%


 \comment{
%
\bibitem{Wegner}
{F.J. Wegner}, Phys. Rev. B {\bf 5}, 4529 (1972); B {\bf 6}, 1891 (1972).
%
\bibitem{Lipa}
{J.A. Lipa},
 {D.R. Swanson}, {J. Nissen}, {Z.K. Geng},
 {P.R. Williamson},
{D.A. Stricker},
{T.C.P. Chui}, {U.E. Israelson}, and
{M. Larson},
{Phys. Rev. Lett.} {\bf 84},
4894 (2000);
H. Kleinert,
 {\em Seven Loop Critical Exponents from Strong-Coupling
$\phi^4$-Theory
 in Three Dimensions\/},
Phys. Rev. D {\bf 60 },  085001 (1999) (hep-th/9812197);
H. Kleinert
{\em Theory and
Satellite Experiment
on Critical Exponent alpha of Specific Heat
 in
Superfluid Helium\/}\\
FU-Berlin preprint 1999 (cond-mat/9906107)
\bibitem{PI2}
H.Kleinert, {\em Five-Loop Critical Temperature Shift in
Weakly Interacting
Homogeneous  Bose-Einstein Condensate\/},
and references quoted in this paper.
%
%
\bibitem{C1MC}
{P.Grueter, D.Ceperley, F.Laloe, Phys. Rev. Lett. {\bf 79}, 3549 (1997).}
%
\bibitem{RenGroup}
{ D.J.Amit,
{\it Field Theory, the Renormalization Group
and Critical Phenomena}, McGRaw-Hill,1978}
%
%
%
\bibitem{LET}
{B. Hamprecht}
and {H. Kleinert},
{\em Summing Non-Borel Tunneling Amplitudes
by Variational Perturbation Theory.
\/}, (hep-th/0302124).
%
\bibitem{HaPe}
W. Janke, A. Pelster, H.-J. Schmidt, and M. Bachmann (eds.),
{\it Fluctuating Paths
and Fields}, (World Scientific,Singapore, 2001), pg. 347.
%
}




\comment{
\bibitem{AVR}
{J.E. Avron}, {I.W. Herbst}, {B. Simon}, Phys. Rev. A
{\bf 20\/}, 2287 (1979).
%
\bibitem{wunner}
H.~Ruder, G.~Wunner, H.~Herold, and F.~Geyer, {\em Atoms in Strong Magnetic Fields\/}
(Springer-Verlag, Berlin, 1994).
%
}









\bibitem{BEWU}
 C.M.~Bender and T.T.~Wu, Phys.~Rev.~{\bf 184}, 1231 (1969);

\bibitem{Coleman}

    {S. Coleman}, Nucl. Phys. B {\bf 298}, 178 (1988).

\bibitem{Neveu}
{{B.Bellet}, {P.Garcia}, {A.Neveu}, Int. J. of Mod. Phys. A {\bf 11}, 5587(1997)}
%
\bibitem{RULES}
{J.-L. Kneur}, {D. Reynaud}, (hep-th/0205133v2).
See also
\cite{Braaten},
\cite{ramospr}, \cite{prb}.
%
\bibitem{Braaten}
{{E. Braaten}, {E. Radescu}, (cond-math/0206186v1)}.
%
\comment{
\bibitem{Ramos}
F.F. de Souza Cruz, M.B. Pinto
and R. O. Ramos,
Phys. Rev. A {\bf 65},
 053613
 (2002)
(cond-mat/0112306);
Phys. Rev. B {\bf 64}, 014515 (2001);
{{J.-L. Kneur}, {M.B. Pinto}, {R.O. Ramos},
(cond-mat/0207295),  (cond-mat/0207295), Phys.Rev.Lett. 89 (2002) 210403}.
}
%
\bibitem{Tunn}
%
The low-order results were first obtained by\\
H. Kleinert,
     Phys.~Lett.~B {\bf 300}, 261 (1993)
({\tt http://www.physik.fu-berlin.de/\~{}kleinert/214}),   \\
and extended by\\
 R.~Karrlein
and H.~Kleinert,
    Phys.~Lett.~A {\bf 187}, 133 (1994) (hep-th/9504048).

\bibitem{ZINNJ}
{{J.~Zinn-Justin}, J.~Math~Phys. {\bf 22}(3), 511 (1981).}
The first 10 coefficients of expansion
(\ref{ZJ})
are calculated.

\bibitem{BelGarNev}
{B.Bellet, P.Garcia, A.Neveu, Int. J. of Mod. Phys. {\bf A11}, 5587(1997)}.
The family structure of optimal variational parameters
emphasized in this paper
was
 discussed in great detail earlier
in Chapter 5 of the textbook
\cite{PI}, but with correct
application rules.



\bibitem{BKP}
{M. Bachmann}, {H. Kleinert}, and {A. Pelster},
Phys. Rev. A {\bf 62\/}, 52509 (2000) (quant-ph/0005074),
Phys. Lett. A {\bf 279}, 23 (2001) (quant-ph/000510).



\bibitem{cizek1}
J.E.~Avron, B.G.~Adams, J.~\v{C}\'\i\v{z}ek, M.~Clay, M.L.~Glasser, P.~Otto, J.~Paldus,
and E.~Vrscay, Phys. Rev. Lett. {\bf 43}, 691 (1979).

\bibitem{landau}
L.D.~Landau and E.M.~Lifschitz, {\em Quantenmechanik}, Sechste Auflage
(Akademie-Verlag Berlin, 1979).


\bibitem{AVR}

{J.E. Avron}, {I.W. Herbst}, {B. Simon}, Phys. Rev. A
{\bf 20\/}, 2287 (1979).
See also the resummation treatment in
J.-C. Le Guillou, J. Zinn-Justin,
Ann. Phys. {\bf 147}, 57 (1983).

\bibitem{wunner}
H.~Ruder, G.~Wunner, H.~Herold, and F.~Geyer,
{\em Atoms in Strong Magnetic Fields\/}
(Springer-Verlag, Berlin, 1994).
%

\comment{\bibitem{Kas}
B. Kastening
(cond-mat/0303486).
%
\bibitem{FIVE}
H.\ Kleinert,
Mod.\ Phys.\ Lett.\ B {\bf 17}, No.\  19, 1011 (2003)
(cond-mat/0210162).
%
 \bibitem{density}
M.~Bachmann, H.~Kleinert, and A.~Pelster, Phys. Rev. A {\bf 60}, 3429 (1999)
(quant-ph/9812063).
}




%
\comment{
\bibitem{refs}
\aut{R.~Seznec} and
\aut{J.~Zinn-Justin}, J.~Math.~Phys.~{\bf 20},
   1398 (1979); \\
\aut{T.~Barnes} and
 \aut{G.I.~Ghandour}, Phys.~Rev.~D {\bf 22},
  924 (1980);\\
\aut{B.S.~Shaverdyan} and
\aut{A.G.~Usherveridze},
 Phys.~Lett.~B {\bf 123}, 316 (1983);\\
\aut{K. Yamazaki},
J. Phys. A {\bf 17}, 345 (1984);\\
\aut{H. Mitter} and
 \aut{K. Yamazaki}, J. Phys. A {\bf 17}, 1215 (1984);\\
\aut{P.M.~Stevenson}, Phys.~Rev.~D {\bf 30}, 1712 (1985);
D {\bf 32}, 1389 (1985);\\
\aut{P.M.~Stevenson} and
\aut{R.~Tarrach}, Phys.~Lett.~B {\bf 176}, 436 (1986);\\
\aut{A.~Okopinska}, Phys.~Rev. D {\bf 35}, 1835 (1987);
D {\bf 36}, 2415 (1987);\\
\aut{W.~Namgung},
\aut{P.M.~Stevenson}, and
\aut{J.F.~Reed},
 Z.~Phys.~C {\bf 45}, 47 (1989);\\
\aut{U.~Ritschel}, Phys.~Lett.~B {\bf 227}, 44 (1989);
    Z.~Phys.~C {\bf 51}, 469 (1991);\\
\aut{M.H.~Thoma}, Z.~Phys.~C {\bf 44}, 343 (1991);\\
\aut{I.~Stancu} and
\aut{P.M.~Stevenson},
  Phys.~Rev.~D {\bf 42}, 2710 (1991);\\
\aut{R.~Tarrach}, Phys.~Lett.~B {\bf 262}, 294 (1991);\\
\aut{H.~Haugerud} and
\aut{F.~Raunda}, Phys.~Rev.~D {\bf 43},
  2736 (1991);\\
\aut{A.N.~Sissakian},
\aut{I.L.~Solovtsov}, and
  \aut{O.Y.~Shevchenko},
  Phys.~Lett.~B {\bf 313}, 367 (1993).
%
\bibitem{Stevenson}
\aut{P.M.~Stevenson}, Phys.~Rev.~D {\bf 30}, 1712 (1985);
D {\bf 32}, 1389 (1985);
\aut{P.M.~Stevenson} and
\aut{R.~Tarrach}, Phys.~Lett.~B {\bf 176}, 436 (1986).
%
\bibitem{Hagen}
 H. Kleinert,
     {\em Path Integrals in Quantum Mechanics, Statistics
and Polymer Physics,\/}
     World Scientific, Singapore 1995,
     Second extended edition, pp. 1--850.
({\tt http://www.physik.fu-berlin.de/{}\~{}kleinert/b3}).
%
 \bibitem{finiteg}
  \aut{I.R.C.~Buckley},
\aut{A.~Duncan},
\aut{H.F.~Jones}, Phys.~Rev.~{\bf D 47},
    2554 (1993);\\
\aut{C.M.~Bender},
\aut{A.~Duncan},
\aut{H.F.~Jones},
 Phys.~Rev.~{\bf D 49},
    4219 (1994).
 %
%
%
%
%
\comment{
\bibitem{KleinertJanke}
   H.~Kleinert and
 W.~Janke,
      {\em Convergence behavior of variational perturbation
       expansion. A method for locating Bender-Wu singularities\/},
      Phys.~Lett.~{\bf A 206}, 283 (1995) (quant-ph/9509005).
\bibitem{Guida}
\aut{R.~Guida}, \aut{K. Konishi}, and
 \aut{H. Suzuki},
Annals Phys. {\bf 249}, 109 (1996)
(hep-th/9505084).
 \bibitem{strong}
H. Kleinert,
{\em Strong-Coupling Behavior of $\phi^4$-Theories and
 Critical Exponents\/},
Phys. Rev. D {\bf 57},  2264 (1998)
({\tt http://www.physik.fu-berlin.de/~kleinert/257a});
Addendum: Phys. Rev. D {\bf 58 },  107702 (1998) (cond-mat/9803268);
 {\em Seven Loop Critical Exponents from Strong-Coupling
$\phi^4$-Theory in Three Dimensions\/},
Phys. Rev. D {\bf 60 },  085001 (1999) (hep-th/9812197);
\bibitem{epsilon}
H. Kleinert,
{\em Strong-Coupling $\phi^4$-Theory in $4- \epsilon$ Dimensions,
and
 Critical
Exponent\/},
Phys. Lett. B {\bf  434},  74 (1998) (cond-mat/9801167).
}
%
%
\bibitem{Wegner}
\aut{F.J. Wegner}, Phys. Rev. B {\bf 5}, 4529 (1972); B {\bf 6}, 1891 (1972).
%
%
%
\bibitem{Verena}
{H.Kleinert, V.Schulte-Frohlinde
{\it Critical Properties of $\Phi^4$-Theories}, World Scientific, Singapore, 2001}.
%
\bibitem{Lipa}
\aut{J.A. Lipa},
 \aut{D.R. Swanson}, \aut{J. Nissen}, \aut{Z.K. Geng},
 \aut{P.R. Williamson},
\aut{D.A. Stricker},
\aut{T.C.P. Chui}, \aut{U.E. Israelson}, and
 \aut{M. Larson},
{Phys. Rev. Lett.} {\bf 84},
4894 (2000);
H. Kleinert,
 {\em Seven Loop Critical Exponents from Strong-Coupling
$\phi^4$-Theory in Three Dimensions\/},
Phys. Rev. D {\bf 60 },  085001 (1999) (hep-th/9812197);
H. Kleinert
{\em Theory and
 Satellite Experiment
on Critical Exponent alpha of Specific Heat in
Superfluid Helium\/},
Phys. Lett. A {\bf 277}, 205 (2000) (cond-mat/9906107).
%
%
%
\bibitem{Braaten}
{E.Braaten, E.Radescu, cond-math/0206108}.
%
\bibitem{BraatenBE}
{E.Braaten, E.Radescu, cond-math/0206186v1}.
%
\bibitem{Ramos}
F.F. de Souza Cruz, M.B. Pinto and R.O. Ramos,
Phys. Rev. A {\bf 65},
 053613
 (2002)
(cond-mat/0112306);
Phys. Rev. B {\bf 64}, 014515 (2001);
{\aut{J.-L. Kneur}, \aut{M. B. Pinto}, \aut{R. O. Ramos},
(cond-mat/0207295),  (cond-mat/0207295), Phys.Rev.Lett. 89 (2002) 210403}.
%
}
\comment{
\bibitem{fermions}
A similar program has been proposed
a long time ago for fermion systems
by\\
 P. Nozi\`eres,
 {\em Theory of Interacting Fermi Systems}
 (Benjamin, New York, 1964);\\
and
 recently been applied by \\
 A. Neumayr, W. Metzner,
{\em Renormalized perturbation theory for Fermi systems: Fermi surface deformation and
 superconductivity in the
two-dimensional Hubbard
    model\/},
(cond-mat/0208431);\\
There are also  rigorous convergence proofs in
this context by\\
 J. Feldman, J. Magnen, V. Rivasseau, and
 E. Trubowitz, in
 {\em The State of Matter}, edited by M. Aizenmann and
 H. Araki,
 Advanced Series in Mathematical Physics Vol.~20
 (World Scientific 1994);\\
 J. Feldman, H. Kn\"orrer, M. Salmhofer, and
 E. Trubowitz,
 J. Stat. Phys. {\bf 94}, 113 (1999).
\\
These authors do not realize, however,
 that the
limit will in general produce the
{\em wrong\/} result
in the
most interesting limit of strong coupling
(which governs the
 critical behavior),
as shown
with the
help of
simple models in the
present paper.
%
%
%
%
%
%
%
\bibitem{Hagen2}
{H.Kleinert,
{\em Five-Loop Critical Temperature Shift in Weakly Interacting
Homogeneous Bose-Einstein Condensate\/},
(cond-mat/0210162), and
 references quoted in this paper.}
%
%
\bibitem{C1MC}
{P.Grueter, D.Ceperley, F.Laloe, Phys. Rev. Lett. {\bf 79}, 3549 (1997).}
%
\bibitem{RenGroup}
{ D.J.Amit,
{\it Field Theory, the
Renormalization Group and
 Critical Phenomena}, McGRaw-Hill,1978}.
%
%
\bibitem{without}
H. Kleinert,
{\em Critical Exponents without $ \beta $-Function\/},
Phys. Lett. B {\bf463 }, 69 (1999) (cond-mat/9906359).
%
}

\comment{
\bibitem{rLipa}J.A. Lipa, D.R. Swanson, J. Nissen,
T.C.P. Chui, and U.E. Israelson, {Phys. Rev. Lett.} {\bf 76}, 944 (1996).
The number in Eq.~(\ref{@spaceshuttal}) is from a corrected analysis of the data
published in a remark in  Ref.~[15]
of J.A. Lipa,
 D.R. Swanson, J. Nissen, Z.K Geng, P.R. Williamson, D.A. Stricker,
T.C.P. Chui, U.E. Israelson, and M. Larson, {Phys. Rev. Lett.} {\bf 84},
4894 (2000).
See also the related papers by\\ {D.R. Swanson, T.C.P.Chui and J.A. Lipa,
{Phys. Rev B} {\bf 46}, 9043 (1992);
D. Marek, J.A. Lipa and D. Philips, {Phys. Rev B} {\bf 38}, 4465 (1988).  }
}

\bibitem{ahl}
G. Ahlers, Phys. Rev. A {\bf 3}, 696 (1971);
K.H. Mueller, G. Ahlers, F. Pobell,  Phys. Rev. B {\bf 14}, 2096 (1976);

\comment{\bibitem{kl}
H. Kleinert,
Phys. Rev. D {\bf 57}, 2264 (1998)
(www.physik.fu-berlin.de/\~{}kleinert/257);
Addendum:
ibid. D {\bf 58}, 1077 (1998) (cond-mat/9803268).
 (also available from
www.physik.fu-berlin.de/\~{}kleinert/klein\_re257).\\
Note that in the journal version,
the expansion for $ \eta _m=2- \nu ^{-1}$ in
Eq.~(61) of the first paper contains a misprinted
sign of
the $\hat g_0^2$-term, which must be alternating.
%
%
\bibitem{seven}
H. Kleinert,
{\em Seven Loop Critical Exponents from Strong-Coupling
$\phi^4$-Theory in Three Dimensions\/},
Phys.
Phys. Rev. D {\bf 60 },  085001 (1999)
(hep-th/9812197).
%
\bibitem{JanKl}
W. Janke and H. Kleinert,
{\em Convergent Strong-Coupling Expansions from Divergent
      Weak-Coupling Perturbation Theory\/},
 Phys. Rev. Lett. {\bf 75}, 2787 (1995)
(quant-ph/9502019).
%
\bibitem{nickel}
B.G.~Nickel, D.I.~Meiron, and G.A.~Baker, Jr., University of Guelph report,
 1977 (unpublished);\\
 G.A.~Baker, Jr., B.G.~Nickel, and D.I.~Meiron, Phys.\
   Rev. B {\bf 17}, 1365 (1978).
%
\bibitem{anton}
S.A. Antonenko and A.I. Sokolov, Phys. Rev. E {\bf 51}, 1894 (1995);
Fiz.~Tverd.~Tela (Leningrad) {\bf 40}, 1284 (1998) [Sov. Phys.\
Sol. State {\bf 40}, 1169 (1998)].
%
}
\bibitem{rgomuha}{L.S. Goldner, N. Mulders and G. Ahlers, J. Low Temp.Phys. 93
(1992) 131.}
%
\bibitem{GZ}
R.~Guida and J.~Zinn-Justin,
{\em Critical exponents of the $N$-vector model\/},
J. Phys. A {\bf 31}, 8130 (1998)
(cond-mat/9803240)




\bibitem{rLGZJi}J.C. Le Guillou and J. Zinn-Justin,
Phys. Rev. Lett. {\bf 39}, 95 (1977);
Phys. Rev. B {\bf 21}, 3976 (1980); J. de Phys. Lett {\bf 46}, L137 (1985).

\bibitem{KVSF}
 H. Kleinert and V. Schulte-Frohlinde,
J. Phys. A {\bf 34}, 1037 (2001)
(cond-mat/9907214).

\bibitem{MN}D.B. Murray and B.G. Nickel, unpublished.




\bibitem{rPelVic}{A. Pellisetto and E. Vicari, preprint IFUP-TH 52/97,
cond-mat/9711078.}


\bibitem{JKL}F. Jasch and H. Kleinert,
Berlin preprint 199  (cond-mat/9907214).

\bibitem{rJanke}{W. Janke, Phys.Lett. A148 (1990) 306.}



\bibitem{rBFMM}H.G. Ballesteros, L.A. Fernandez, V. Martin-Mayor and A. Munoz
Sudupe, Phys. Lett. B {\bf 387}, 125 (1996).


\bibitem{WOR}M. Ferer, M.A. Moore, and M. Wortis,
Phys. Rev. B {\bf 65}, 2668 (1972).

\bibitem{rBUCOM}P. Butera and M. Comi, {Phys. Rev.} B {\bf 56}, 8212 (1997) (hep-lat/9703018).




\end{thebibliography}

\end{document}